\documentclass[11pt]{article}
\usepackage[letterpaper,top=2cm,bottom=2cm,left=3cm,right=3cm,marginparwidth=1.75cm]{geometry}
\usepackage{hyperref}       
\usepackage{url}            
\usepackage{booktabs}      
\usepackage{authblk}
\usepackage{amsthm,amsmath,amsfonts, amssymb}
\usepackage{mathrsfs}
\usepackage{mathtools}
\usepackage{nicefrac}      
\usepackage{xcolor}        
\usepackage{natbib}
\usepackage{cancel}
\usepackage{graphicx}
\usepackage{caption}
\usepackage{booktabs}
\usepackage{subcaption}
\usepackage{makecell}
\usepackage{bm}
\usepackage{blindtext}
\usepackage{lineno}
\usepackage{multirow}
\usepackage{float} 
\usepackage{comment} 
\usepackage{algorithm}
\usepackage{algorithmic}
\usepackage{bbm}
\usepackage{subcaption}
\DeclareCaptionSubType * [alph]{table}
\DeclareCaptionLabelFormat{subtable}{#2.\quad}
\newcommand{\ab}[1]{{\leavevmode\color{black}#1}}

\RequirePackage[normalem]{ulem} 
\RequirePackage{color}\definecolor{RED}{rgb}{1,0,0}\definecolor{BLUE}{rgb}{0,0,1} 

\title{Spatio-temporal Hawkes point processes: statistical inference and simulation strategies}
\author[1]{Alba Bernabeu}{}{}
\author[1]{Jorge Mateu}{}{}

\affil[1]{Department of Mathematics, University Jaume I, Castellón, Spain 

bernabeu@uji.es; mateu@uji.es}

\begin{document}

\maketitle

\begin{abstract}


Spatio-temporal Hawkes point processes are a particularly interesting class of stochastic point processes for modeling self-exciting behavior, in which the occurrence of one event increases the probability of other events occurring. These processes are able to handle complex interrelationships between stochastic and deterministic components of spatio-temporal phenomena. However, despite its widespread use in practice, there is no common and unified formalism and every paper proposes different views of these stochastic mechanisms. With this in mind, we implement two simulation techniques and three unified, self-consistent inference techniques, which are widely used in the practical modeling of spatio-temporal Hawkes processes. Furthermore, we provide an evaluation of the practical performance of these methods, while providing useful code for reproducibility.

\end{abstract}

{\bf Keywords}: Acceptance-rejection; EM algorithm; Hawkes point processes; Self-exciting processes; Spatio-temporal point processes

\section{Introduction}

In the context of point processes, Hawkes processes, also known as self-exciting point processes, provide a powerful tool for modeling dependent events where clustered event sequences are observed (\citealt{reinhart2018review, bernabeu2024spatio}). The Hawkes model was introduced by Alan Hawkes in the early seventies (\citealt{hawkes1971spectra}), initially focused on one-dimensional frameworks, where the dependence structure was defined exclusively in terms of time. This initial approach allowed researchers to capture temporal dependencies between events, laying the groundwork for subsequent advancements in modeling dependent data. Unlike traditional models, where the probability of an event is influenced by external factors, a Hawkes process is a model with a self-exciting nature, where the occurrence of one event can stimulate and increase the probability of future events.

However, the need to extend these models to capture both spatial and temporal dependencies simultaneously has been recognized in recent decades. Spatio-temporal Hawkes models are considered important because they allow to capture complex dependencies amongst events while achieving accurate predictions, and thus enabling a more realistic representation of phenomena by incorporating both dimensions. Additionally, they can provide more detailed information about the dynamics of events, including how the occurrence of an event at a specific time and place can drive the occurrence of subsequent events in the same or nearby locations. For these reasons, Hawkes processes are widely used in a broad range of fields including economics (\citealt{bacry2014,Hawkes01022018, Chen03052022}), epidemiology (\citealt{CHIANG2022505, Schoenberg2023, alaimo2024bayesian}), seismology (\citealt{ebrahimian2017robust,rotondi2019,grimm2022}), criminology (\citealt{MOHLER2018431,alexmateu2023,isabel2024}) and social phenomena such as traffic accidents (\citealt{LI2018312, kalair2021nonparametric}), among others.

Hawkes processes have been the subject of study for decades (\citealt{worrall2022fifty}), and an increasing amount of research is being done on the topic nowadays. Despite the extensive literature, there is little consensus on simulation and estimation techniques, particularly in the context of spatio-temporal processes, as each study has focused adhoc on specific aspects related to its particular application or development. In \citet{bernabeu2024spatio}, an exhaustive and updated review of spatio-temporal Hawkes processes is presented, focusing on the simulation and estimation methods currently used in practice. In this article, we have selected some of the simulation and estimation methods with the aim of offering resources for their implementation in the \texttt{R} programming language, which can serve as a guide for future researchers. For the proposed methods, we also provide an explanation of the methodology used to address the various challenges that may arise during their implementation. Furthermore, this paper includes a simulation study where we apply each of the selected inference methods to compare them in terms of parameter estimation accuracy and computational cost. For this purpose, we explore different ways of modeling a Hawkes process to evaluate the adaptability of the methods based on the data structure. The \texttt{R} codes related to these simulation and estimation methods are available in the open Github repository {\small \url{https://github.com/Koroxal/Spatio-temporal-Hawkes-point-processes}}.

By simulating Hawkes processes, we can replicate the temporal and spatial dynamics of events, explore different hypothetical scenarios by fitting model parameters, validate inference methods through the building of synthetic data, study theoretical properties such as convergence, and generate datasets to test algorithms. We focus on two simulation methods that offer a practical and efficient approach to study and understand the behavior of spatio-temporal Hawkes processes in different contexts, the parents-offspring method and the acceptance-rejection method. However, alternative approaches exist in the literature for simulating Hawkes processes (\citealt{moller2005perfect, moller2006approximate, 10.1214/ECP.v18-2717, rizoiu2017tutorial}).

Regarding inference, we focus on parametric approaches, as we assume we know some characteristics of the underlying stochastic process and this allows us to provide a unified and general code for its application. When prior information about the data is available, parametric methods are preferred to non-parametric methods because they are less computationally complex, which translates into less computational time, especially when there is a large volume of data. In addition, they offer greater ease of interpretation of the results. Specifically, we address the methods of likelihood, Expectation-Maximization (EM) and Bayesian inference using INLA. For information on non-parametric methods, which may be more suitable for complex and unstructured data beyond the scope of this study, the reader is referred to \citet{mohler2011selfexciting}, \citet{ADELFIO2015119}, \citet{fox2016spatially}, \citet{chiodi2017mixed}, \citet{ZhuangMateu2019}, \citet{yuang2019} \citet{kalair2021nonparametric}, \citet{dangelo2024self}, \citet{AlaimoDiLoro2024}, \citet{KwonZhengJun2023} and \citet{nonparametricexample}. Other related methods and techniques can be found in \citet{diggle2010partialb}, \citet{ross2021bayesian}, \citet{molkenthin2022gp}, \citet{muir2023deep}, \citet{stockman2023forecasting} and \citet{choiruddin2024algorithms}.
\ab{We have not included an application, as we see this paper as one of expository nature with the aim of unifying methods, presenting their performance and sharing code.}

The paper is organized as follows. Section 2 provides a general overview of spatio-temporal Hawkes processes and describes the triggering functions used throughout the paper. Section 3 is dedicated to two simulation approaches: parents-offspring and acceptance-rejection methods. In Section 4, we explore three estimation strategies: likelihood, Expectation-Maximization and Bayesian inlabru. Section 5 presents an intensive and extensive simulation study. Finally, Section 6 presents some conclusions and a discussion.

\section{Spatio-temporal Hawkes point processes} \label{hawkes_section}

A spatio-temporal point process $\mathbf{X}$ on $\mathbb{R}^2 \times \mathbb{R}$ is a locally finite random subset of $\mathbb{R}^2 \times \mathbb{R}$. That is, the random number of points of $\mathbf{X}$ in $S \times T$, $N(S \times T)$, is almost surely finite whenever $S \times T \subset \mathbb{R}^2 \times \mathbb{R}$ is bounded. In practice, $\mathbf{X}$ is observed within a spatio-temporal window $W= S \times T$, where $W \subset \mathbb{R}^2 \times \mathbb{R}$ is a bounded region of area $|S|>0$ and $T$ is a bounded interval within $\mathbb{R}^+$. A realization of $\mathbf{X}$ in $S \times T$ is a set $\mathbf{x}=\{(s_i,t_i)_; i=1,\ldots,m\}$, where $s_i=(x_i,y_i)$ and $t_i$, respectively, represent the spatial and temporal coordinates for the occurrence of the $i$-th event, and where $m$ is the number of points in $S \times T$. For further details, we refer the readers to \cite{banerjee2003hierarchical}, \citet{diggle2013statistical} and \citet{gonzalez2016spatio}.
As we assume $N(S \times T)$ is almost surely finite, then $\{N(S \times T): S \times T \subset \mathbb{R}^2 \times \mathbb{R} \}$ characterizes a point process (see e.g. p.205 of \cite{cressie2015statistics}).

Hawkes point processes are a self-exciting spatio-temporal stochastic point model that assumes each event has the capacity to generate new events in its temporal and spatial proximity, resulting in clusters with a hierarchical branching structure. This approach enables the study of how an initial event can trigger a cascade of secondary events and facilitates the analysis of propagation patterns in both time and space. \ab{These self-exciting dynamics} of spatio-temporal processes are specified by the conditional intensity function, which models the expected event rate based on the history of the process, capturing both temporal and spatial interactions between events through excitation functions, also known as triggering functions. This conditional intensity function, denoted by $\lambda(s, t| \mathcal{H}_t)$, is generally defined for a location $s \in S \subseteq \mathbb{R}^2$ and a time point $t \in T \subseteq \mathbb{R}^+ $ as

\begin{equation}
    \lambda(s, t| \mathcal{H}_t) = \mu(s,t) + \ab{k} \sum_{(s_i,t_i) \in \mathcal{H}_t}
g(s - s_i , t - t_i ),
\label{eqhawkst}
\end{equation}
where $\mathcal{H}_t$ stands for the history of the process up to time $t$, defined as  $\mathcal{H}_t =\{(s_i, t_i) \in S \times T : t_i < t\}$, $\mu(s,t)$ is called background or baseline rate of the process,  \ab{$k$ is the
reproduction number representing the average number of events induced by a single event}, and $g(s, t)$ represents the triggering function. To be more specific, the background rate $\mu(s,t)$ represents the baseline level of event occurrences and can be modeled in various ways, such as in separable components, i.e., $\mu(s,t)=\mu_{\raisebox{-0.3ex}{\ab{\scriptsize S}}}(s)\mu_{\raisebox{-0.3ex}{\ab{\scriptsize T}}}(t)$ (\citealt{ZhuangMateu2019}), as a constant, i.e., $\mu(s,t)=\mu$ (\citealt{Schoenberg2023}), or by incorporating additional covariates (\ab{\citealt{parkcovariates}}). Furthermore, the triggering function $g(s,t)$ describes how an event at a given point influences the probability of additional events occurring at other points within the spatio-temporal domain. This can also be modeled in a separable form (\citealt{miscouridou2023coxhawkes}) or in a non-separable form (\citealt{KwonZhengJun2023}). 

The intensity function of a Hawkes process is built upon several critical assumptions. One of these is that the intensity of any point process must be non-negative\ab{, which implies that both the baseline function and the triggering function should be defined appropriately to ensure this}. The baseline function encapsulates the inherent patterns in both temporal and spatial domains that promote the clustering of events at specific times and spatial coordinates. \ab{It is essential that its integral over the space-time domain satisfies $\int_S \int_T \mu(s, t) dt \ ds > 0$.} 
Regarding the triggering function $g$, the sum of the effects of all past events (which contribute to the future intensity) must be finite to ensure that the total intensity is well-defined. Moreover, independence of event arrival times is assumed only for background events and the offspring of each individual parent, not for the entire realization of the Hawkes process. A fundamental characteristic of this process is the branching ratio \ab{$k$, which must be between 0 and 1 to ensure the stationarity of the process (assuming that the triggering function is properly normalized, $\int_S \int_T g(s, t) \, dt \, ds=1$)}, a essential prerequisite for well-defined properties of the Hawkes process \ab{(\citealt{hawkes2,hawkes1971spectra, daley2003introduction})}. In the case of $k = 0$, the process corresponds to a non-homogeneous Poisson process, while for $k \geq 1$, an explosive behavior occurs. To delve into the details of how this explosive phenomenon arises, one can refer to \citet{grimmett2001probability}. Further insights into the impact of different values of $k$ are available in \citet{asmussen2003applications}.

\subsection{Triggering functions} \label{subsec2}

As discussed, the triggering function determines how past events influence the intensity of future events. Therefore, choosing the appropiate triggering function is essential for accurately modeling spatio-temporal patterns. In many applications, it is common to decompose the triggering function into separable components that consider temporal and spatial influence separately (\ab{\citealt{OGATA200613, Patel2021SpatiotemporalEE, https://doi.org/10.1002/sta4.558, dangelo2024self, zhou2024bayesianinferenceaggregatedhawkes}}). This allows for a simpler and more structured modeling of space-time interactions, as in many practical cases, spatial and temporal dependencies may have different effects or may be more easily understood when modeled separately. Furthermore, it makes interpretation easier and reduces computational complexity. For this reason, we use separable triggerings as they are more computationally adaptable to the methods presented here. We have chosen two spatial triggering functions and two temporal triggering functions commonly used in the literature, along with their combinations, in order to capture various and common possibilities of spatio-temporal relationships in point pattern data. 

Regarding temporal triggerings, we first use the exponential function, represented as $g_{\raisebox{-0.3ex}{\ab{\scriptsize T}}}(t)=\alpha \cdot \exp\ab{\{}-\alpha(t-t_i)\ab{\}}$. This function stands out as one of the most commonly used for capturing the temporal influence of previous events. Its distinctive characteristic lies in its exponential decay form as time passes after the previous event $(t-t_i)$. Second, we use a reparametrization of the power-law function, defined as $g_{\raisebox{-0.3ex}{\ab{\scriptsize T}}}(t)=(\gamma-1)\cdot (1+t-t_i)^{-\gamma}$, where $\gamma$ is a parameter that affects the amplitude and decay rate. The power-law function exhibits a slower decay as time passes since the last event. This can be useful for capturing more complex patterns of temporal dependence, as power-law decay allows the model to account for events that have lasting effects over extended periods, a characteristic commonly observed in real-world processes such as earthquakes, financial transactions or epidemics (\ab{\citealt{ogata1998, pwlw2014, zhang2016power, Chen03052022}}). 

In relation to the spatial domain, one of the spatial triggering functions we use here is the Gaussian function. This function, denoted as $g_{\raisebox{-0.3ex}{\ab{\scriptsize S}}}(s)=\frac{1}{2\pi \sigma_x \sigma_y}\cdot  \exp\{-\frac{(x - x_i)^2}{2\sigma_x^2} - \frac{(y - y_i)^2}{2\sigma_y^2}\}$, assigns greater influence to events in close proximity, while the effect diminishes with a squared exponential decay as the distance between events becomes larger. The scale parameters $\sigma_x$ and $\sigma_y$ control the width of the Gaussian distribution in the $x$ and $y$ directions, respectively. For simplicity, we use $\sigma_x=\sigma_y=\sigma$, meaning that the Gaussian distribution has the same width in both directions. Alternatively, we also make use of an exponential function, written as $g_{\raisebox{-0.3ex}{\ab{\scriptsize S}}}(s)=\frac{1}{2\pi \beta^2} \cdot \exp\{-\frac{||(x, y) - (x_i, y_i)||}{\beta}\}$, which models a rapid reduction in spatial influence, where the parameter $\beta$ determines how quickly the influence of a past event decreases as the distance between the event and the location being considered increases.

\begin{figure}[h!]
  \begin{minipage}[b]{0.5\textwidth}
    \centering
    \includegraphics[width=\textwidth]{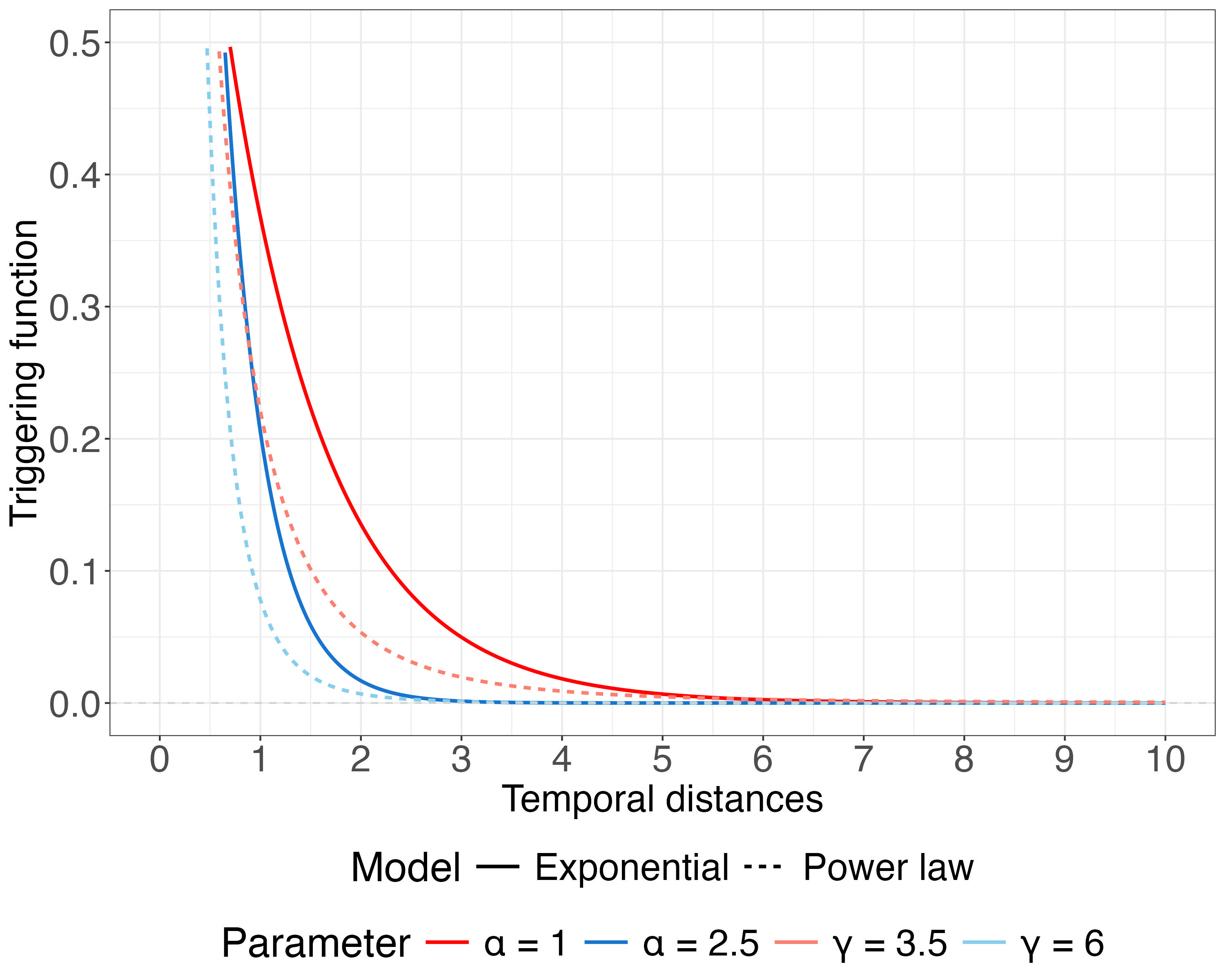}
    \caption*{Temporal triggerings}
  \end{minipage}%
  \begin{minipage}[b]{0.5\textwidth}
    \centering
    \includegraphics[width=\textwidth]{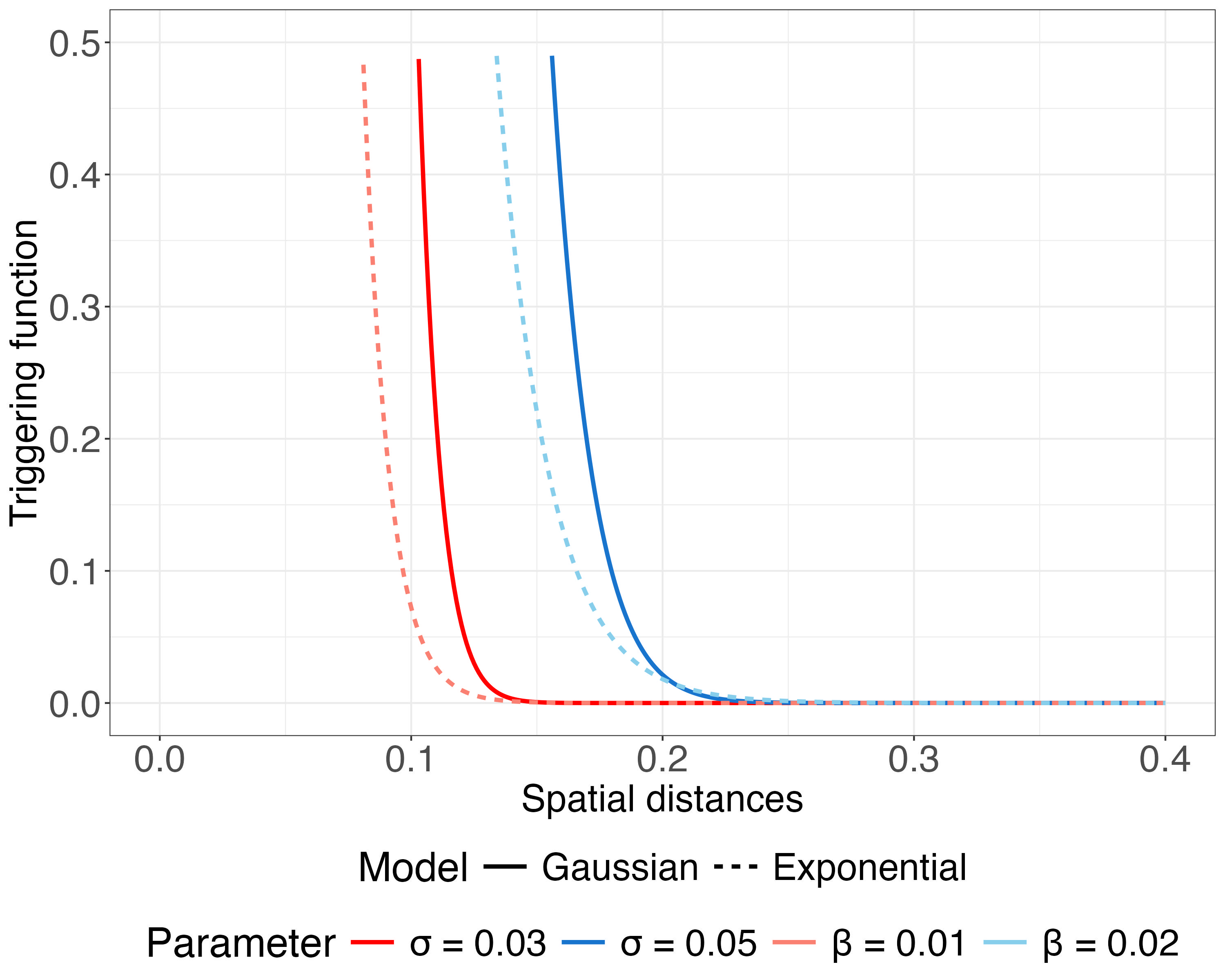}
    \caption*{Spatial triggerings}
  \end{minipage}
    \caption{Examples of temporal and spatial triggering functions under different parameter settings. For the temporal domain, the exponential triggering function with $\alpha = 1, 2.5$ is shown using solid lines, while the power-law triggering function with $\gamma = 3.5, 6$ is illustrated using dashed lines. In the spatial domain, the Gaussian triggering function is plotted for $\sigma = 0.03, 0.05$ using solid lines, and the exponential spatial triggering function is shown for $\beta = 0.01, 0.02$ using dashed lines.}
  \label{fig:figuras}
\end{figure}

These functions were selected not only for their ability to adequately model different types of interactions but also for their flexibility and efficiency in parameterization, allowing them to adapt well to a variety of spatial and temporal behaviors (\ab{\citealt{mohler2011selfexciting, pmlr-v31-zhou13a, mohler2014marked, yuang2019, 9177186, siviero2024flex}}). However, it is important to emphasize that the selection of triggering functions by the researcher should be based on the specific features of the available data, ensuring that the model is the most suitable for the phenomenon under question. Figure \ref{fig:figuras} shows some examples of the triggering functions considered. \ab{In the case of temporal triggering functions, larger parameter values result in a faster decay of the function. Conversely, for spatial triggering functions, smaller parameter values lead to a faster spatial decay. The parameter values are chosen so that the influence of the triggering decays within equivalent temporal and spatial ranges. This decision allows for a comparison between the different types of triggering functions across the scenarios considered in the simulation study presented in this paper.}

\section{Simulation strategies} \label{simul}

The process of modeling is inherently linked to simulation, as the practical interest of a model depends on its ability to replicate the behavior of the underlying process. In this section, we present two techniques for simulating Hawkes point processes and provide a detailed description of their implementation. In the absence of readily available resources developed in the same programming language for these two techniques, our aim is to provide an implementation in \texttt{R}, with the goal of unifying the available resources in the literature and offering an accessible tool in a single language.

\subsection{Parents-offspring method}  \label{sec:parents-offspring}

The parents-offspring method is a simulation technique that is particularly useful for modeling Hawkes point processes due to its ability to capture the self-exciting nature of these processes. A self-exciting process, such as Hawkes processes, can be interpreted through a cluster representation, originally introduced by \citet{hawkes1974cluster}. In this representation, events are hierarchically organized around an initial set of events, referred to as ``parents", which follow a Poisson distribution. These parent events generate subsequent ``offspring'' events, which in turn can act as new parents, producing additional offspring and creating a cascading structure governed by a conditional intensity function. This hierarchical structure is essential for modeling phenomena such as the ``contagiousness" of events in spatial and temporal systems. To effectively employ this methodology, it is crucial to consider both the base rate of event generation and the specific characteristics of the conditional intensity function, as these factors determine the cascade dynamics and the size of the clusters. To explain this process explicitly, we can break it down into several steps:

\begin{enumerate}
    \item \textbf{Generate the initial parent events.} The process begins by generating the number of parent events, which follows a Poisson distribution with parameter $n_0=\int_W \mu(s, t) \ ds \ dt$ (the expected number of background events). This means that $n_0$ indicates how many parent events occur, on average, per unit of time and space. Then, both the temporal and spatial locations of these events are determined according to particular distributions selected by the user, allowing the events to occur at specific times and locations within the defined time interval and spatial region. Such a distribution is $\bar{\mu}(\cdot, \cdot)$  (i.e., the normalized background) for the background events. Additionally, the parent events are independent of each other, meaning that the occurrence of one event does not influence the probability of occurrence of other events. This step establishes the initial events on which future events will be generated, thus forming the basis for all subsequent behavior of the process.

    \item \textbf{Generate offspring events.} For each parent event determine the number of offspring it produces. The number of offspring can be sampled from a Poisson distribution with mean $n_p=\int_{W} g(s, t) \ ds \ dt$, which represents the average number of offspring each parent event produces (the branching ratio). Each offspring is then assigned a temporal and spatial location, which are determined according to the triggering function of the Hawkes process. Specifically, these locations follow the distribution $\bar{g}(\cdot -s_i, \cdot - t_i)$ (i.e., the normalized excitation) for the offspring of event $i$.    
    If the offspring falls within the determined spatial region and time interval, we keep it; otherwise, we discard it and do not include it in the process. To generate values from a specific distribution, we use the Inverse Transform Theorem, which consists of inverting the cumulative distribution function (CDF) of the target distribution (\ab{\citealt{gentle2003random}, pp. 103}). However, the parents-offspring method could be inefficient or unfeasible if the inverse of the CDF or the conditional distributions are not analytically available. \ab{This is the main limitation of this method, ensuring that there is a way to simulate according to the interaction function $g$.} For unknown or complex distributions, there are other methods to generate values from a specific distribution, such as the Gibbs Sampling method (MCMC), the Fourier Transform method, or the Markov Chain method, among others.

    \item \textbf{Recursive process generation.} Repeat this process recursively for each new offspring event, treating it as a new parent to generate more offspring, until a stopping criteria defined by the user or a maximum number of iterations is reached. \ab{The most commonly used stopping criteria are the number of simulated points or fixing the simulation window beforehand.}
\end{enumerate}

After completing the three steps, several independent clusters are formed, with their centers being the parent events. The entire process includes all the data, involving both parents and offspring, which must be sorted chronologically to obtain the complete event sequence of the Hawkes process. \ab{In addition to the commonly used stopping criteria (\citealt{YOUSSEF2001167}), it is important to consider that each criterion has its own convergence and effectiveness, which may vary depending on the specific characteristics of the simulated process (\citealt{AYTUG2000662,bremaud2002hawkes, 10.1007/978-3-540-28645-5_41}). Some additional examples include a maximum sample size for the simulation, stability of the event rate, convergence in the spatial or temporal distribution (e.g., when the distribution becomes constant), simulation time or computational resources.} The pseudo-code for this algorithm can be found in Section 3.1 of \citet{bernabeu2024spatio} \ab{or in Appendix \ref{pseudo_append} of this paper (see Algorithm \ref{alg1})}. The code provided here in the \texttt{R} language is based on \citet{mohler2021modified}, which implements similar code in another programming language. The data shown in Figure \ref{data_extended} has been generated using this method.

\subsection{Acceptance-rejection method}

The acceptance-rejection method\ab{, introduced by \citet{ogata_acceptrej},} is a probabilistic approach \ab{to generate samples that begin by simulating a dominating Poisson process with intensity $\lambda_{\text{max}}$, which has the particularity that $\lambda_{\text{max}} \geq \lambda(t)$ for any $t$, and then evaluates whether they should be accepted or rejected.} This decision to accept or reject randomly generated points is known as the thinning process, which involves, once a point is generated, evaluating whether it should be accepted as part of the Hawkes process. The acceptance probability depends on comparing the intensity of the process at that point (which describes how likely an event is to occur at that point according to the intensity function of the Hawkes process) with the intensity of the proposed distribution (the expected intensity of the distribution being used to generate the points). In our case, we simulate random data by generating points uniformly in both space and time, and we use the intensity function of the Hawkes process to determine which of these points will be accepted as part of the process. The acceptance-rejection method suggests two main steps, each with its respective substeps:

\begin{enumerate}
    \item \textbf{Generation of an homogeneous Poisson point pattern.}
    \begin{enumerate}
        \item Simulate the number of events $n = N(S)$ occurring in $S$ according to a Poisson distribution with mean $\lambda_{\text{max}} \times |S \times T|$ (average event rate of the Poisson process in the spatio-temporal space $S \times T$), where $\lambda_{\text{max}}$ is an upper bound defined for the intensity function $\lambda(s, t)$.
        \item Sample each of the $n$ locations according to a uniform distribution in $S\times T$.
    \end{enumerate}
    \item \textbf{Thinning of the homogeneous Poisson process.}
    \begin{enumerate}
        \item Calculate $p = \frac{\lambda(s, t)}{\lambda_{\text{max}}}$ for each point $(s, t)$ of the homogeneous Poisson process.
        \item Generate a sample $u$ from the uniform distribution on $(0, 1)$.
        \item Retain the locations for which $u \leq p$.
    \end{enumerate}
\end{enumerate}

The pseudo-code for this method can be found in Section 3.2 of \citet{bernabeu2024spatio} \ab{or in Appendix \ref{pseudo_append} of this paper (see Algorithm \ref{alg2})}. \ab{This approach offers the key advantage of being easier to implement, particularly when the distribution is complex or when the inversion method cannot be applied due to the lack of an analytical expression for the distribution function. However, this method may prove to be less efficient than the parent–offspring approach in certain cases, as it requires the value of the intensity $\lambda$ to be computed at each simulated point in order to compute the probability $p$ of retaining the point. This involves evaluating the integral in the intensity, which requires to sum over each point simulated in the past. This sum contains $N(t)$ terms at any given time $t$. The only way of reducing this is either by considering exponential separable interactions, making it possible to sequentially update the intensity (this is why such triggerings are commonly used in practice), or by considering bounded interactions. In contrast, the parent–offspring method  does not need to compute this sum, avoiding this updating step throughout the simulation. Another} limitation of the acceptance-rejection approach lies in the choice of the upper bound for the maximum intensity ($\lambda_{\text{max}}$). Since the data has not yet been generated, it is difficult to establish a threshold to avoid an excessive or insufficient density of points in the resulting Poisson process. This challenge can complicate the initial parameter setup and affects the accuracy and efficiency of the acceptance-rejection method in generating spatio-temporal events. \ab{The algorithm proposed by Ogata \citealt{ogata_acceptrej} constructs the upper bound for the intensity function adaptively at each step through approximation. However, when the parameters and the intensity function are known and fixed, it is possible to compute a constant and safe upper bound that can be used directly within the thinning algorithm (\citealt{lemaire2018exact}). In such cases, the resulting algorithm is straightforward to implement and can produce exact simulations, provided that additional assumptions are satisfied (for instance, it is intuitive that a constant bound may be inefficient when the intensity function exhibits significant temporal or spatial variability). Then, the thinning theorem remains valid under a constant upper bound, which simplifies the simulation of the Poisson point process. Importantly, Ogata emphasises that the upper bound does not need to be exact, it merely needs to be a valid upper bound at all points in time.} To address this limitation \ab{and estimate the maximum intensity ($\lambda_{\text{max}}$), we construct a three-dimensional spatio-temporal grid, and using the fixed parameters of the Hawkes process we aim to simulate, we compute the intensity at each grid point and take the maximum over all cells (\citealt{andral2024}).} 

Figures \ref{fig:figura2a} and \ref{fig:figura2b} show simulated spatio-temporal point patterns using the acceptance-rejection method for the combination of parameters used in the simulation study. To perform these simulations, the code proposed in \citet{zhu_stpp_simulator} has been adapted to the \texttt{R} programming language, which is available in the GitHub repository associated with this paper.

\section{Methods for statistical inference}

Inference in point processes is essential for understanding how and why events occur, as well as for accurately modeling their behavior. It involves estimating the parameters of the model from the observed data. To perform inference in spatio-temporal Hawkes processes, various statistical techniques are employed. In this paper, we have selected three parametric methods that we consider sufficiently flexible and versatile to be applied in various scenarios and discuss their similarities and differences in terms of estimation accuracy and computational speed. We focus on the classical full likelihood, the Expectation-Maximization (EM) method, and the Integrated Nested Laplace Approximations (INLA). All three strategies involve the likelihood, though they approach it differently. We explain step by step how these strategies have been implemented, applying them to a specific example of an intensity function. Each of these methods has its own advantages and limitations, which are also discussed in this section. Additionally, this paper provides generalized code in \texttt{R} for estimating spatio-temporal Hawkes processes using the three proposed methods, with the aim of contributing to the development of future research in this field.

\subsection{Full likelihood approach}

The full likelihood function is a central concept in statistics, which evaluates the probability that a given observed data comes from a specific model, given certain hypotheses and parameters. The method known as maximum likelihood involves the direct maximization of the likelihood function, seeking the parameter values that achieve this goal.

Instead of defining the likelihood explicitly, we directly define the log-likelihood for the conditional intensity of the Hawkes process. The log-likelihood is often preferred in statistical inference as it simplifies the mathematical computations and avoids numerical instability when dealing with very small or very large probabilities. This function can be expressed as

\begin{equation}
\log L(\theta) = \sum_{i=1}^{n} \log \lambda(s_i, t_i | \mathcal{H}_t) - \int_{T}\int_{S} \lambda(s, t | \mathcal{H}_t) \, ds \, dt,
\label{lk}
\end{equation}
where $\theta$ represents the vector of unknown parameters on which the intensity depends on, and the data \(\{(s_i, t_i)\}_{i=1}^n \subseteq S \times T\) are ordered by time (\citealt{diggle2010a, diggle2010partialb}).

Looking at (\ref{lk}), the conditional intensity function must be integrable over the entire region. If the intensity function has a known and manageable explicit analytical expression, the computation of the log-likelihood (including both the first and second terms) becomes more straightforward, as its integral can be calculated exactly or the terms can be simplified analytically. Conversely, if the intensity function is complex or does not have a known explicit analytical form, calculating its exact integral (which appears in the second term of the log-likelihood) can be analytically intractable. In such cases, numerical methods must be used to approximate the integral, which increases computational cost.

The strategy we consider for computing the log-likelihood follows two steps. Firstly, we focus on the sum of the logarithm of the conditional intensities. This part of the calculation can be quite simple, as long as we have an explicit expression for \(\lambda(s, t | \mathcal{H}_t)\). By using this expression, we can directly evaluate the logarithm of the intensity for each event in our observed data and sum them up. This provides a measure of how well our model fits the data. In a second step, we face the challenge of computing the double integral of the intensity function. This integral is crucial for the total log-likelihood, but it can be difficult to compute analytically, especially in complex situations. To overcome this challenge, we propose a strategy based on discretizing the space-time domain using a grid system and calculate the intensity in each of these cells. \ab{Specifically, the integral of the conditional intensity function over the entire region is approximated by summing the intensity values at the centre of each cell multiplied by the cell volume, which is the product of the spatial and temporal resolutions. Formally, this can be expressed as 
\begin{equation*}
    \int_T\int_S \lambda(s, t \mid \mathcal{H}_t) \, ds \, dt \approx \sum_{k} \lambda(s_k, t_k \mid \mathcal{H}_{t_k}) \Delta s \Delta t,
\end{equation*}
where \( (s_k, t_k) \) represents the centre of the \(k\)-th cell, and \( \Delta s \) and \( \Delta t \) are the spatial and temporal dimensions of each cell, respectively. This discretisation reduces the integral to a finite sum, which is often more computationally tractable than attempting to evaluate the integral analytically. Moreover, due to the additive structure of the model, this sum can be further decomposed into the sum of the background intensity and the triggering function evaluated over the grid cells, which simplifies the computation.} In our simulations, we have used a grid of size $25\times 25\times 25$ to discretize our space-time domain. This cell size in the grid has worked well in our case. As the number of divisions in the grid increases (i.e., as the cell size decreases), we obtain greater precision in the value of the integral, but with a higher computational cost. On the other hand, a grid with larger cells (fewer divisions) could result in less accurate approximations. Therefore, it is important to find a balance between precision and computational efficiency. To maximize this log-likelihood function, we use the \texttt{optim} function in \texttt{R}.

\subsection{Expectation-Maximization (EM) algorithm}

The Expectation-Maximization (EM) algorithm serves as a powerful optimization tool for estimating unknown parameters within statistical models. Specifically within the realm of Hawkes processes, the EM algorithm iteratively maximizes the likelihood function, unlike the previous method which did so directly. This algorithm unfolds in two main phases, the Expectation (E-step) phase, which computes the expected values of latent variables based on the current model parameters and given conditions, and the Maximization (M-step) phase, which updates the model parameters to maximize the conditional likelihood computed in the E-step. This iterative process persists until the likelihood converges to a local maximum. However, although the EM algorithm is valuable in scenarios where analytical maximization of likelihood is impractical or latent variables are present, its convergence to a global maximum is not guaranteed, except under certain hypotheses (\citealt{dempster1977maximum}). Moreover, its efficacy may heavily rely on parameter initialization and model structure.

To explain how the EM method works in practice, we consider a general form of the conditional intensity in (\ref{eqhawkst}). \ab{This model can be estimated using the EM algorithm, leveraging its representation as a branching process. With this in mind, we define the variable $p$, where $p_{ij}$ represents the probability that $i$ is a background event when $i=j$, or $i$ is triggered by $j$ when $j<i$. Then, the log-likelihood in (\ref{lk}) is given by the expectation of the log-likelihood, considering the defined probabilities, as follows:}

\begin{equation}
\begin{aligned}  
    \mathbb{E}[\log L(\theta)] = \sum_{i} p_{ii} \log(\mu(s_i, t_i)) - \int_T\int_S \mu(s, t)\, ds\, dt \\
 + \sum_{j<i} p_{ij} \log(\ab{k} \ g(s_i - s_j, t_i - t_j)) - \sum_{j} \int_T\int_S \ab{k} \ g(s - s_i, t - t_i)\, ds\, dt .
\label{expect}
\end{aligned}
\end{equation}

The mathematical details can be found in \ab{\citet{Veen01062008}}. Therefore, at this stage, we have separated the background function from the triggering component. To maximize the expectation of the log-likelihood $\mathbb{E}[\log L(\theta)]$ in a Hawkes process, the partial derivatives of the log-likelihood with respect to each parameter of the model (background intensity and triggering function) must be calculated. These derivatives are then set to zero to find the parameter values that maximize the log-likelihood. The resulting values are the point estimates of the parameters, representing the best approximations for the model parameters given the observed data. This process is carried out using the maximum likelihood method and is crucial for optimally fitting the model parameters. 

For completeness, we explain step by step how the algorithm works, by considering a particular conditional intensity function of the form:

\begin{equation}
    \lambda(s, t|\mathcal{H}_t)=\mu + \sum_{(s_i,t_i) \in \mathcal{H}_t} k \cdot \alpha \cdot \exp\{-\alpha (t-t_i)\} \cdot \frac{1}{2\pi \sigma^2} \cdot \exp\{-\frac{(x-x_i)^2+(y-y_i)^2}{2\sigma^2}\},
    \label{ejemplohwk}
\end{equation}
where $s\in S = [X_1, X_2] \times [Y_1, Y_2]$ and $t\in T=[T_1, T_2]$.
\citet{schoenberg2013facilitated} noted that in the case of certain conditional intensities, integrating analytically over 
$\mathbb{R}^2 \times \mathbb{R}$ might be significantly more straightforward than over an arbitrary set $S\times T$. As a result, the following \ab{bound} could be used:
\begin{equation*}
    \int_T\int_S g(s - s_i, t - t_i)\, ds\, dt \leq \int_0^{\infty}\int_{\mathbb{R}^2} g(s - s_i, t - t_i)\, ds\, dt .
\end{equation*}
The approximation can reduce the integral to a form that can be evaluated directly and it is exact when the effect of self-excitation is entirely contained within $S$ and before $t=T_2$, and overestimates otherwise (\citealt{mohler2014marked}). Note that as overestimation decreases the calculated likelihood, \citet{schoenberg2013facilitated} argued that likelihood maximization will avoid parameter values where overestimation is large. Consequently, in the situation we are considering,

\begin{equation*}
    \int_{\mathbb{R}^2} \frac{1}{2\pi \sigma^2} \cdot \exp\{-\frac{(x-x_i)^2+(y-y_i)^2}{2\sigma^2}\} \ dx \ dy=1,
\end{equation*}
given that the integral of the probability density function (PDF) of a Gaussian distribution must be equal to 1 to ensure that the total probability is normalized. And

\begin{equation*}
    \int_{0}^{\infty} \alpha \cdot \exp\{-\alpha (t-t_i)\} \, dt=[-\exp\{-\alpha(t-t_i)\}]_0^{\infty}=1.
\end{equation*}
Therefore, assuming equality \ab{of the approximation} for simplicity,
\begin{equation*}
    \int_T\int_S \ab{k} \ g(s - s_i, t - t_i)\, ds\, dt = k \int_T\int_S  \exp\{-\alpha (t-t_i)\} \cdot \frac{1}{2\pi \sigma^2} \cdot \exp\{-\frac{(x-x_i)^2+(y-y_i)^2}{2\sigma^2}\} \, dx \, dy \, dt = k.
\end{equation*}
As a result, the term of the sum of integrals remains as

\begin{equation*}
     \sum_j\int_T\int_S\frac{k\alpha}{2\pi \sigma^2}\cdot \exp\{-\alpha(t-t_i)\}\cdot \exp\{-\frac{(x-x_i)^2+(y-y_i)^2}{2\sigma^2}\} \, dx \, dy \, dt = \sum_j k = kn, 
\end{equation*}
where $n$ is the number of points. Based on Equation (\ref{expect}), we can write

\begin{equation*}
\begin{split}
     \mathbb{E}[\log L(\theta)] & =\log(\mu)\sum_ip_{ii}-\mu(T_2-T_1)(X_2-X_1)(Y_2-Y_1)+\log(\frac{k\alpha}{2\pi \sigma^2})\sum_{i>j}p_{ij} \\ &
     - \alpha\sum_{i>j}p_{ij}(t_i-t_j)-\frac{1}{2\sigma^2}\sum_{i>j}p_{ij}((x_i-x_j)^2+(y_i-y_j)^2) + kn,
\end{split}
\end{equation*}
where 

\begin{equation*}
p_{ii} = \frac{\mu(s_i, t_i)}{\lambda(s_i, t_i)} \quad \text{and} \quad p_{ij} = \frac{ \ab{k} \ g(s_i - s_j, t_i - t_j)}{\lambda(s_i, t_i)}.
\end{equation*}

Note that, in our case, we assume $\mu(s, t) = \mu$, constant. To find the parameters that maximize the function, the next step is to calculate the partial derivatives with respect to each parameter, set them equal to zero, and solve to obtain the maximum likelihood estimator, as follows,

\allowdisplaybreaks

{\small
\begin{flalign*}
& \frac{\partial \mathbb{E}[L(\theta)]}{\partial \mu} = \frac{1}{\mu}\sum_i p_{ii} - (T_2 - T_1)(X_2 - X_1)(Y_2 - Y_1) = 0 \rightarrow \hat{\mu} = \frac{\sum_i p_{ii}}{(T_2 - T_1)(X_2 - X_1)(Y_2 - Y_1)} && \\
& \frac{\partial \mathbb{E}[L(\theta)]}{\partial \alpha} = \frac{1}{\alpha}\sum_{i>j} p_{ij} - \sum_{i>j}p_{ij}(t_i - t_j) = 0 \rightarrow \hat{\alpha} = \frac{\sum_{i>j}p_{ij}}{\sum_{i>j}p_{ij}(t_i - t_j)} && \\
& \frac{\partial \mathbb{E}[L(\theta)]}{\partial k} = \frac{1}{k}\sum_{i>j} p_{ij} - n = 0 \rightarrow \hat{k} = \frac{\sum_{i>j}p_{ij}}{n} && \\
& \frac{\partial \mathbb{E}[L(\theta)]}{\partial \sigma} = -\frac{2}{\sigma}\sum_{i>j} p_{ij} + \frac{1}{\sigma^3}\sum_{i>j}p_{ij}((x_i - x_j)^2 + (y_i - y_j)^2) = 0 \rightarrow \hat{\sigma} = \sqrt{\frac{\sum_{i>j}p_{ij}((x_i - x_j)^2 + (y_i - y_j)^2)}{2 \sum_{i>j}p_{ij}}} && 
\stepcounter{equation}\tag{\theequation}\label{estimatesbyhand}
\end{flalign*}
}

The procedure begins with an initial guess for the parameters $\mu, k, \alpha, \sigma$. It iterates through two main steps: the Expectation step, where $p_{ij}$ is computed for each pair  $j\leq i$ using the current parameter values, and the Maximization step, which updates the parameters by maximizing the likelihood function based on the calculated $p_{ij}$. This process continues iteratively until convergence, determined by a desired tolerance $\varepsilon$, or simply when the obtained parameter estimates stabilize. Regarding the stopping criterion, which is a rule that indicates when the EM algorithm should stop iterating, we have used a fixed number of iterations and assessed whether the difference between the intensity functions with the obtained estimates and previous estimates is small enough. This allowed us to identify if the parameter estimates had stabilized. However, a tolerance could have been used, which is an acceptable threshold of change between the parameter estimates in each iteration. This would have allowed us to save computational time, but it has the limitation that it is important to choose an appropriate tolerance, as if it is too strict, the algorithm could keep iterating longer than necessary, and if it is too relaxed, it might stop before reaching an accurate solution. It is also worth noting that to calculate the parameter estimates of the log-likelihood, we have used the mathematical and analytical expressions of the triggerings to manually calculate the expressions of the parameters that maximize the likelihood function in (\ref{estimatesbyhand}). However, functions do not always have a simple analytical expression to solve for, and in such cases, we would need to calculate the expectation of the log-likelihood and use a computational function to find where the log-likelihood is maximized, making this process more computationally expensive.
The motivation for including this method is that, although it also evaluates the likelihood and has limitations similar to the previous method, its implementation facilitates the exploration of the parameter space by augmenting the data with the unknown branching structure, leading to the complete and expected log-likelihood, which generally behaves better than the full log-likelihood. \ab{For additional information on the technical aspects of this estimation method, we refer the reader to \citet{WHEATLEY2016120}, \citet{xueying}, \citet{mohler2021modified}, \citet{AlaimoDiLoro2024} and \citet{bernabeu2024spatio}.}

\subsection{Bayesian approach: INLA methodology}

The Bayesian methodology is based on Bayes' theorem, which provides a way to update beliefs about a model or parameters as new information provided by the data is obtained. One of the main advantages of the Bayesian approach is that it allows the incorporation of prior knowledge about the model parameters through prior distributions, before observing the data. After observing the data, the posterior distribution of the parameters is calculated, which is the conditional distribution of the parameters given the data. Thus, instead of treating the parameters as fixed and unknown values, the Bayesian methodology treats them as random variables and allows modeling the uncertainty about these parameters.

INLA (Integrated Nested Laplace Approximation) methodology (\citealt{ruehavard2017}) is a powerful and efficient approach for estimating the posterior distribution in Bayesian inference, particularly for complex models such as spatial and spatio-temporal processes. \citet{serafini2023approximation} presents an efficient method for estimating the parameters of Hawkes processes using the \texttt{R} package \texttt{inlabru}, although that article focuses exclusively on the temporal dimension and disregards spatial aspects. It is worth mentioning that there is a package called \texttt{ETAS.inlabru} that is also developed for the temporal temporal Epidemic-Type Aftershock Sequence (ETAS) model (\citealt{OGATA200613}), which is used to model earthquake occurrences (\citealt{naylor2022bayesianmodellingtemporalevolution}). This methodology, also called inlabru, is designed to provide fast and accurate approximations to the marginal likelihood and posterior distributions, making it a suitable method for large-scale problems where traditional Markov Chain Monte Carlo (MCMC) methods may be computationally expensive. INLA uses a combination of numerical integration and Laplace approximation to estimate the posterior, often leading to significant computational savings. 

We here extend this methodology to the spatio-temporal domain by incorporating the spatial component throughout the modeling process to broaden its applicability.  We then illustrate the technique for approximating the log-likelihood of the Hawkes process in a spatio-temporal context. This approximation method enables to represent the Hawkes process log-likelihood as a sum of linear functions of the parameters. Let us consider the scenario where we have observed $n$ events over a spatio-temporal region, denoted by $\mathbf{x}=\{(s_1, t_1),...,(s_n, t_n) | (s_i, t_i) \in S \times [T_1, T_2] \  \forall i =1,..., n\}$, where $0\leq T_1 < T_2 < \infty$ and $S\subseteq \mathbb{R}^2$. Here, $T_1$ and $T_2$ denote the lower and upper bounds of the time interval, respectively, and $S$ is a subset of the two-dimensional Euclidean space. We use a rectangular region $S=[X_1, X_2] \times  [Y_1, Y_2]$ and denote $\mathcal{D}=S\times [T_1, T_2]$. We also use the intensity function defined in Equation (\ref{ejemplohwk}) to explain the method below. The general form of the parametric log-likelihood function of a spatio-temporal point process model, given the observations, is given by
\begin{equation*}
 \log L(\theta|\mathbf{x}) = -\Lambda(\mathcal{D}|\mathbf{x}
) + \sum_{i=1}^{n} \log \lambda(s_i, t_i|\mathcal{H}_{t_i}),
\end{equation*}
where $\mathcal{H}_{t_i}$ is the history up to time $t_i$, and

\begin{equation}
\Lambda(\mathcal{D}|\mathbf{x}) =\int_{\mathcal{D}}\lambda(s, t|\mathcal{H}_t) \, ds \ dt = \int_{T_1}^{T_2} \int_S \lambda(s, t|\mathcal{H}_t) \, ds \ dt
    \label{integraltrigg1}
\end{equation}
is the integrated conditional intensity corresponding to the expected number of points in the study region $\mathcal{D}$. The integrated conditional intensity can be broken down by leveraging the hierarchical nature of Hawkes processes. Essentially, we can conceptualize the expected count of events within a region as consisting in both the expected count of background events and the expected count of events triggered by each past observation. To formalize this, given the observation of $n$ events, 

\begin{equation*}
    \Lambda (\mathcal{D}|\mathbf{x})=\Lambda_0(\mathcal{D})+\sum_{i=1}^n \Lambda_i(\mathcal{D}),
\end{equation*}
where 
\begin{equation*}
    \Lambda_0 (\mathcal{D}) = \int_{\mathcal{D}} \mu \ ds \ dt = \mu (T_2-T_1) (X_2-X_1) (Y_2-Y_1)
\end{equation*}
is the background rate that represents the expected count of background events. The other quantity is expressed as

\begin{equation}
    \Lambda_i(\mathcal{D}) = \int_{\mathcal{D}}  \ab{k} \ g(s, t) \ ds \ dt =  \ab{k} \int_S \int_{T_1}^{T_2} g_{\raisebox{-0.3ex}{\ab{\scriptsize T}}}(t - t_i) g_{\raisebox{-0.3ex}{\ab{\scriptsize S}}}(s - s_i) \ dt \ ds,
    \label{integraltrigg}
\end{equation}
and represents the expected number of points generated by each observation, $(s_i, t_i)$, in the process.

In practice, we divide the integration region into multiple subregions, with the aim of more accurately approximating the integral. In this approach, both the time interval and the spatial domain are partitioned, dividing the time interval into subintervals and the space into circular subareas.
This results in a better estimation of the Hawkes process intensity within each subinterval and spatial subregion. The novelty here is that we use circular bands to divide the space, instead of rectangular bins, as \citet{serafini2023modelling} did. This provides greater computational speed and more accurate estimates of the integral. This strategy has been used in other methods by other authors (\citealt{ogata1998, ADELFIO2015119, jalilian2019etas, AlaimoDiLoro2024}).

More explicitly, for the temporal case, we consider, as described in \citet{naylor2022bayesianmodellingtemporalevolution}, a partition of the integration interval $[\max(T_1, t_i), T_2]$
in $B_i$ bins, $t_0^{(b_i)}, ..., t_{B_i}^{(b_i)}$, such that $t_0^{(b_i)}$ = $\max(T_1 , t_i )$, $t_{B_i}^{(b_i)} = T_2$ and $t_j^{(b_i)} < t_k^{(b_i)}$ if $j < k$.  This binning approach is characterized by three parameters, $\Delta_t$, $\delta_t > 0$, and $n_{\text{max}} \in \mathbb{N}^+$. This strategy divides the time region around the observed point $t_i$ into bins, where the bin boundaries are given by 

\begin{equation*}
    t_i, t_i + \Delta_t, t_i + \Delta_t(1+\delta_t), t_i + \Delta_t(1+\delta_t)^2, ..., t_i + \Delta_t(1+\delta_t)^{n_i},T_2,
\end{equation*}
with $n_i \leq n_{\text{max}}$ ensuring that $t_i + \Delta_t(1 + \delta_t)^n$ does not exceed $T_2$ for the maximum allowable $n \in \mathbb{N}^+$. Parameter $\Delta_t$ determines the length of the initial bin, while $\delta_t$ regulates how this length increases in subsequent bins, and $n_{\text{max}}$ sets the maximum number of bins.

This approach offers two key advantages. First, it results in shorter bins near the point $t_i$ and longer bins as we move away from it. \ab{This binning strategy assumes that the triggering function varies more rapidly close to $t_i$ and then flattens out as time increases, but it should be noted that this assumption may not hold for triggering functions with heavy tails or those featuring a set delay, which could affect the accuracy of the binning approach in such cases.} Second, each bin (whether the first, second, third, or any other except the last one) has a consistent length across all points. This uniformity proves beneficial since the integral within a bin depends solely on its length and not its position. Consequently, we only need to compute the integral value once for each bin and this value remains constant across all observations. As a result, computational time is significantly reduced.

\begin{figure}[h!]
    \centering
    \begin{minipage}[t]{0.49\linewidth}
        \centering
 \vspace{-4cm}
        \includegraphics[width=\linewidth]{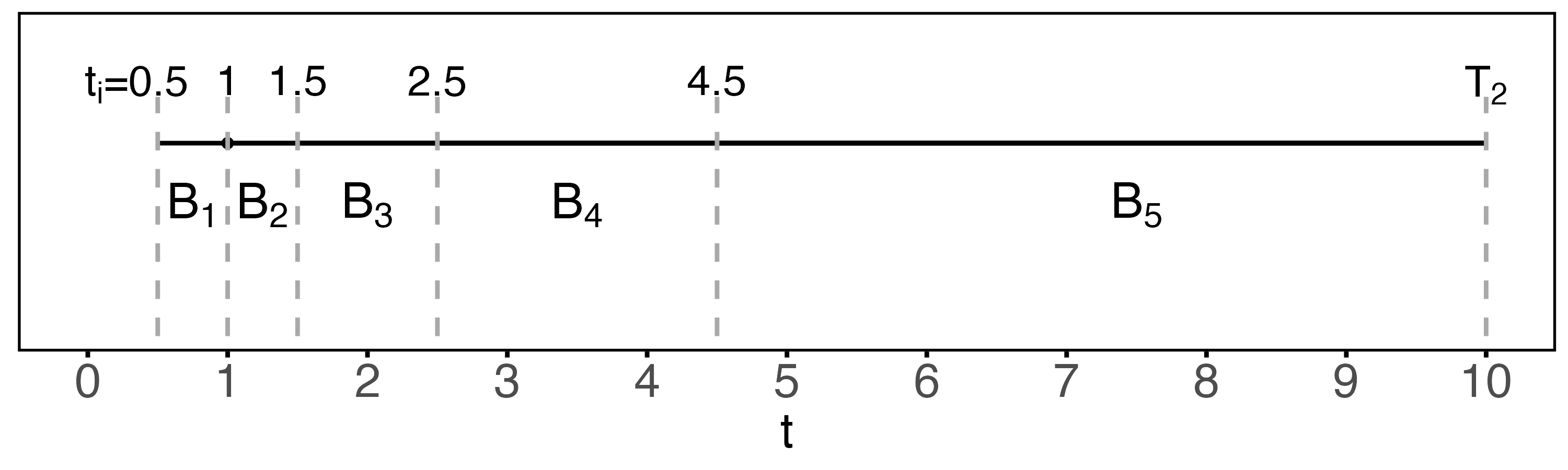}
    \end{minipage}
    \hfill
    \begin{minipage}[t]{0.49\linewidth}
        \centering
        \includegraphics[width=\linewidth]{ 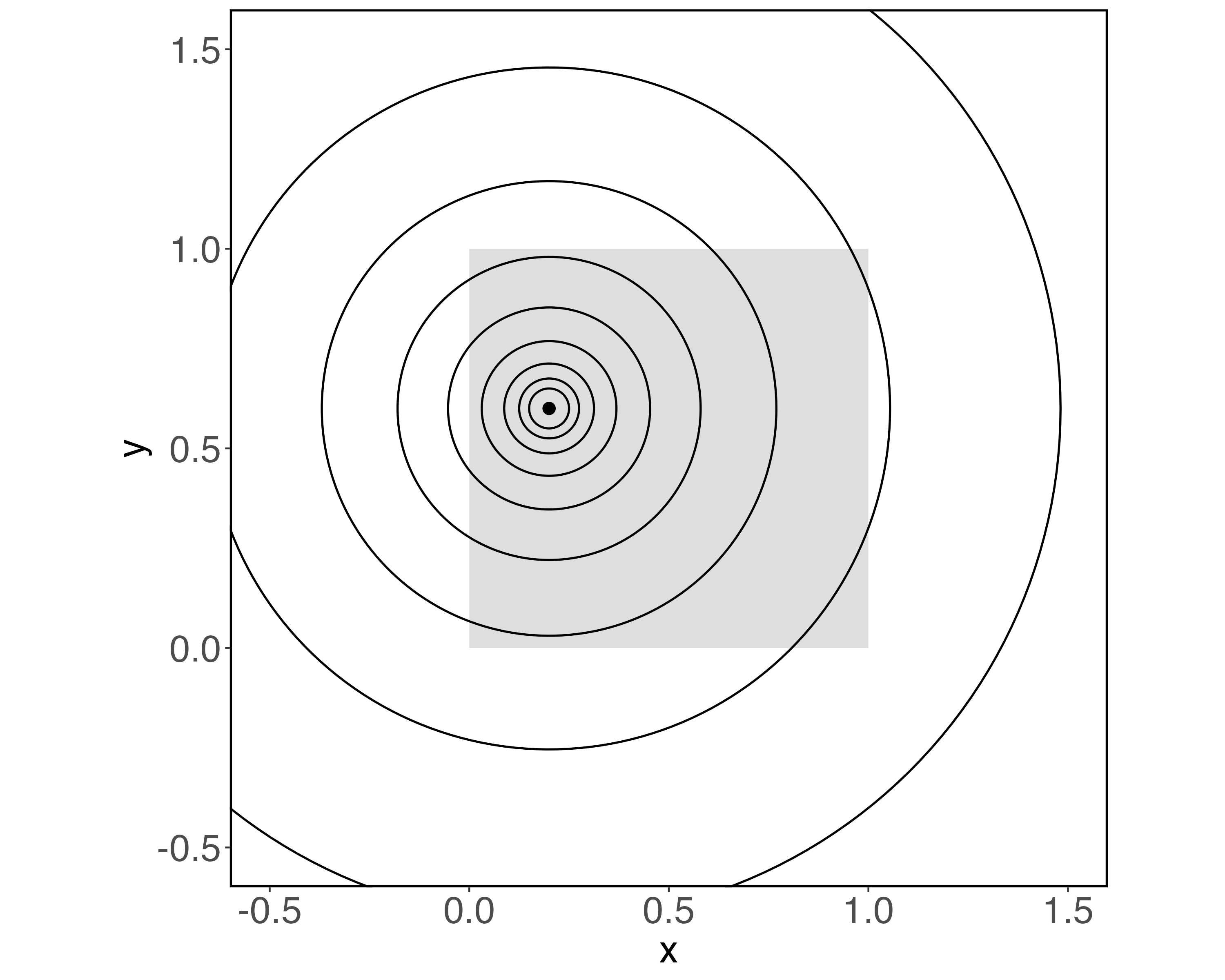}
    \end{minipage}
    \caption{On the left, an example of the binning strategy for the temporal interval defined by $\Delta_t = 0.5$, $\delta_t = 1$, and $n_{\max} = 3$. On the right, an example of circular partitioning for the spatial region with $\Delta_s = 0.05$, $\delta_s = 0.5$, and $\text{nmax}_c = 8$ (the study region is shaded in gray).}
    \label{circularbands}
\end{figure}

For the spatial case, our goal is to divide the area into circular bands, such that the size of each one increases gradually as we move away from a reference point, following the same principle as in the temporal case. In our scenario, this spatial area corresponds to the unit square. To achieve this, we also define three parameters, $\Delta_s$, which determines the size of the first circle, $\delta_s$, which controls how the radius of each circle increases in relation to the previous one, and $\text{nmax}_c$, which sets the maximum number of circles we want to create. So, the circles relative to an observed point are given by $C_1, C_2, C_3, \ldots, C_{\text{nmax}_c}$, where the radius of circle $C_i$ is calculated as
\begin{equation*}
\text{radius}_i = \Delta_s \times (1 + \delta_s)^i,   
\end{equation*}
where $i$ ranges from 0 to $\text{nmax}_c - 1$. This ensures that the size of the circle increases gradually as we move away from the reference point, similar to how time intervals increase gradually in the time partitioning strategy. It is important to note that the last circle must cover the entire area. \ab{Figure~\ref{circularbands} shows a representation of the construction of the temporal bins and the circular spatial bands for a point.}

Once we have divided the area into circular bands and determined the radii of each band, we proceed to calculate the integral in 
(\ref{integraltrigg1}) or (\ref{integraltrigg}) within each circular band. Then we multiply this value by the corresponding weight, which is the proportion of the area of the circular band that lies within our study region. The weight can be any value associated with the circular band, such as its area, point density, or another relevant factor for the analysis at hand. In our case, we estimate the proportion of each circular band within the domain by uniformly simulating many points and determining the proportion of those that lie within the study region. This process allows to obtain an estimate of the contribution of each circular band to the total integral over the region of interest. This approach allows to approximate the integral over the region efficiently by dividing the calculation into smaller and manageable parts. The key advantage is that by calculating the integral only once for each circle band and then multiplying it by each weight depending on the location of our spatial point, we significantly reduce the time and computational resources required to perform the calculation. Doing this, the integral becomes

\begin{equation*}
    \Lambda(\mathcal{D}| \mathbf{x})=\Lambda_0(\mathcal{D}) +\sum_{(s_i,t_i)\in \mathbf{x}} \sum_{j=0}^{B_i-1} \sum_{p=0}^{C_i-1} \Lambda_i (t_j^{(b_i)}, t_{j+1}^{(b_i)},s_p^{(c_i)}, s_{p+1}^{(c_i)}).
    \label{integral_bins}
\end{equation*}
We note that binning is crucial for the accuracy of the approximation and the convergence of the algorithm. Different partitioning strategies can be employed and their performance depends on the shape of the triggering function. The Hawkes process log-likelihood can now be expressed as

\begin{equation}
    \log L(\theta|\mathbf{x})=-\Lambda_0(\mathcal{D})-\sum_{i=1}^n \sum_{j=0}^{B_i-1} \sum_{p=0}^{C_i-1} \Lambda_i (t_j^{(b_i)}, t_{j+1}^{(b_i)},s_p^{(c_i)}, s_{p+1}^{(c_i)})+ \sum_{i=1}^n \log \ \lambda(s_i, t_i|\mathcal{H}_{t_i} ).
    \label{integral_bins2}
\end{equation}

The approximation method consists of linearizing the logarithm of the individual elements of the summations with respect to the posterior mode $\theta^*$. Then, the approximate log-likelihood is given by

\begin{equation*} \label{eq1}
\begin{split}
\log \overline{L}(\theta|\mathbf{x}) & =-\exp\left\{\overline{\log \Lambda_0}(\mathcal{D})\right\} - \sum_{i=1}^n \sum_{j=0}^{B_i-1} \sum_{p=0}^{C_i-1} \exp\left\{\overline{\log \Lambda_i} (t_j^{(b_i)}, t_{j+1}^{(b_i)},s_p^{(c_i)}, s_{p+1}^{(c_i)})\right\}  \\
 & + \sum_{i=1}^n \overline{\log  \lambda}(s_i, t_i|\mathcal{H}_{t_i}),
\end{split}
\end{equation*}
where, for a generic function $f(\theta)$ with argument $\theta\in \Theta \subset \mathbb{R}^m$, its linearized version around a point $\theta^*$ is given by

\begin{equation*}
    \bar{f}(\theta) = f(\theta^*) + \sum_{k=1}^{m} (\theta_k - \theta_k^*) \frac{\partial f(\theta)}{\partial \theta_k} \Bigg|_{\theta=\theta^*}.
\end{equation*}

Upon reaching this point, we need to define and provide the three functions, one per each term of Equation (\ref{integral_bins2}). This approach enables to improve each component individually, thereby ensuring that modifications do not compromise the overall system performance. To better understand this, we continue using the previously mentioned example with the conditional intensity function defined in Equation (\ref{ejemplohwk}). Then, we need to provide these three components

\begin{equation*}
    \log \Lambda_0(\mathcal{D}) = \log(\mu)+\log(T_2-T_1)+\log(X_2-X_1)+\log(Y_2-Y_1),
\end{equation*}

\begin{equation*}
\begin{split}
\log \ \Lambda_i (t_j^{(b_i)}, t_{j+1}^{(b_i)},s_p^{(c_i)}, s_{p+1}^{(c_i)}) & = \log(k)-\log(2\pi \sigma^2) + \log(\exp\{-\alpha(t_j^{b_i}-t_i\}-\exp\{-\alpha(t_{j+1}^{b_i}-t_i\}) \\ & + \log\int_{s_{p}^{(c_i)}}^{s_{p+1}^{(c_i)}} \exp\{-\frac{(x-x_i)^2+(y-y_i)^2}{2\sigma^2}\} dx \ dy,
\end{split}
\end{equation*}
and

\begin{equation*}
    \log\lambda(s, t| \mathcal{H}_t) = \log (\mu + \sum_{t_i \in \mathcal{H}_t} k \cdot \alpha \cdot  \exp\{-\alpha (t-t_i)\}\cdot \frac{1}{2\pi \sigma^2}\cdot  \exp\{-\frac{\sqrt{(x - x_i)^2+(y - y_i)^2}}{2\sigma^2}\}).
\end{equation*}

The next step involves setting prior distributions for each parameter and defining their initial values, which is key in Bayesian analysis. It is important to choose appropriate prior distributions that reflect prior knowledge or expectations about the parameters, while also allowing the data to influence the final estimates. The initial values of the parameters serve as starting points for the inference process and can affect the convergence and stability of the algorithm, as they have a strong influence on the posterior distribution. There are several types of priors and the choice between them depends on the specific problem. We have non-informative or flat priors, which are used when we have little or no prior information about the parameter, assigning equal probability to all possible values within a certain range and allowing the data to speak for itself. On the other hand, informative priors are used when we have prior knowledge about the parameters and assign higher probabilities to values close to the known information, and lower probabilities to values that contradict the prior belief. \ab{A conjugate prior is such that for a given likelihood function, both the prior and posterior distributions belong to the same family.} Finally, hierarchical priors are useful in multi-level models where the parameters themselves may have prior distributions. In addition to the priors and initial values, it is also necessary to choose the coefficients involved in the generation and size of the temporal intervals and spatial circular bands because these coefficients also affect the precision and efficiency of the inference process. If the intervals and bands are too large, we may lose important details about the evolution of the process over time and the spatial variability could be lost. In contrast, if they are too small, there may not be enough information to make accurate inferences, in addition to increasing the computation time without improving the results.

\section{Simulation study}

We conduct a simulation study to illustrate a comparative performance of the simulation and estimation methods. \ab{While the parents-offspring method requires an easily invertible analytical expression, the acceptance-rejection method allows for more flexibility in the form of the intensity function, which may make it preferable when working with more complex or diverse triggering forms. However, this method requires setting an upper bound for the process intensity, which can be challenging. Despite their differences, with both methods we can obtain data that adequately represent a Hawkes process, and the differences in simulation should only depend on the choice of parameters. In Section \ref{sec_simul}, we present a comparison of both simulation methods in terms of computation time and memory usage.} In Section \ref{sec_estim}, simulated data are estimated using the different selected inference methods in order to compare them in terms of accuracy of estimation and computational time. To do this, 300 simulations are generated and then estimated using each of these techniques, presenting their mean absolute error (MAE), mean and standard deviation, along with the mean computational time. Additionally, different scenarios are used to model the intensity function in order to observe their versatility, both in the simulation and estimation methods. We use simulated data with the acceptance-rejection strategy and constant background rates, although we also provide an extended case using the parents-offspring strategy and a non-constant background rate \ab{in Section \ref{sec_extend}} to illustrate the adaptability of the methods in this case as well. All experiments in this section were run in a workstation with 128GB of RAM splitted into 4 cores of 32GB at 3.20GHz.

In this study, we use the following general form of the conditional intensity

\begin{equation*}
    \lambda(s, t|\mathcal{H}_t)=\mu + k \sum_{t_i \in \mathcal{H}_t} g_{\raisebox{-0.3ex}{\ab{\scriptsize T}}}(t) g_{\raisebox{-0.3ex}{\ab{\scriptsize S}}}(s),
\end{equation*}
where we consider a constant background and a separable triggering in both time and space. For these triggering functions, we select two for the time domain and two for the spatial domain, along with their combinations, as discussed in Section \ref{subsec2}, resulting in four scenarios, as described in Table \ref{tabla_scenarios}.
Explicitly, we consider the exponential and power law triggerings for time and the Gaussian and exponential triggerings for space. \ab{All triggering functions are normalized to ensure they integrate to one over their respective domains, which simplifies the stability condition to $k<1$ (as discussed in Section \ref{hawkes_section}).}

\begin{table}[h]
\centering
\renewcommand{\arraystretch}{2} 
\setlength{\tabcolsep}{12pt} 
\begin{tabular}{|c|c|c|}
\hline
\textbf{Scenario} & \textbf{Temporal triggering,} \bm{$g_{\raisebox{-0.3ex}{\ab{\scriptsize T}}}(t)$} & \textbf{Spatial triggering,} \bm{$g_{\raisebox{-0.3ex}{\ab{\scriptsize S}}}(s)$} \\
\hline
\hline
1 & $\alpha \cdot \exp\{-\alpha (t-t_i)\}$ 
& $\frac{1}{2\pi \sigma^2} \cdot \exp\{-\frac{(x-x_i)^2+(y-y_i)^2}{2\sigma^2}\}$ \\
\hline
2 & $\alpha \cdot \exp\{-\alpha (t-t_i)\}$ & $\frac{1}{2\pi \beta^2} \cdot \exp\{-\frac{\sqrt{(x-x_i)^2+(y-y_i)^2}}{\beta}\}$ \\
\hline
3 & $(\gamma-1) \cdot (1+(t-t_i))^{-\gamma}$ & $\frac{1}{2\pi \sigma^2} \cdot \exp\{-\frac{(x-x_i)^2+(y-y_i)^2}{2\sigma^2}\}$ \\
\hline
4 & $(\gamma-1) \cdot (1+(t-t_i))^{-\gamma}$ & $\frac{1}{2\pi \beta^2} \cdot \exp\{-\frac{\sqrt{(x-x_i)^2+(y-y_i)^2}}{\beta}\}$ \\
\hline
\end{tabular}
\caption{Triggering functions considered in the simulation study.}
\label{tabla_scenarios}
\end{table}

\subsection{Comparative analysis of simulation methods} \label{sec_simul}

This section presents a comparative analysis of the acceptance-rejection and parents-offspring simulation methods. The aim is to evaluate their computational performance in terms of execution time and memory usage. To this end, 300 simulations are carried out for each method under two parameter configurations for scenario 1 (specified in Table \ref{tabla_scenarios}). Table \ref{compar_simul} reports the mean and standard deviation of the number of simulated events, execution time, and memory usage in the simulations.

\begin{table}[H]
\begin{minipage}{\textwidth}
\centering

\begin{tabular}{llccc}
\hline
\textbf{Method} & \boldsymbol{$\hat{\mathbb{E}}(N)$} & \textbf{Time (s)} & \textbf{Memory (kB)} \\
\hline
Acceptance-rejection & $ 956\pm  59.772$ & $ 1.740 \pm 0.254 $ & $23.141\pm  7.989$ \\
Parents-offspring  & $ 994 \pm 60.684$ & $ 0.005 \pm   0.001$ & $26.102 \pm 1.606 $ \\
\hline
\end{tabular}
\end{minipage}
\caption*{(i) $\mu=5, k=0.5, \alpha=2.5, \sigma=0.03$}

\vspace{0.5cm}

\begin{minipage}{\textwidth}
\centering
\begin{tabular}{llccc}
\hline
\textbf{Method} & \boldsymbol{$\hat{\mathbb{E}}(N)$} & \textbf{Time (s)} & \textbf{Memory (kB)} \\
\hline
Acceptance-rejection & $1938 \pm 298.681$ & $1.464 \pm  0.383$ & $34.269 \pm 5.453$ \\
Parents-offspring  & $2379 \pm 477.857$ & $0.024 \pm 0.005$ & $ 67.577 \pm 11.857$ \\
\hline
\end{tabular}
\caption*{(ii) $\mu=3, k=0.9, \alpha=1, \sigma=0.05$}
\end{minipage}
\caption{Mean and standard deviation of the number of simulated events, execution time (in seconds) and memory usage (in kilobytes), based on 300 simulations per method.}
\label{compar_simul}
\end{table}

In general terms, the parents-offspring method generates a larger number of data points (under the same set of parameters). This is partly due to the nature of each method and may vary depending on the stopping criterion of the parents-offspring method or the upper bound of the acceptance-rejection method. Furthermore, the parents-offspring method is faster than the acceptance-rejection method because the latter relies on generating samples from a dominating Poisson process and then checking whether they should be accepted or not, which increases computational cost. In particular, the efficiency of the acceptance-rejection method decreases as the upper bound used for the intensity function increases (which grows with parameters that make the intensity higher), since a higher bound leads to a greater number of candidate events. In contrast, the parents-offspring method is more efficient as it employs a hierarchical structure that allows for a more direct generation of samples. However, the acceptance-rejection method uses less memory, which may be due to the fact that only the accepted events need to be stored. Conversely, the parents-offspring method may require storing more data related to the generation of offspring, as well as additional relationships during the simulation.

\subsection{Comparative analysis of inference methods} \label{sec_estim}

\begin{figure}[H]
    \centering
    \begin{subfigure}[t]{0.45\textwidth}
        \centering
        \includegraphics[width=0.9\linewidth]{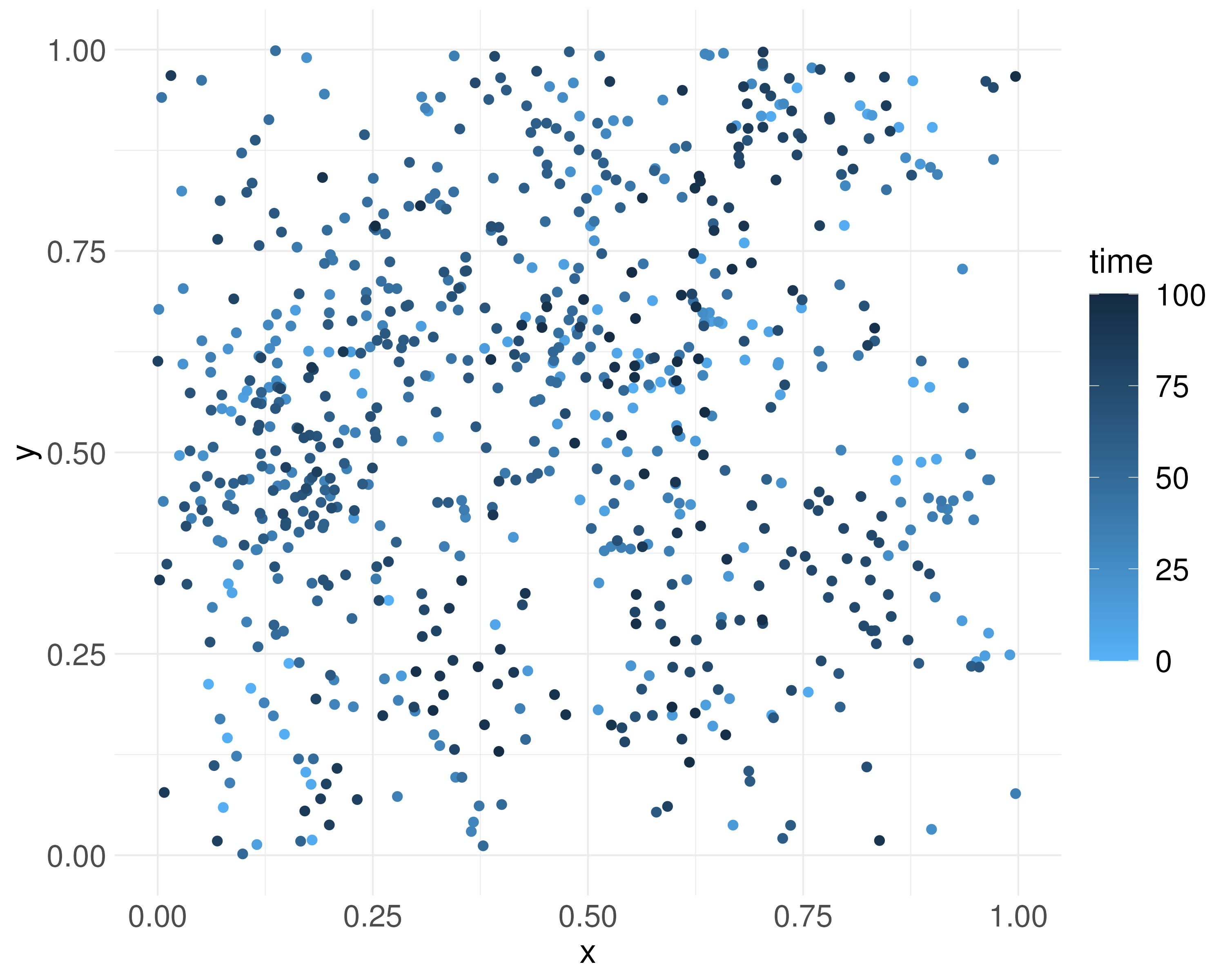}
    \end{subfigure}
    \hfill
    \begin{subfigure}[t]{0.45\textwidth}
        \centering
        \includegraphics[width=0.9\linewidth]{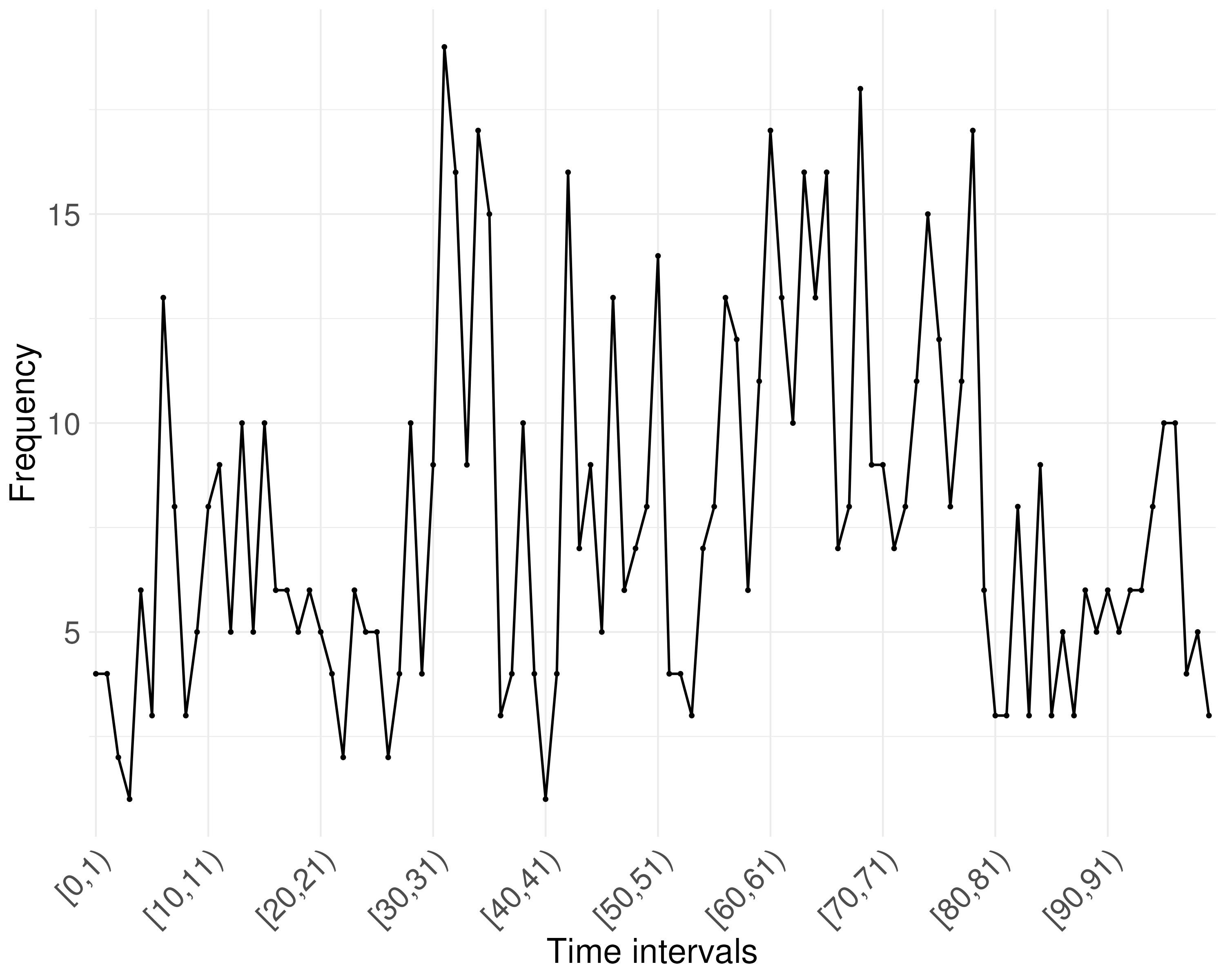}
    \end{subfigure}
    \vspace{-0.3cm}
    \caption*{Scenario 1 (a): $\mu=2, k=0.85, \alpha=1, \sigma=0.05$}
 \centering
    \begin{subfigure}[t]{0.45\textwidth}
        \centering
        \includegraphics[width=0.9\linewidth]{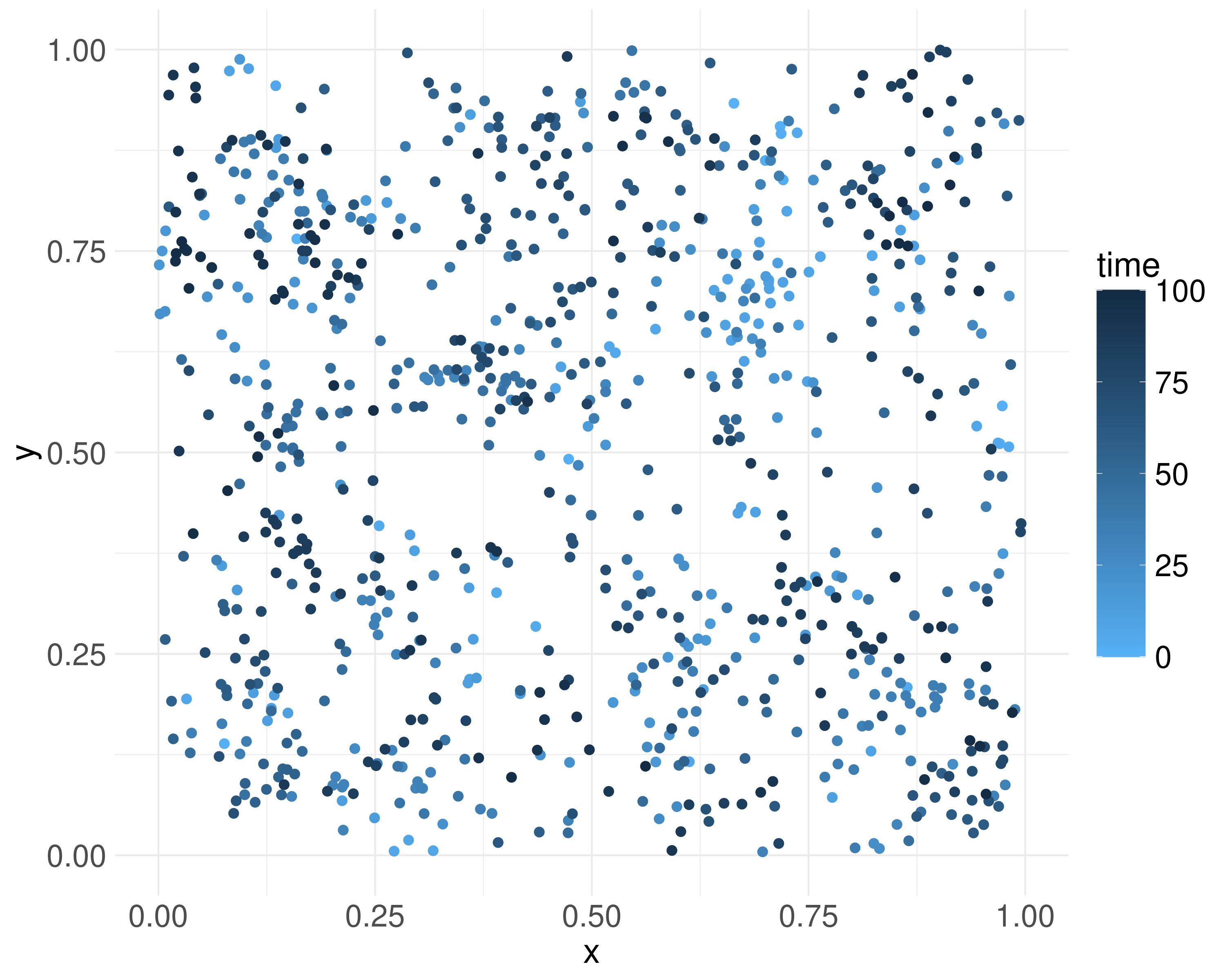}
    \end{subfigure}
    \hfill
    \begin{subfigure}[t]{0.45\textwidth}
        \centering
        \includegraphics[width=0.9\linewidth]{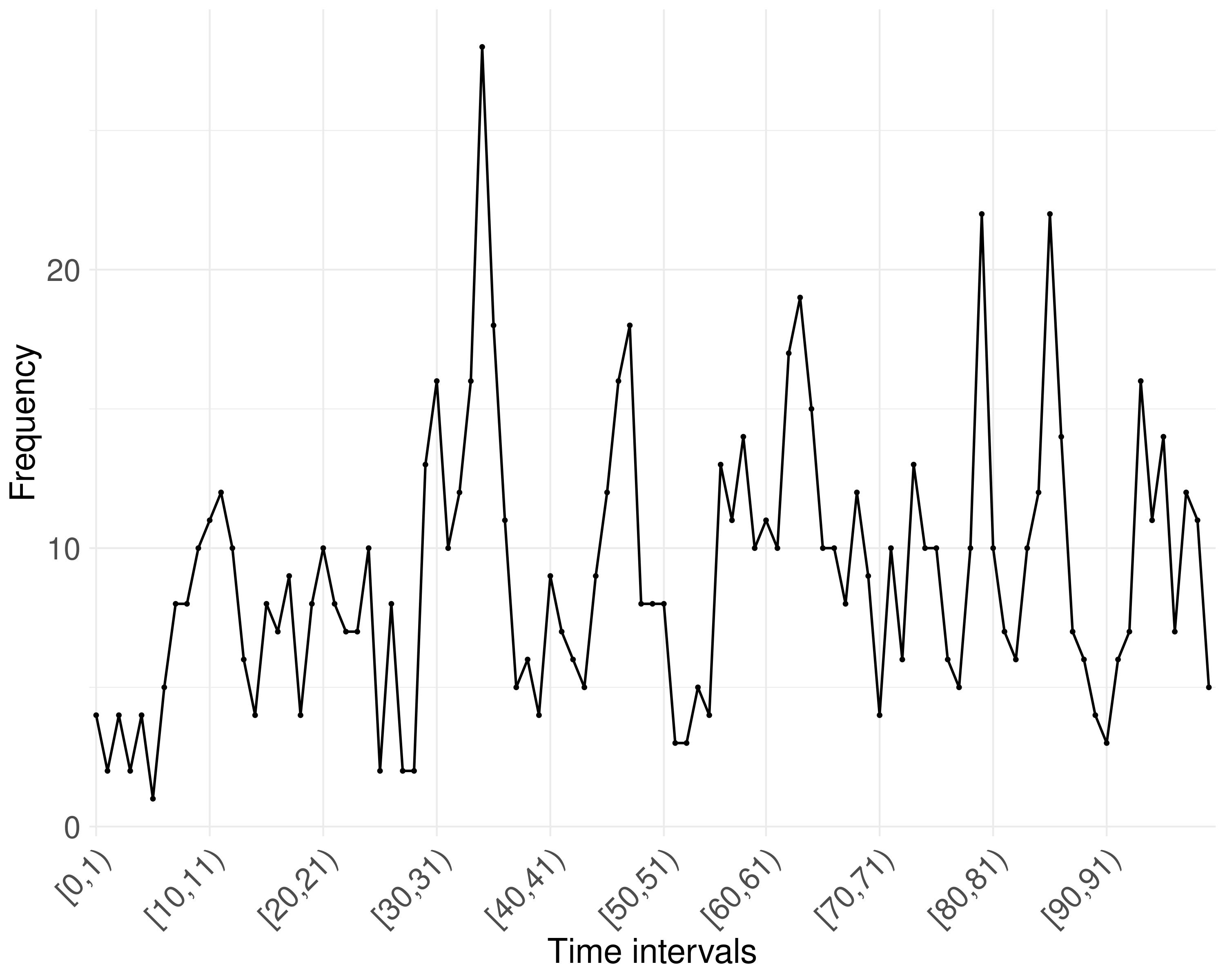}
    \end{subfigure}
        \vspace{-0.3cm}
    \caption*{Scenario 1 (b): $\mu=4, k=0.6, \alpha=2.5, \sigma=0.03$}
     \centering
    \begin{subfigure}[t]{0.45\textwidth}
        \centering
        \includegraphics[width=0.9\linewidth]{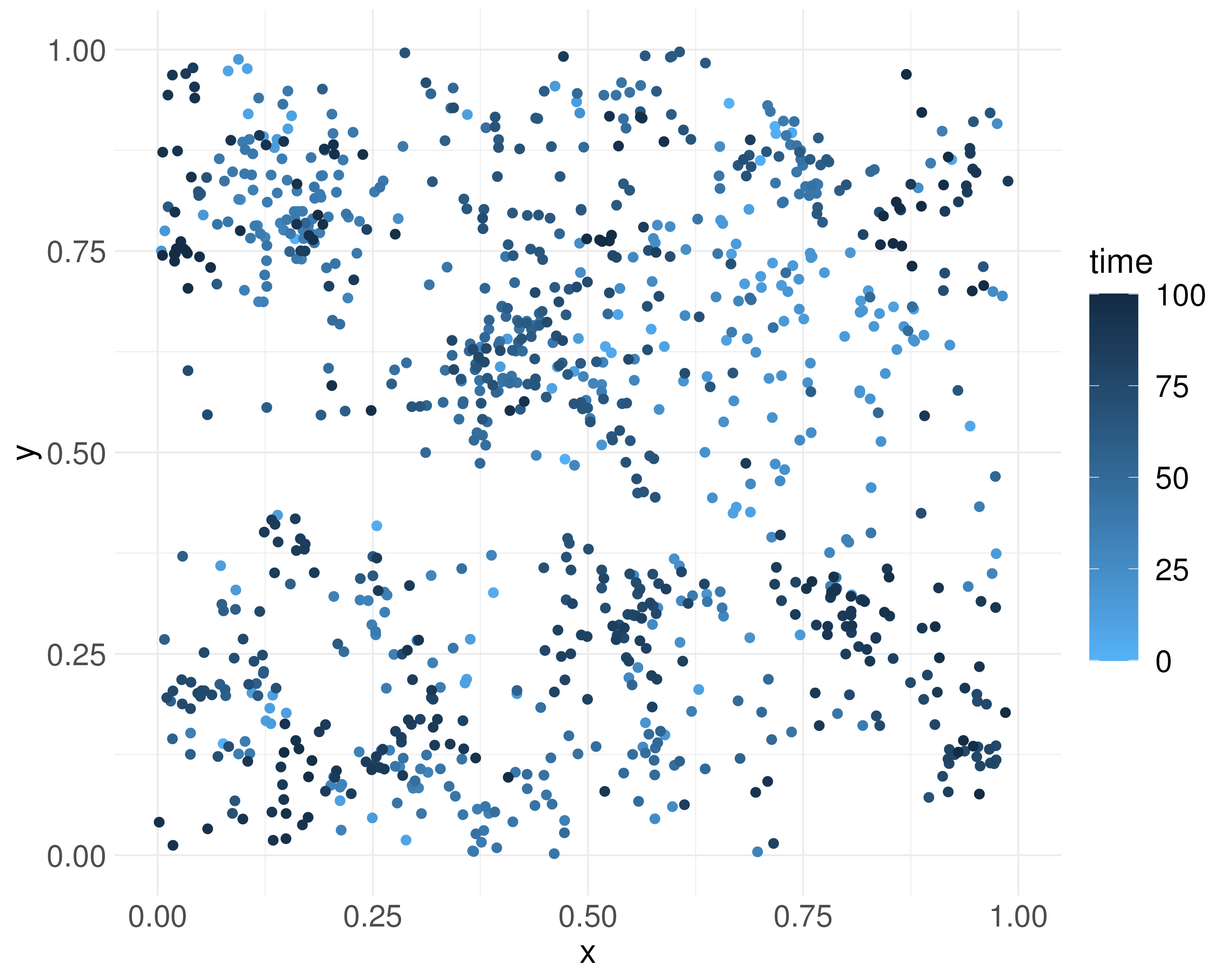}
    \end{subfigure}
    \hfill
    \begin{subfigure}[t]{0.45\textwidth}
        \centering
        \includegraphics[width=0.9\linewidth]{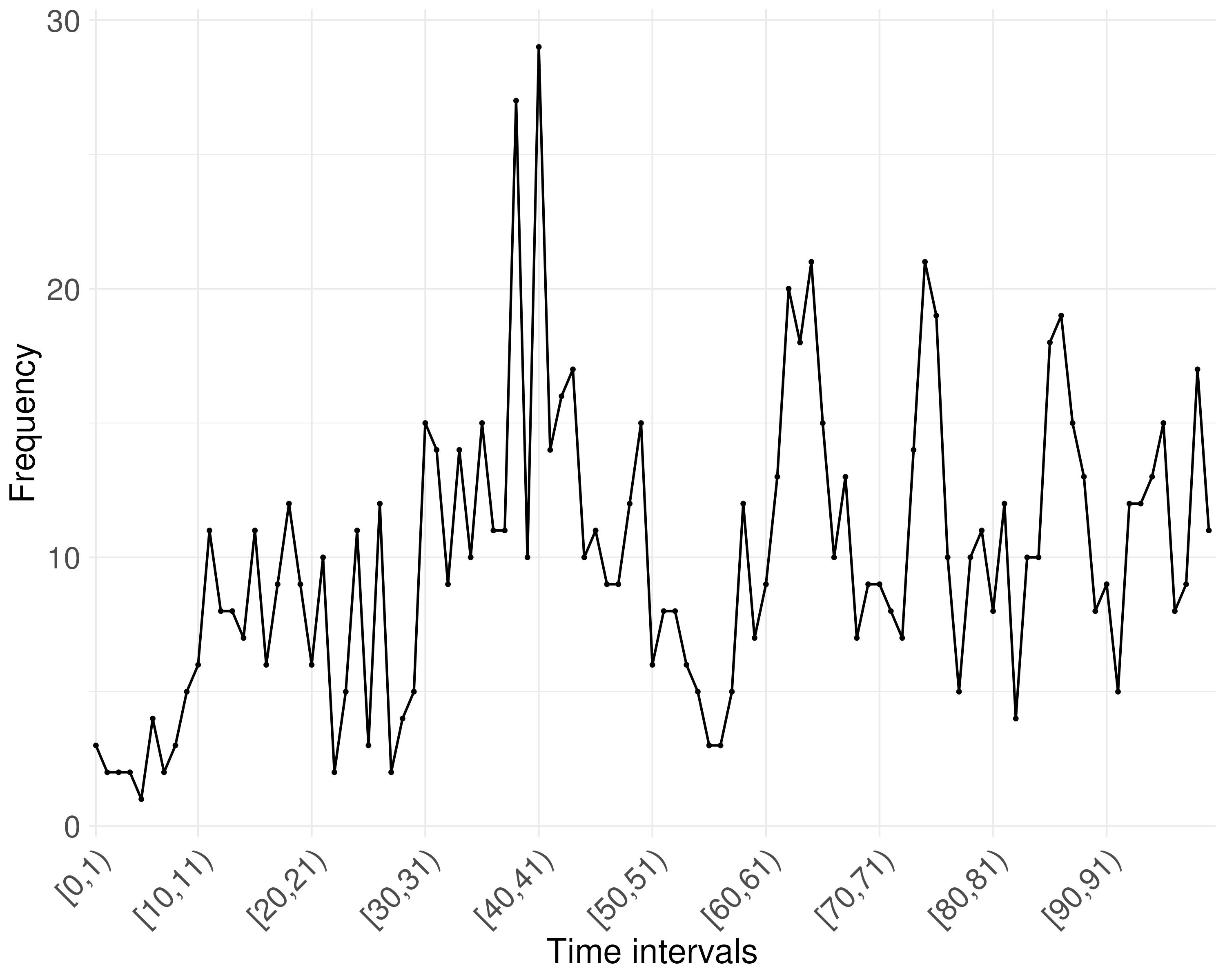}
    \end{subfigure}
        \vspace{-0.3cm}
    \caption*{Scenario 2 (a): $\mu=2, k=0.85, \alpha=1, \beta=0.02$}
     \centering
    \begin{subfigure}[t]{0.45\textwidth}
        \centering
        \includegraphics[width=0.9\linewidth]{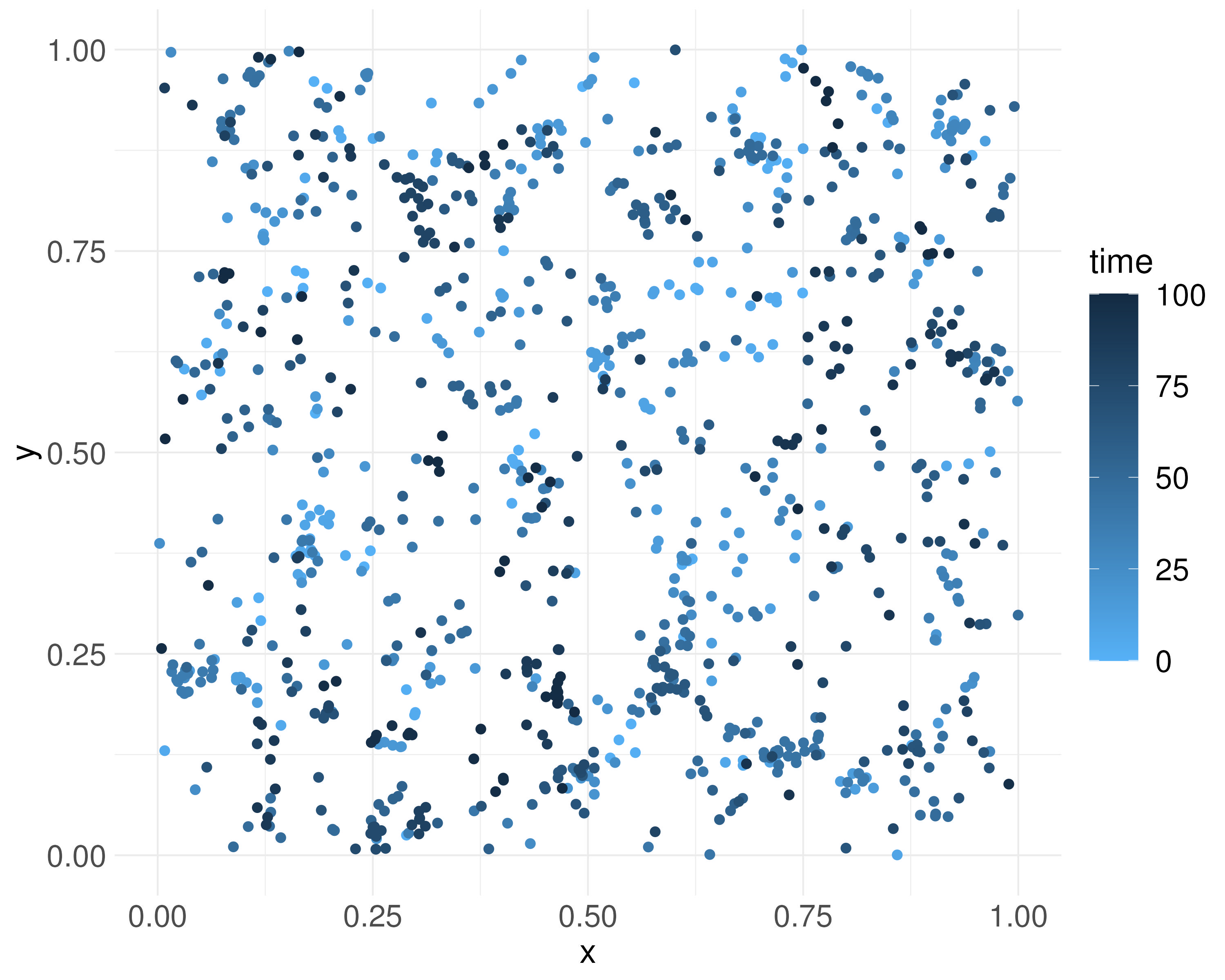}
    \end{subfigure}
    \hfill
    \begin{subfigure}[t]{0.45\textwidth}
        \centering
        \includegraphics[width=0.9\linewidth]{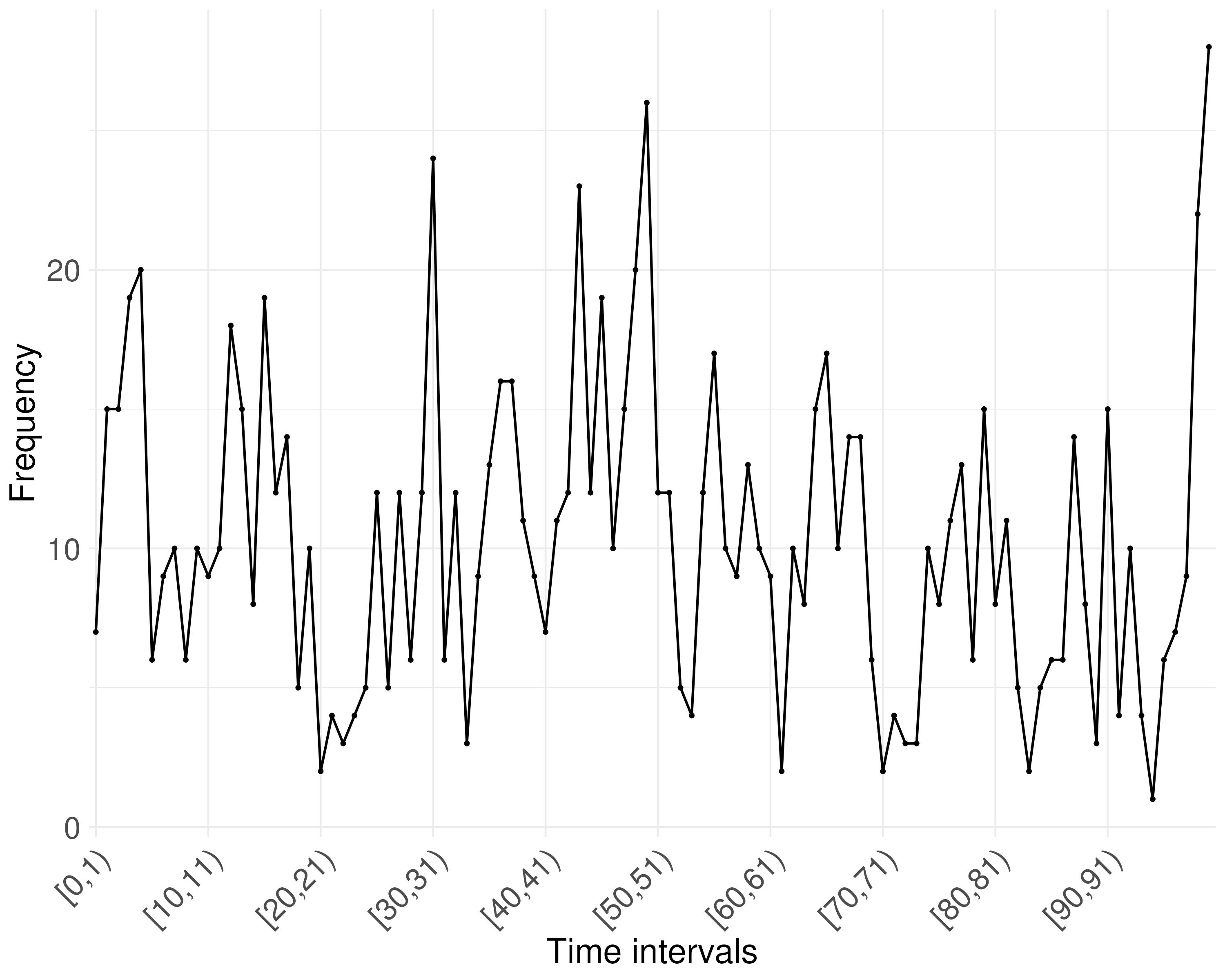}
    \end{subfigure}
        \vspace{-0.3cm}
    \caption*{Scenario 2 (b): $\mu=4, k=06, \alpha=2.5, \beta=0.01$}
            \vspace{-0.2cm}
\caption{Examples of simulated data using the acceptance-rejection method for scenarios 1 and 2 (as described in Table \ref{tabla_scenarios}), with the parameter configuration specified for each row. On the left, the point patterns are shown, and on the right, timelines display the frequency of data within unit time intervals.}
\label{fig:figura2a}
\end{figure}

\begin{figure}[H]
    \centering
    \begin{subfigure}[t]{0.45\textwidth}
        \centering
        \includegraphics[width=0.9\linewidth]{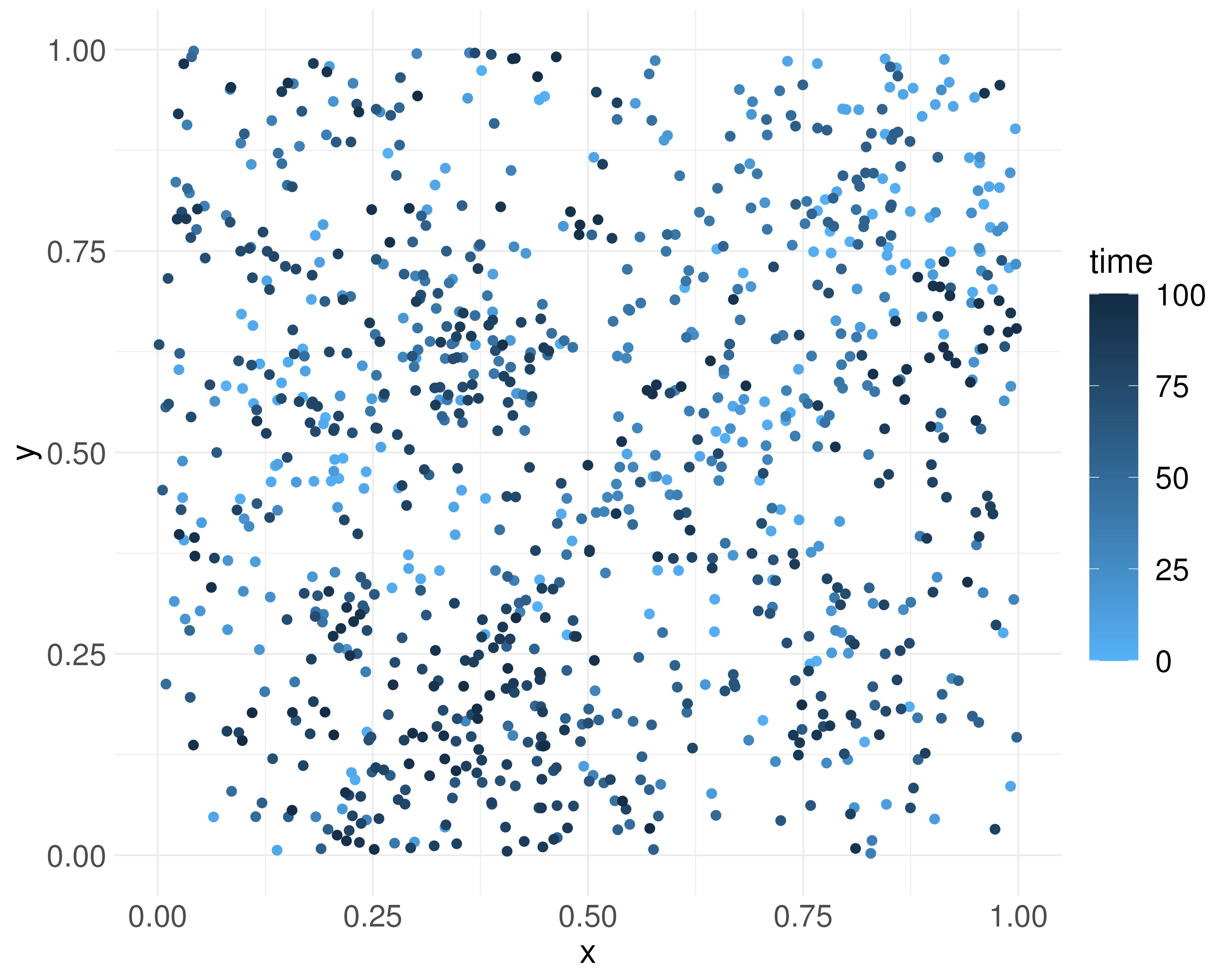}
    \end{subfigure}
    \hfill
    \begin{subfigure}[t]{0.45\textwidth}
        \centering
        \includegraphics[width=0.9\linewidth]{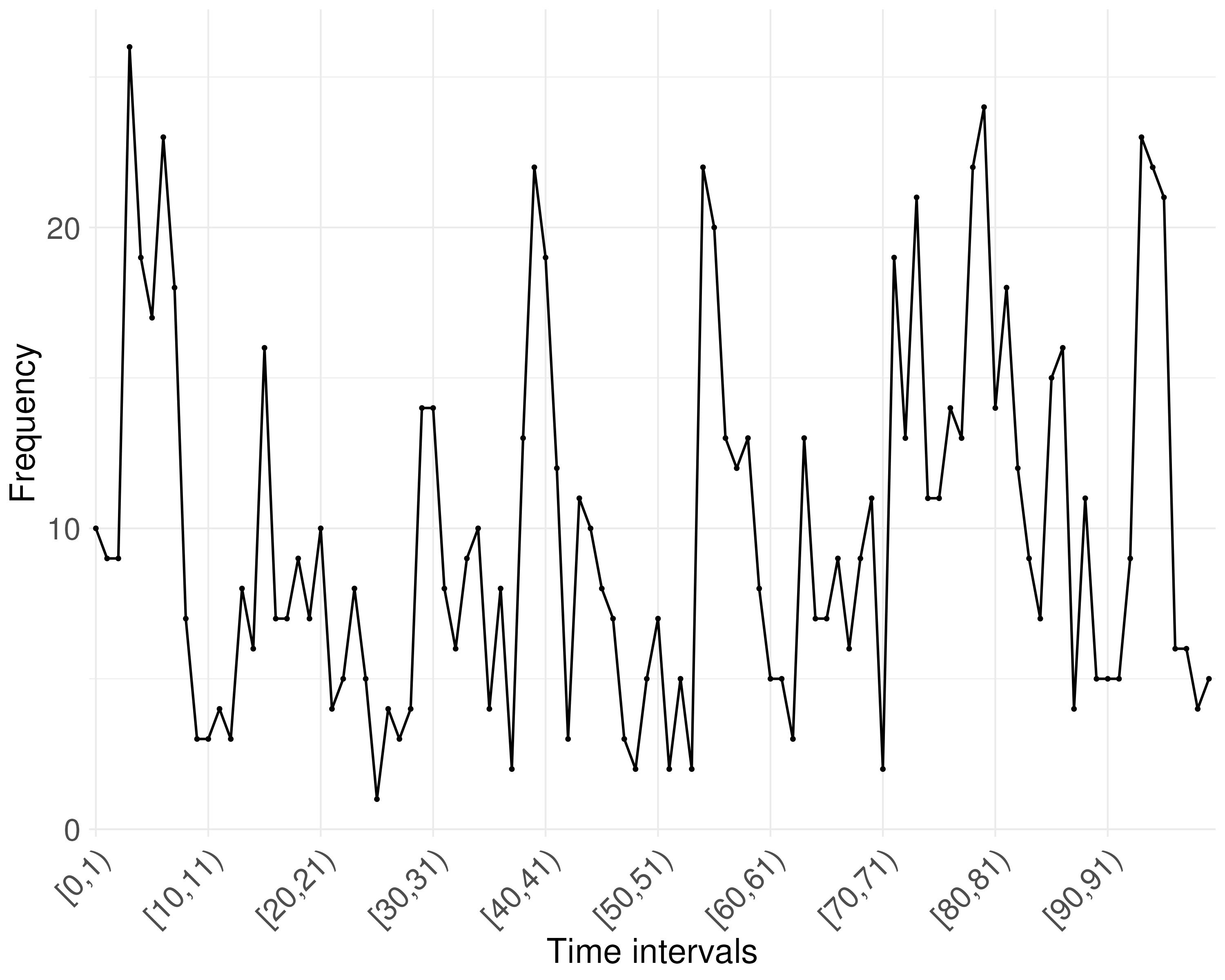}
    \end{subfigure}
        \vspace{-0.3cm}
    \caption*{Scenario 3 (a): $\mu=3, k=0.75, \gamma=3.5, \sigma=0.05$}
 \centering
    \begin{subfigure}[t]{0.45\textwidth}
        \centering
        \includegraphics[width=0.9\linewidth]{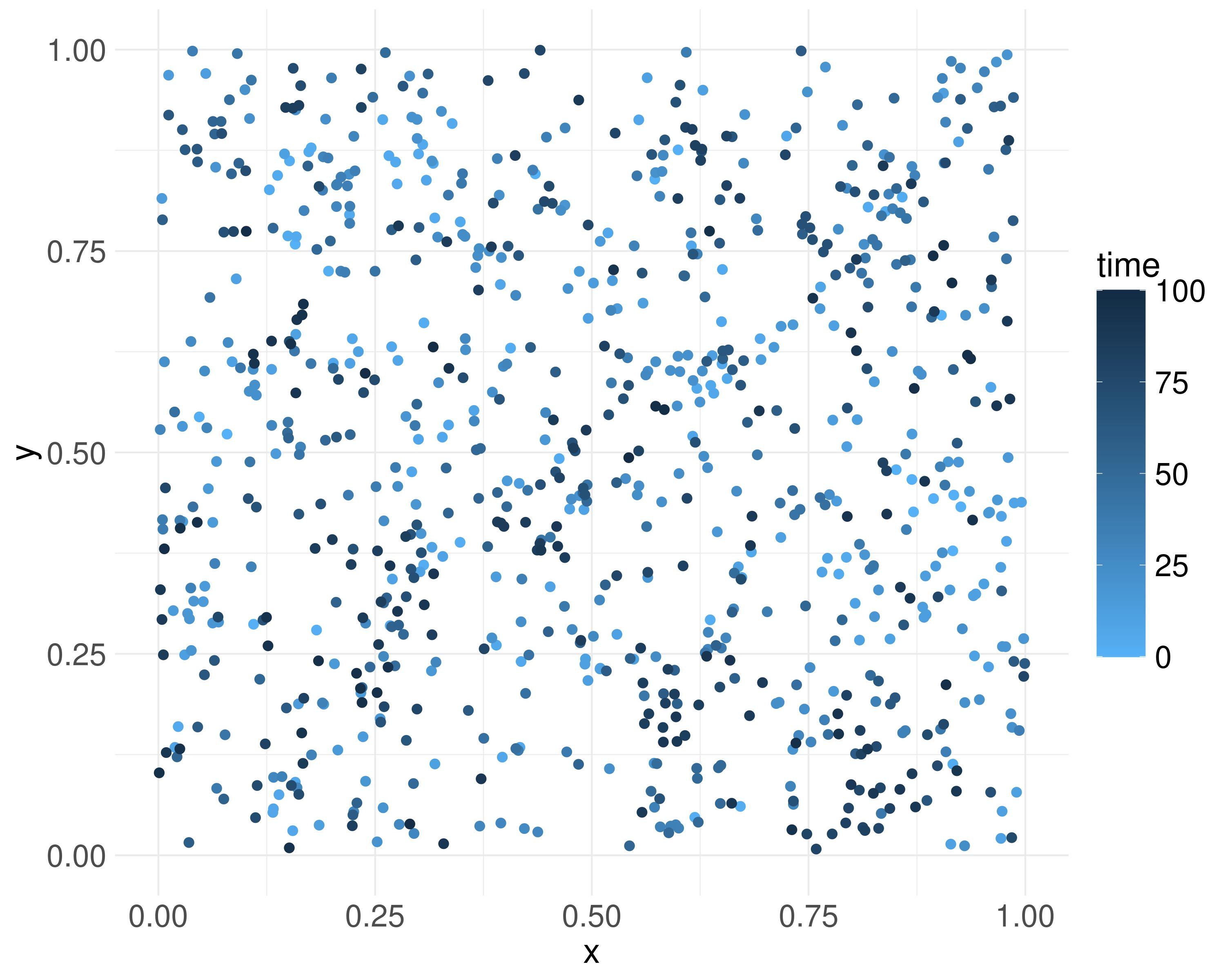}
    \end{subfigure}
    \hfill
    \begin{subfigure}[t]{0.45\textwidth}
        \centering
        \includegraphics[width=0.9\linewidth]{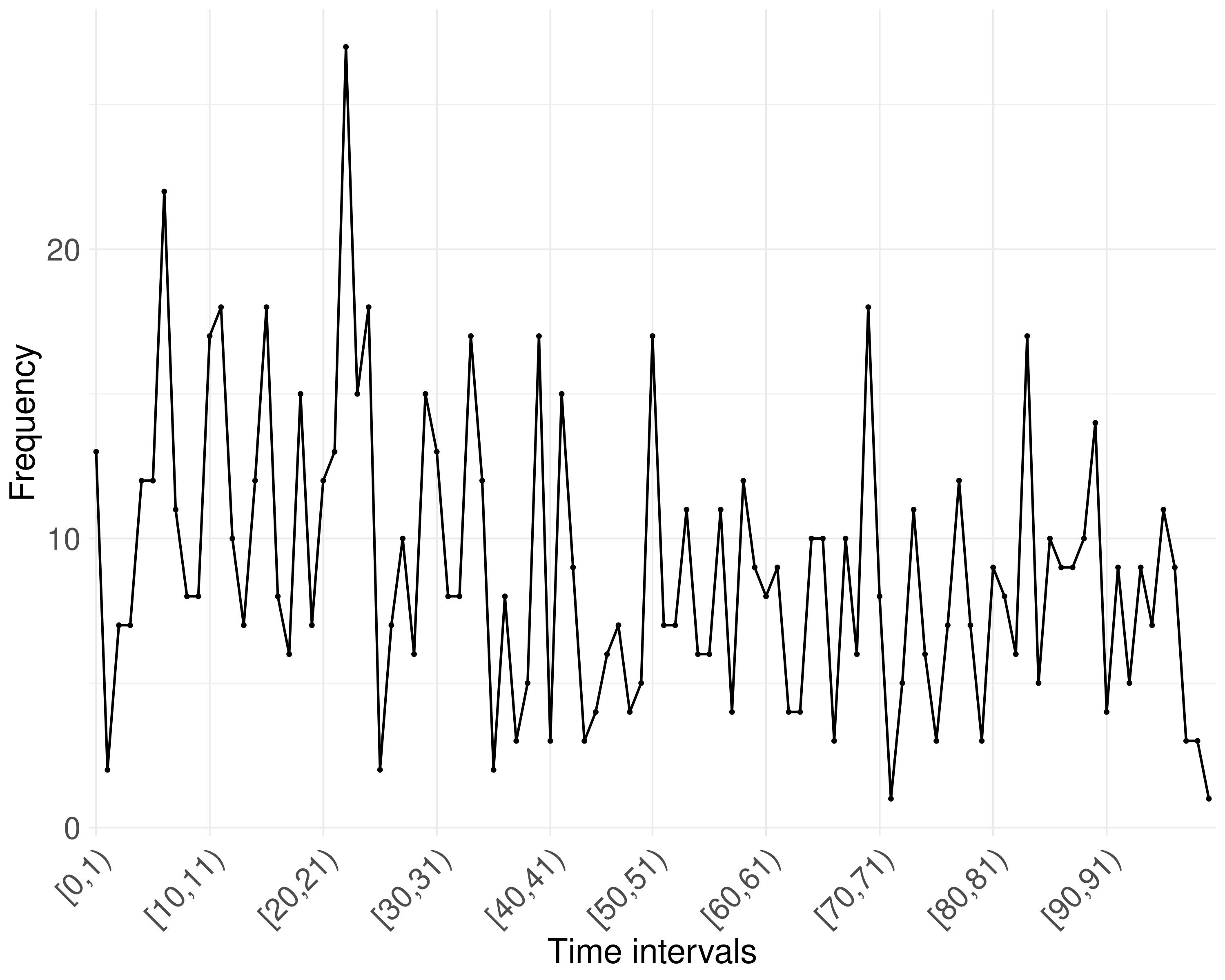}
    \end{subfigure}
            \vspace{-0.3cm}
    \caption*{Scenario 3 (b): $\mu=5, k=0.5, \gamma=6, \sigma=0.03$}
     \centering
    \begin{subfigure}[t]{0.45\textwidth}
        \centering
        \includegraphics[width=0.9\linewidth]{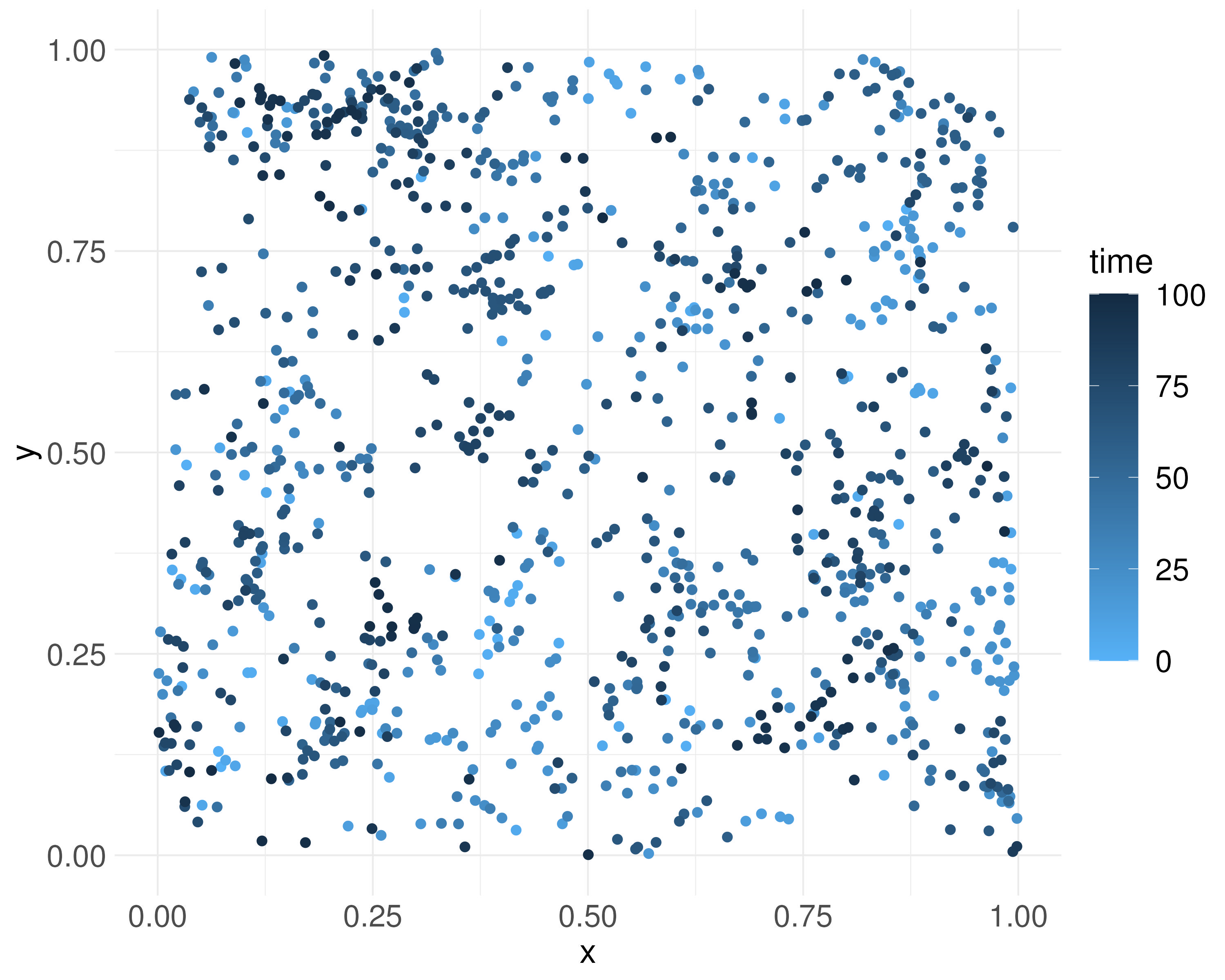}
    \end{subfigure}
    \hfill
    \begin{subfigure}[t]{0.45\textwidth}
        \centering
        \includegraphics[width=0.9\linewidth]{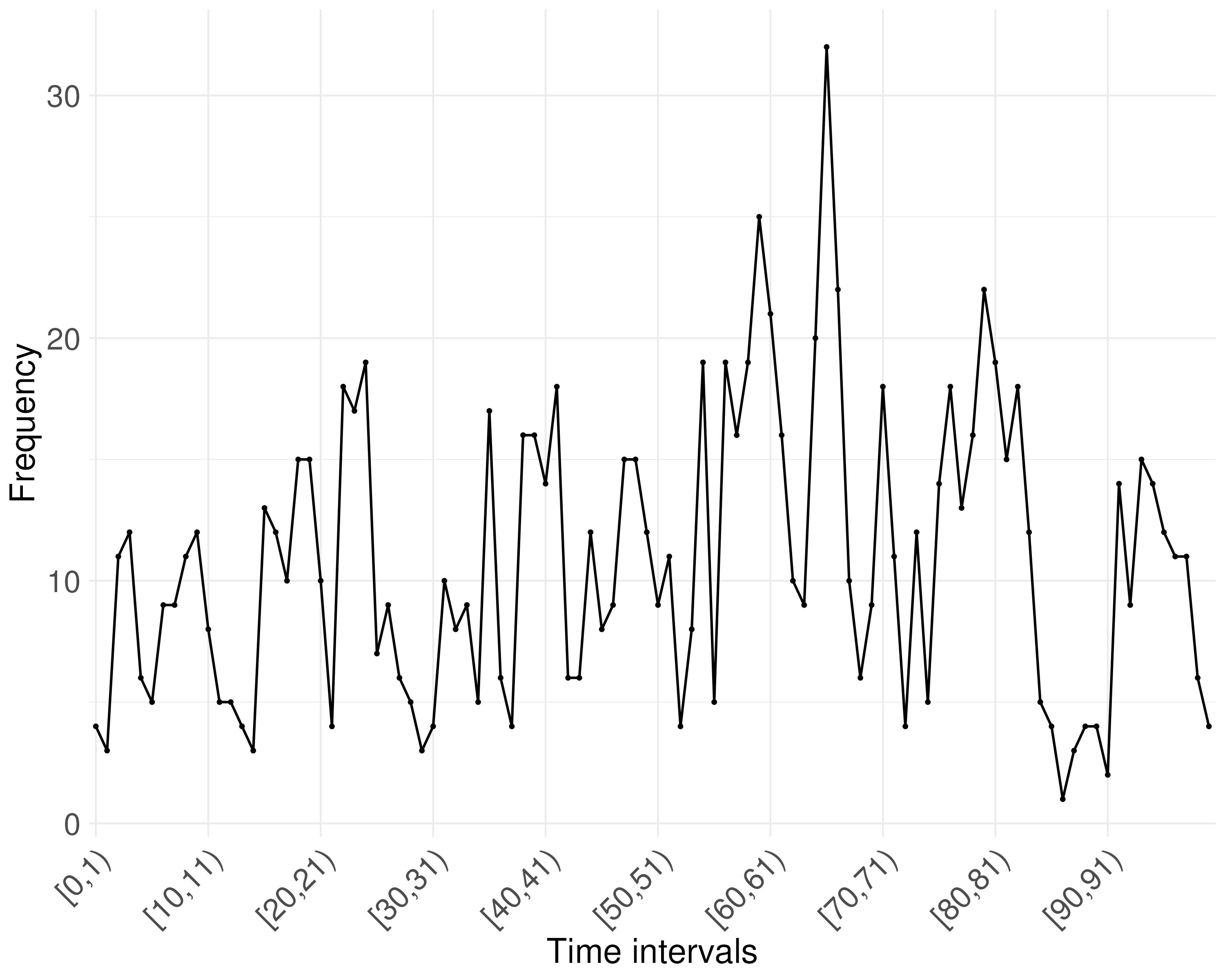}
    \end{subfigure}
            \vspace{-0.3cm}
    \caption*{Scenario 4 (a): $\mu=3, k=0.75, \gamma=3.5, \beta=0.02$}
     \centering
    \begin{subfigure}[t]{0.45\textwidth}
        \centering
        \includegraphics[width=0.9\linewidth]{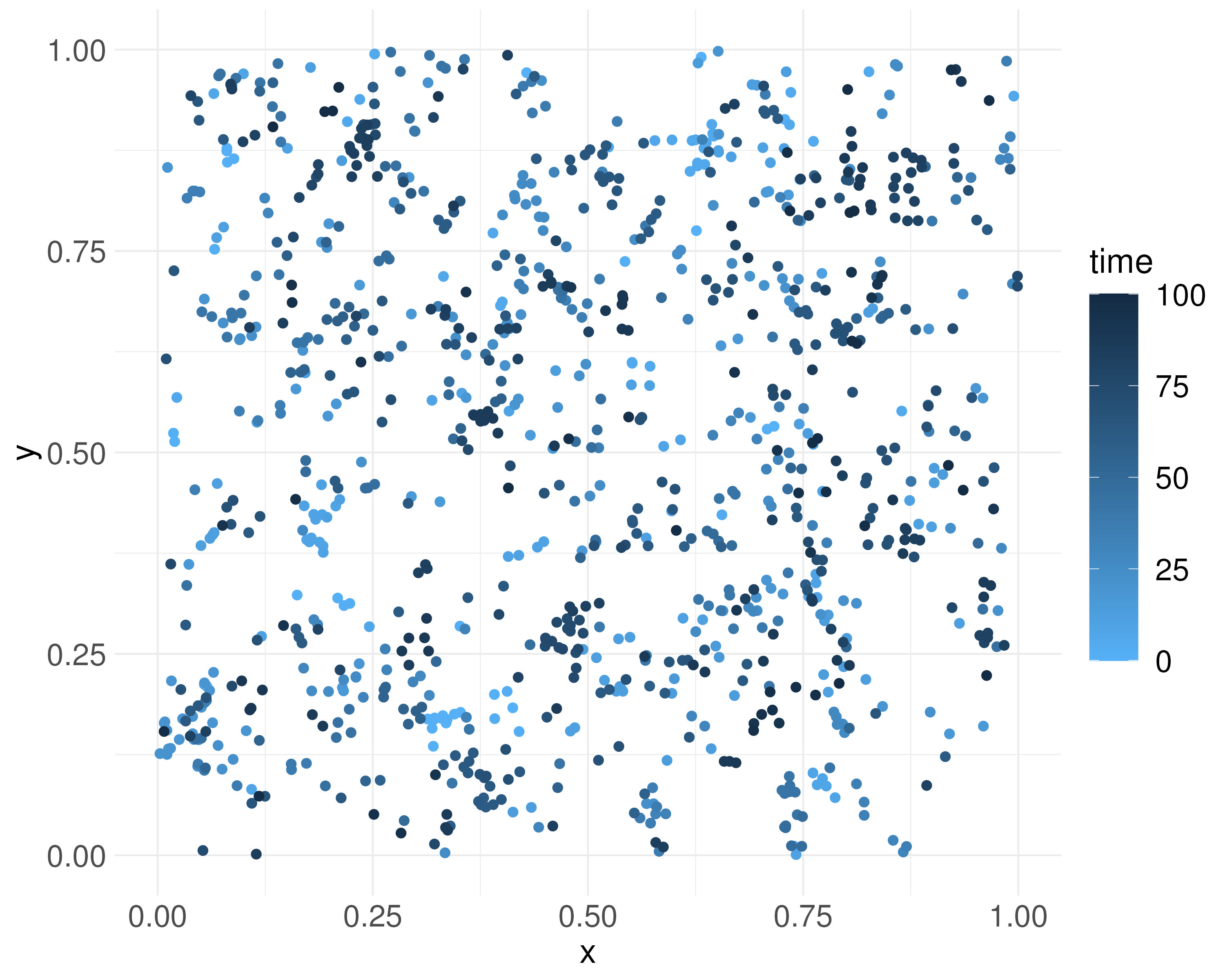}
    \end{subfigure}
    \hfill
    \begin{subfigure}[t]{0.45\textwidth}
        \centering
        \includegraphics[width=0.9\linewidth]{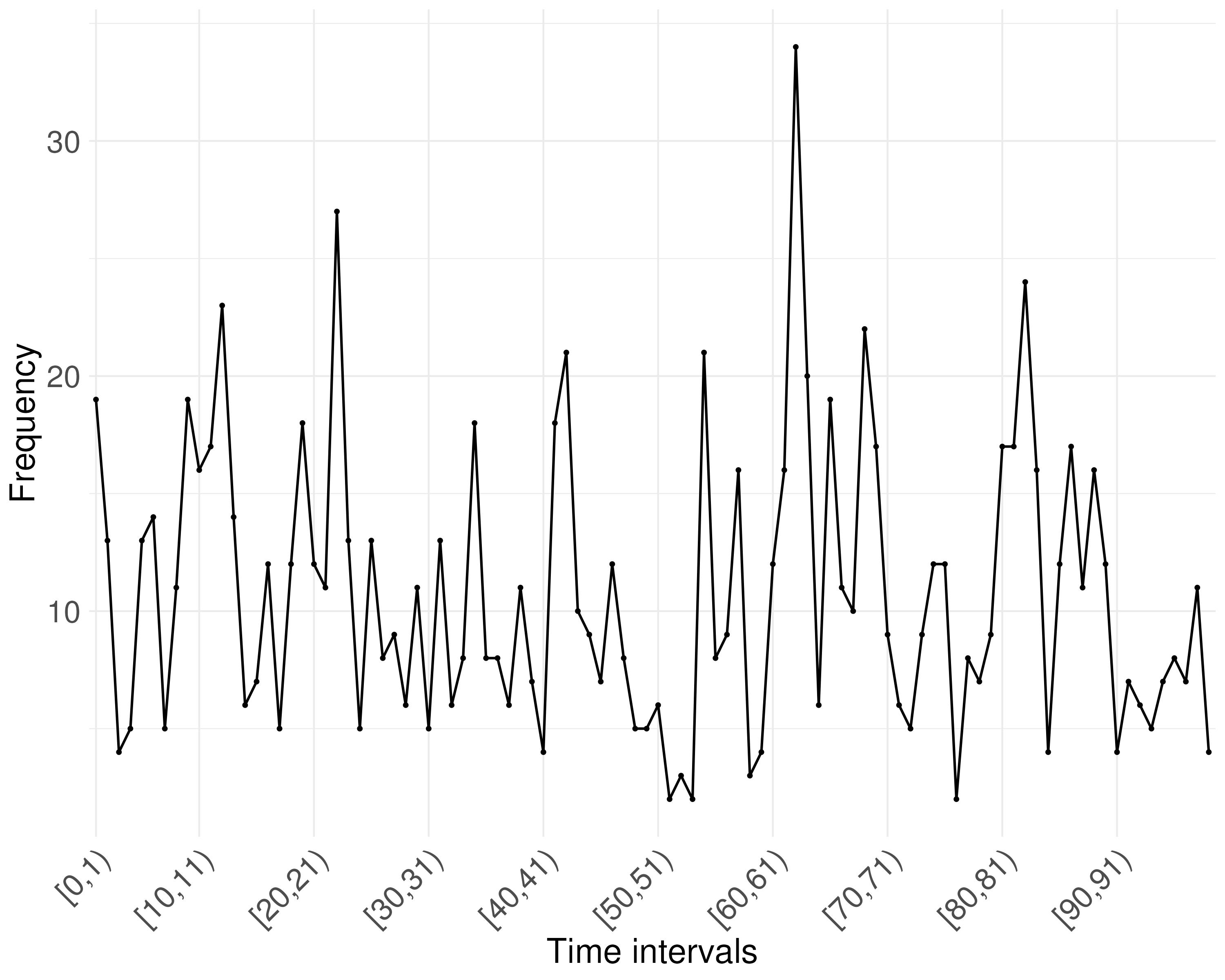}
    \end{subfigure}
            \vspace{-0.3cm}
    \caption*{Scenario 4 (b): $\mu=5, k=0.5, \gamma=6, \beta=0.01$}
            \vspace{-0.2cm}
\caption{Examples of simulated data using the acceptance-rejection method for scenarios 3 and 4 (as described in Table \ref{tabla_scenarios}), with the parameter configuration specified for each row. On the left, the point patterns are shown, and on the right, timelines display the frequency of data within unit time intervals.}
\label{fig:figura2b}
\end{figure}

For each of the scenarios in Table \ref{tabla_scenarios}, two sets of parameter configurations are considered, with which data sets are generated, as depicted in Figures \ref{fig:figura2a} and  \ref{fig:figura2b}. Regarding the effect of the parameters, the background rate $\mu$ controls the amount of independent events in the generated data. When $\mu$ is large, the data may appear more uniform because background events dominate. Conversely, when $\mu$ is low, the events generated by the triggering (clusters) become more evident. As for the value of $k$, this parameter controls how many secondary events are generated by each event. Therefore, a higher $k$ tends to generate a greater number of points, although other parameters also influence this aspect. The spatial parameters ($\sigma$ for the Gaussian triggering or $\beta$ for the exponential triggering) control the level of dispersion, and with smaller values, the data tends to be more spatially clustered due to the rapid decay of the triggering functions. Moreover, the Gaussian triggering decays faster than the exponential at short distances, which explains the denser clusters. Regarding the parameter for exponential temporal triggering ($\alpha$), the larger its value, the faster the temporal influence decays, causing events to cluster more closely in time. The power law, on the other hand, has a slower decay controlled by $\gamma$ (longer tail), allowing events to continue influencing subsequent occurrences over a longer period, thereby creating a more extended temporal dependency.

\begin{figure}[H]
  \begin{minipage}[b]{0.45\textwidth}
    \centering
    \includegraphics[width=\textwidth, height=0.75\textwidth]{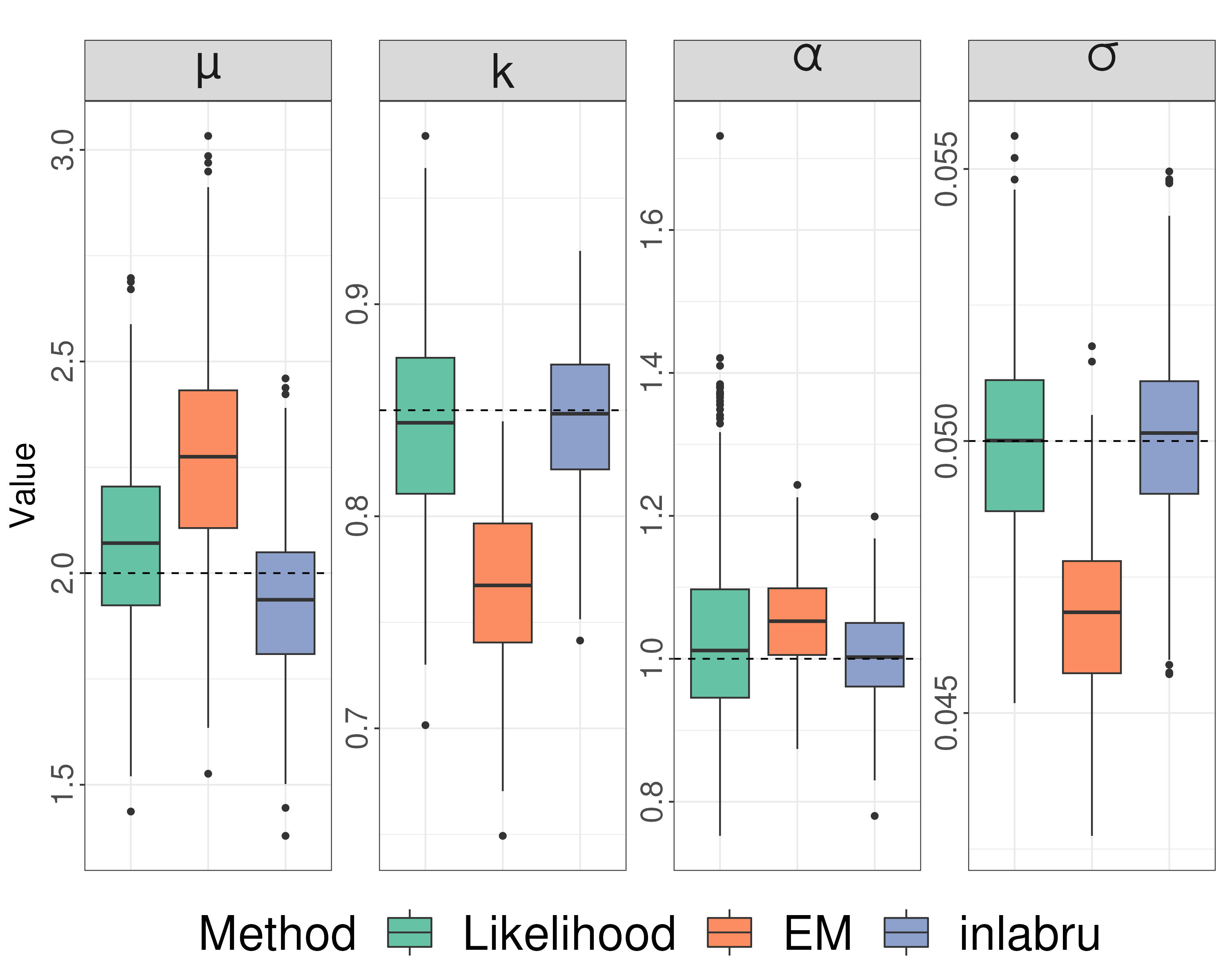}
    \caption*{1 (a): $\mu=2, k=0.85, \alpha=1, \sigma=0.05$}
  \end{minipage}%
  \begin{minipage}[b]{0.45\textwidth}
    \centering
    \includegraphics[width=\textwidth, height=0.75\textwidth]{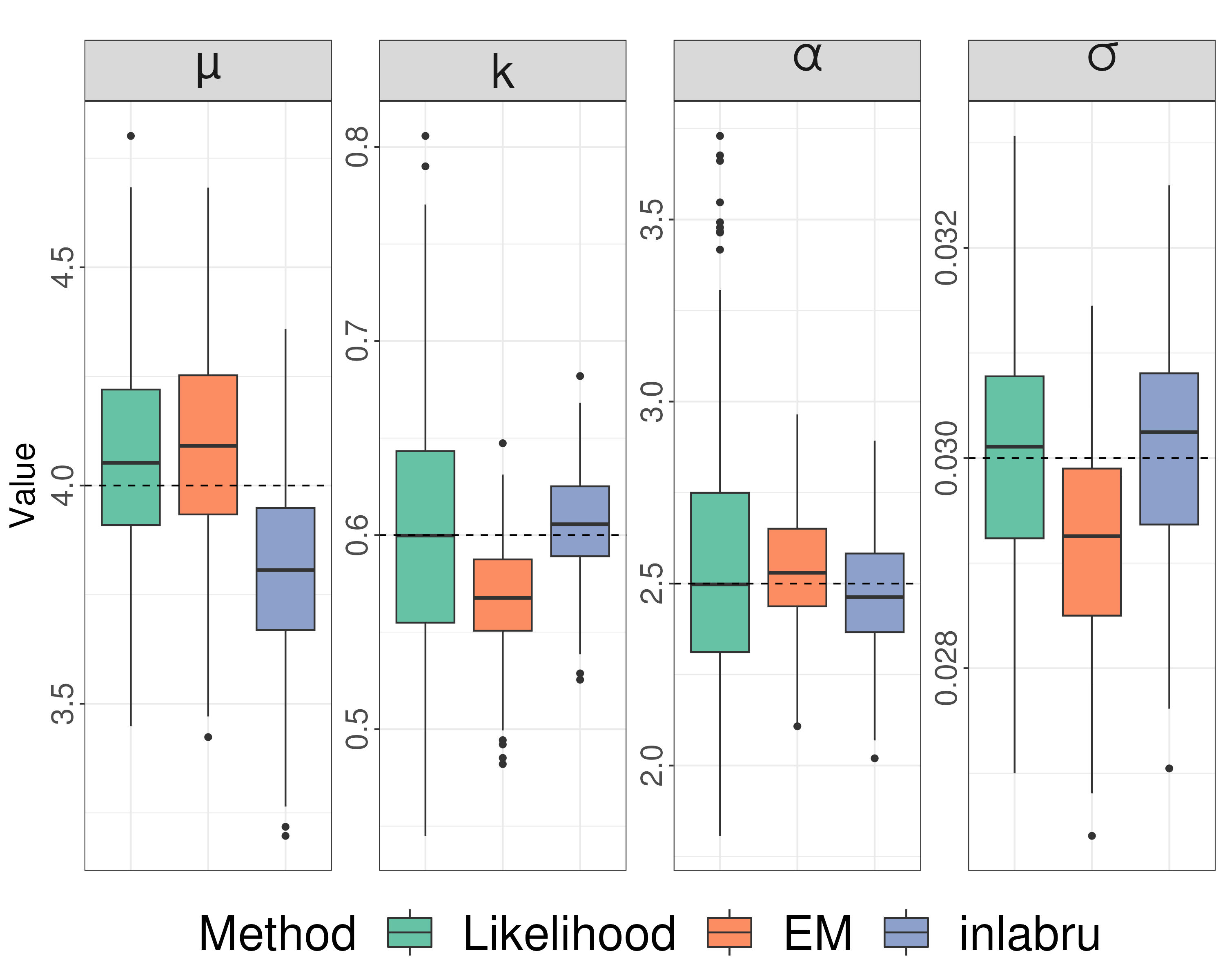}
    \caption*{1 (b): $\mu=4, k=0.6, \alpha=2.5, \sigma=0.03$}
  \end{minipage}
    \medskip
  \begin{minipage}[b]{0.45\textwidth}
    \centering
    \includegraphics[width=\textwidth, height=0.75\textwidth]{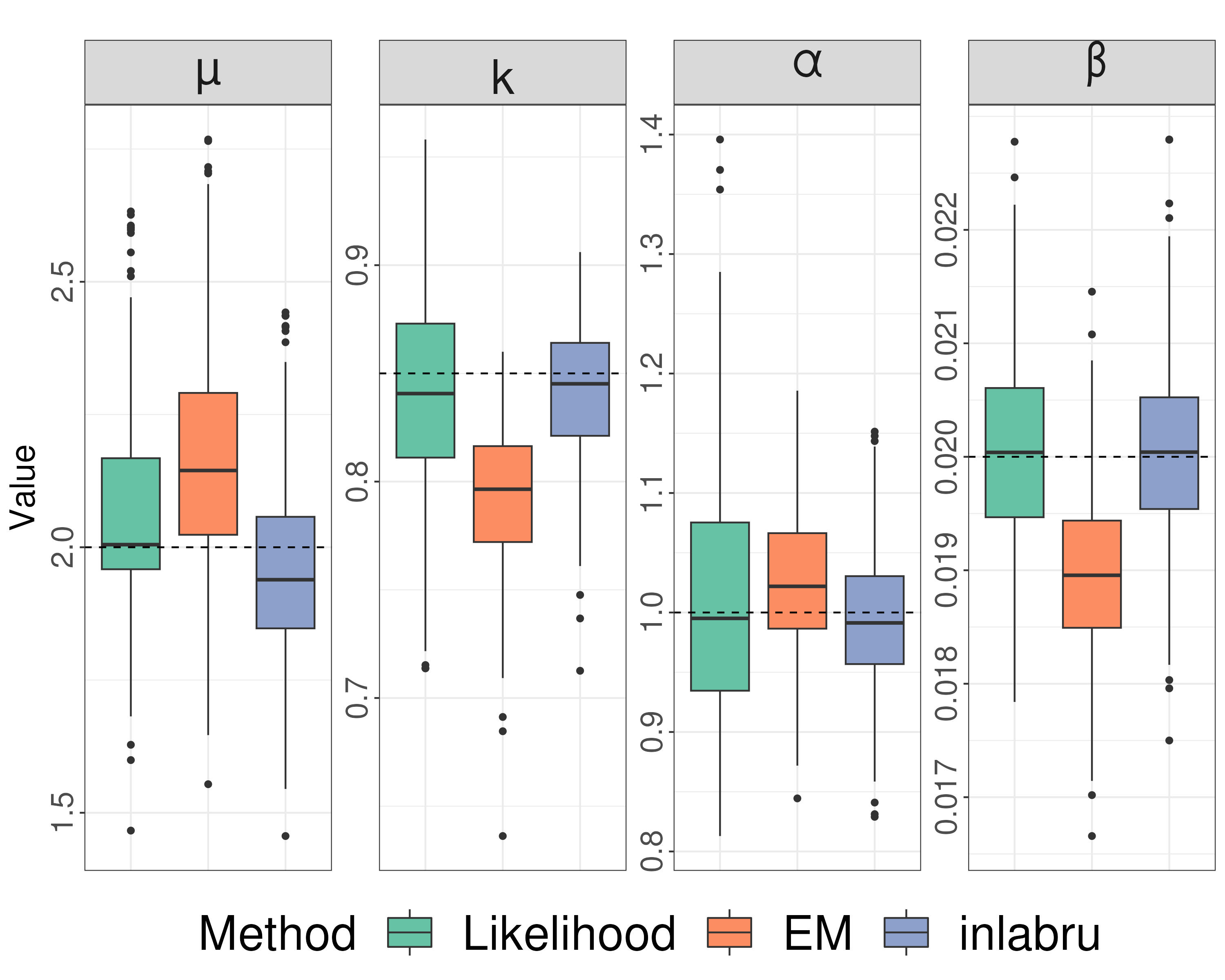}
    \caption*{2 (a): $\mu=2, k=0.85, \alpha=1, \beta=0.02$}
  \end{minipage}%
  \begin{minipage}[b]{0.45\textwidth}
    \centering
    \includegraphics[width=\textwidth, height=0.75\textwidth]{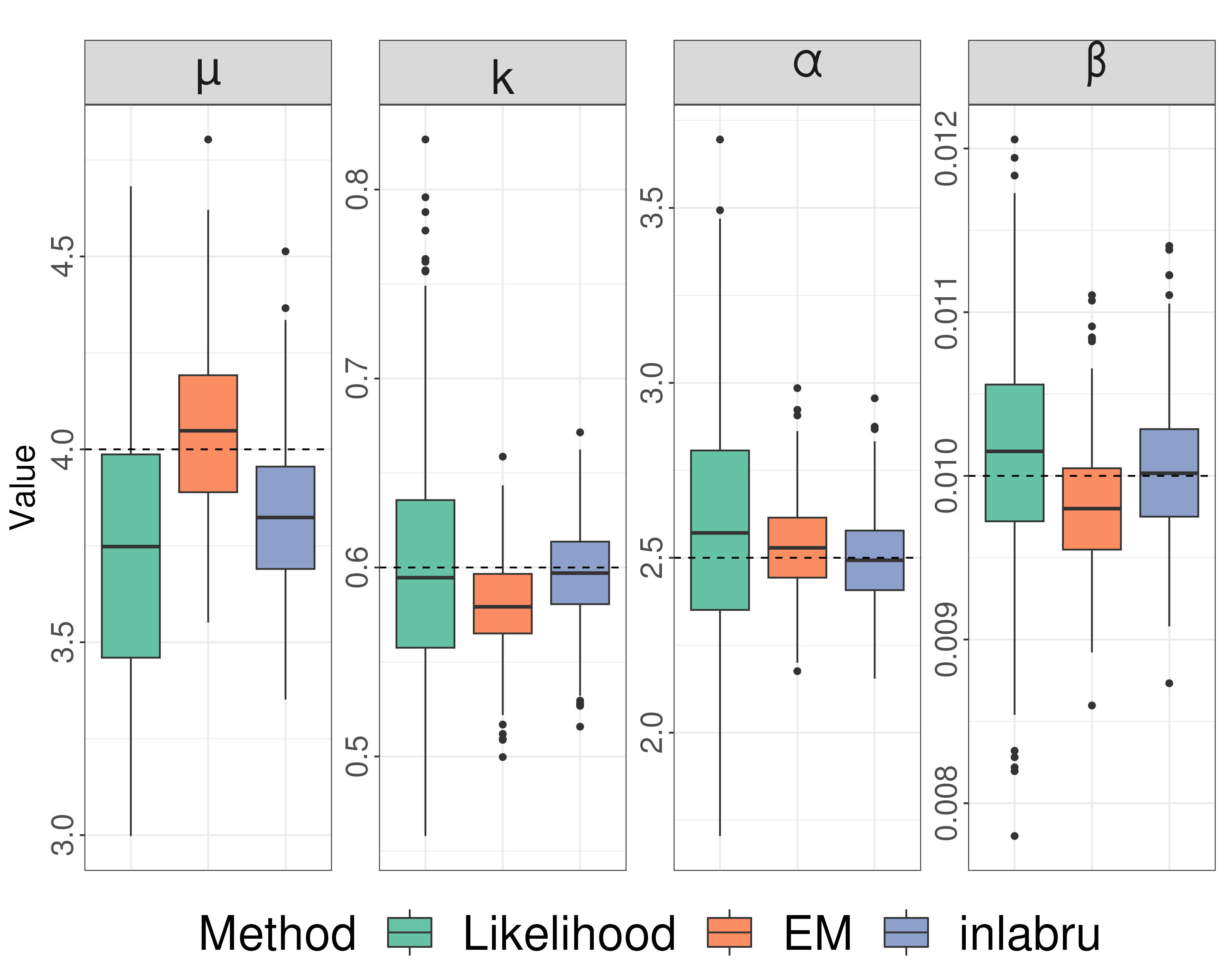}
    \caption*{2 (b): $\mu=4, k=0.6, \alpha=2.5, \beta=0.01$}
  \end{minipage}
  \vspace{-0.2cm}
    \caption{Boxplots by method of parameter estimation based on 300 simulations for each scenario and parameter configuration. The dashed line represents the true value of the parameters.}
  \label{fig-boxplot-1}
\end{figure}

\begin{figure}[H]
  \begin{minipage}[b]{0.45\textwidth}
    \centering
    \includegraphics[width=\textwidth, height=0.75\textwidth]{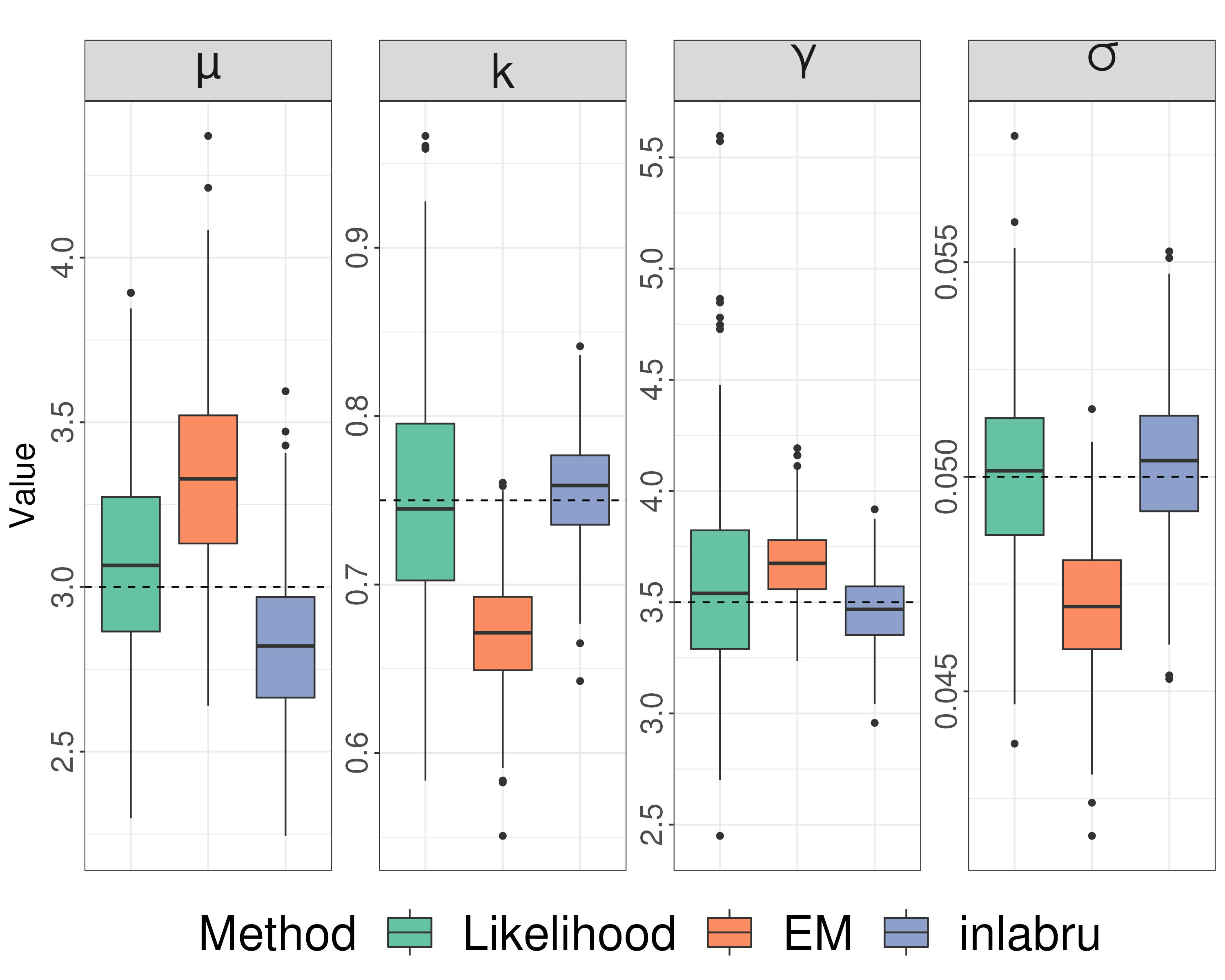}
    \caption*{3 (a): $\mu=3, k=0.75, \gamma=3.5, \sigma=0.05$}
  \end{minipage}%
  \begin{minipage}[b]{0.45\textwidth}
    \centering
    \includegraphics[width=\textwidth, height=0.75\textwidth]{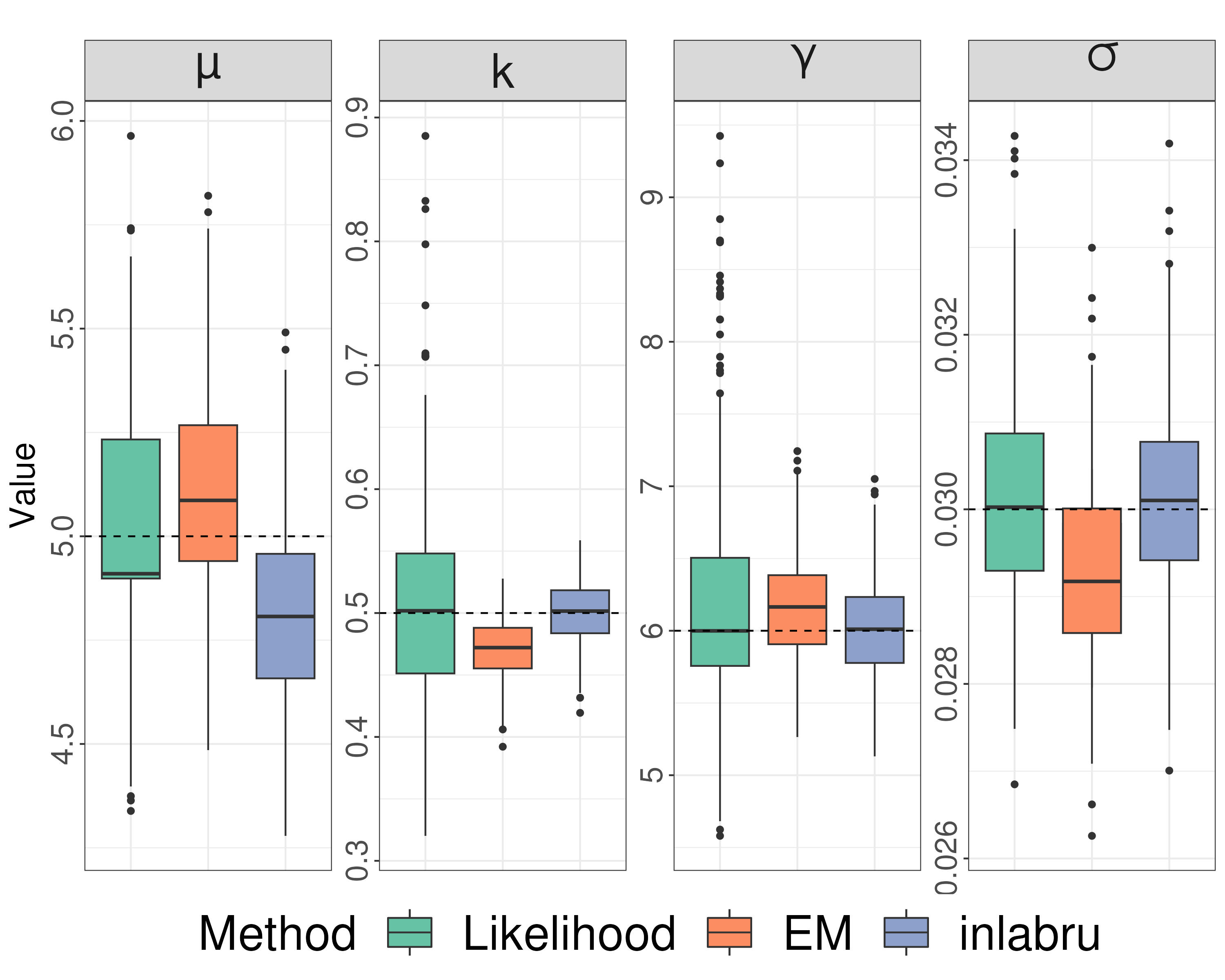}
    \caption*{3 (b): $\mu=5, k=0.5, \gamma=6, \sigma=0.03$}
  \end{minipage}
 \begin{minipage}[b]{0.45\textwidth}
    \centering
    \includegraphics[width=\textwidth, height=0.75\textwidth]{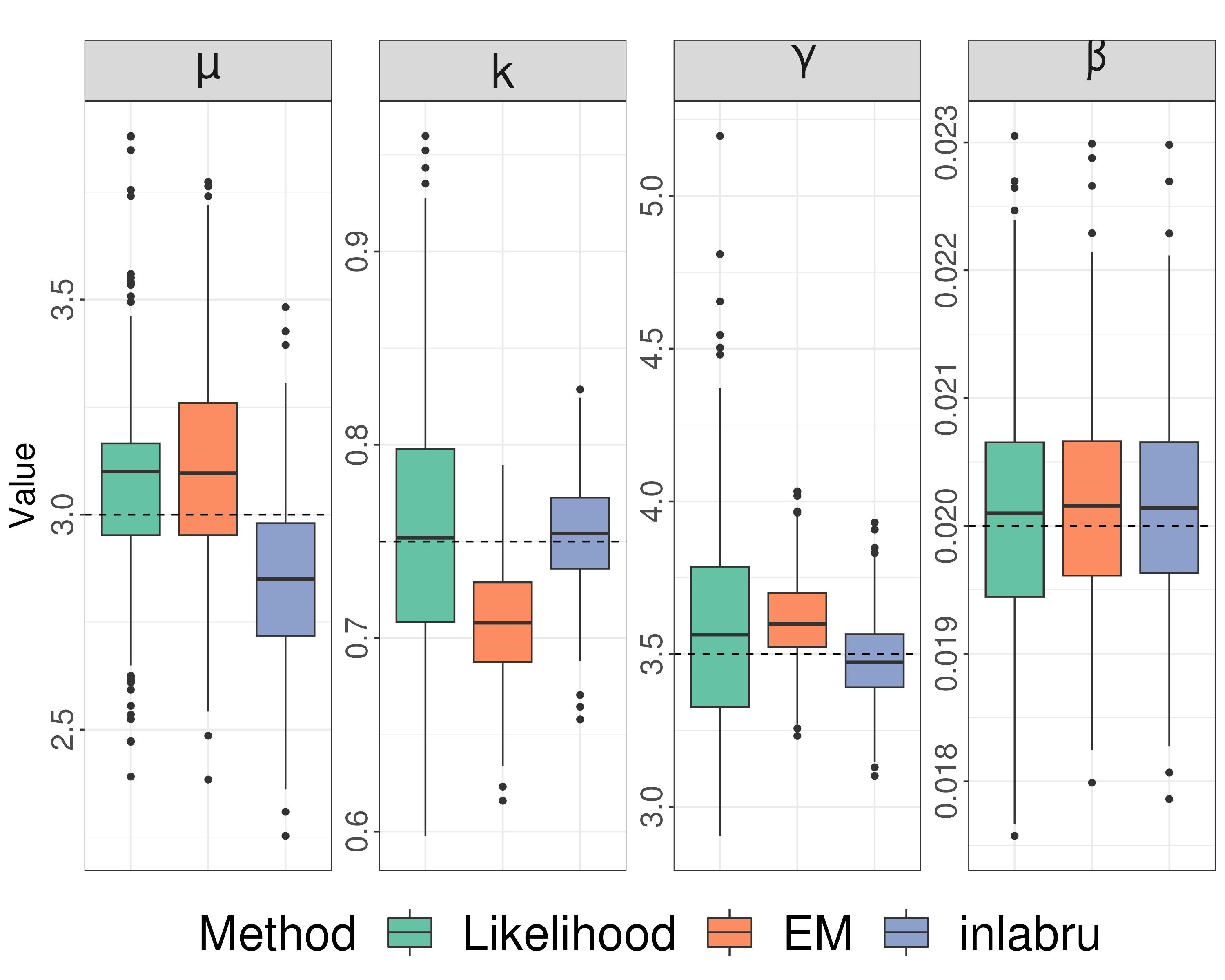}
    \caption*{4 (a): $\mu=3, k=0.75, \gamma=3.5, \beta=0.02$}
  \end{minipage}%
  \begin{minipage}[b]{0.45\textwidth}
    \centering
    \includegraphics[width=\textwidth, height=0.75\textwidth]{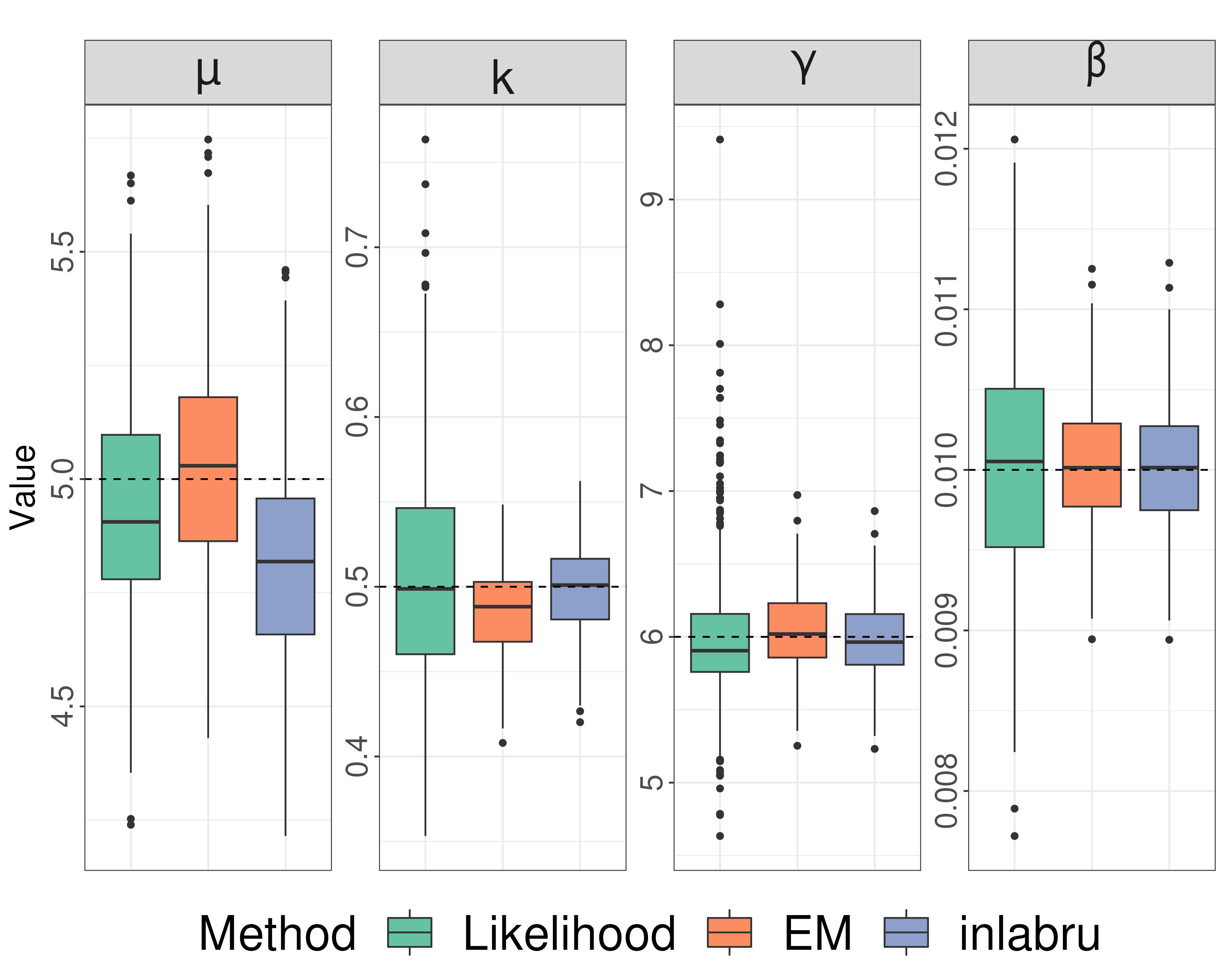}
    \caption*{4 (b): $\mu=5, k=0.5, \gamma=6, \beta=0.01$}
  \end{minipage}
    \vspace{-0.1cm}
   \caption{Boxplots by method of parameter estimation based on 300 simulations for each scenario and parameter configuration. The dashed line represents the true value of the parameters.}
  \label{fig-boxplot-2}
\end{figure}

For each scenario and combination of parameters, 300 repetitions are considered. The point patterns are simulated within the unit square and over the time interval $[0,100]$. Focusing on the comparison and evaluation of the accuracy and efficiency of the proposed methods, we have chosen to keep a simple and straightforward domain, avoiding introducing additional complexity that could distract from the main objectives of the study. We use the three presented estimation methods, full likelihood, EM and Bayesian inlabru. Tables \ref{tabla-scen-1} to \ref{tabla-scen-4} present the results for each particular scenario. The mean absolute error (MAE), along with the mean and standard deviation of the parameter estimates across simulations, are shown for each method. This allows for assessing the uncertainty of the results and enables an effective comparison between the different inference methods. The true values of the parameters are also included for reference. For the inlabru method, we present a summary of the means of the posterior distribution; specifically, we extract the posterior mean from each repetition and calculate the MAE, the mean and the standard deviation of these values. Additionally, the average computation times (in seconds) are included, which facilitates the evaluation of the computational efficiency of each approach. We also calculate the average number of points generated across all patterns ($\hat{\mathbb{E}}(N)$). Finally, we present in Figures \ref{fig-boxplot-1} and \ref{fig-boxplot-2} the boxplots corresponding to the parameter estimation values for all the simulations performed. 


\begin{table}[H]
\begin{subtable}[t]{\textwidth}
\centering
\begin{tabular}{l|ccc|c}
\hline
\multirow{2}{*}{\textbf{True value}} & \multicolumn{3}{c|}{\textbf{MAE} \scriptsize \textbf{(Mean $\pm$ SD)}} & \multirow{2}{*}{\bm{$\hat{\mathbb{E}}(N)$}} \\
 & \textbf{Likelihood} & \textbf{EM} & \textbf{Inlabru} & \\
\hline
$\mu = 2$ &
  \makecell{0.179 \\[-0.4em] \scriptsize (2.071 $\pm$ 0.214)} &
  \makecell{0.307 \\[-0.4em] \scriptsize (2.269 $\pm$ 0.252)} &
  \makecell{0.157 \\[-0.4em] \scriptsize (1.934 $\pm$ 0.184)} &
  \multirow{6}{*}{990} \\
\cline{1-4}
$k = 0.85$ &
  \makecell{0.039 \\[-0.4em] \scriptsize (0.843 $\pm$ 0.048)} &
  \makecell{0.083 \\[-0.4em] \scriptsize (0.767 $\pm$ 0.037)} &
  \makecell{0.028 \\[-0.4em] \scriptsize (0.846 $\pm$ 0.034)} &
  \\
\cline{1-4}
$\alpha = 1$ &
  \makecell{0.105 \\[-0.4em] \scriptsize (1.032 $\pm$ 0.138)} &
  \makecell{0.070 \\[-0.4em] \scriptsize (1.052 $\pm$ 0.068)} &
  \makecell{0.053 \\[-0.4em] \scriptsize (1.002 $\pm$ 0.066)} &
  \\
\cline{1-4}
$\sigma = 0.05$ &
  \makecell{0.001 \\[-0.4em] \scriptsize (0.050 $\pm$ 0.002)} &
  \makecell{0.003 \\[-0.4em] \scriptsize (0.047 $\pm$ 0.002)} &
  \makecell{0.001 \\[-0.4em] \scriptsize (0.050 $\pm$ 0.002)} &
  \\
\hline
\textbf{Average time (s)} &
  73.291 & 6.217 & 100.113 & \\
\hline
\end{tabular}
\caption{Scenario 1 (a)}
\label{s1a}
\end{subtable}
\hfill
\begin{subtable}[t]{\textwidth}
\centering
\begin{tabular}{l|ccc|c}
\hline
\multirow{2}{*}{\textbf{True value}} & \multicolumn{3}{c|}{\textbf{MAE} \scriptsize \textbf{(Mean $\pm$ SD)}} & \multirow{2}{*}{\bm{$\hat{\mathbb{E}}(N)$}} \\
 & \textbf{Likelihood} & \textbf{EM} & \textbf{Inlabru} & \\
\hline
$\mu = 4$ &
  \makecell{0.188 \\[-0.4em] \scriptsize (4.058 $\pm$ 0.231)} &
  \makecell{0.199 \\[-0.4em] \scriptsize (4.092 $\pm$ 0.228)} &
  \makecell{0.233 \\[-0.4em] \scriptsize (3.800 $\pm$ 0.208)} &
  \multirow{6}{*}{956} \\
\cline{1-4}
$k = 0.6$ &
  \makecell{0.049 \\[-0.4em] \scriptsize (0.601 $\pm$ 0.063)} &
  \makecell{0.035 \\[-0.4em] \scriptsize (0.568 $\pm$ 0.028)} &
  \makecell{0.022 \\[-0.4em] \scriptsize (0.606 $\pm$ 0.027)} &
  \\
\cline{1-4}
$\alpha = 2.5$ &
  \makecell{0.269 \\[-0.4em] \scriptsize (2.548 $\pm$ 0.346)} &
  \makecell{0.127 \\[-0.4em] \scriptsize (2.541 $\pm$ 0.155)} &
  \makecell{0.125 \\[-0.4em] \scriptsize (2.471 $\pm$ 0.153)} &
  \\
\cline{1-4}
$\sigma = 0.03$ &
  \makecell{$8.471 \cdot 10^{-4}$ \\[-0.4em] \scriptsize ($0.030 \pm 0.001$)} &
  \makecell{$9.943 \cdot 10^{-4}$ \\[-0.4em] \scriptsize ($0.029 \pm 9.192\cdot 10^{-4}$)} &
  \makecell{$8.178\cdot 10^{-4}$ \\[-0.4em] \scriptsize ($0.030 \pm 9.920 \cdot 10^{-4}$)} &
  \\
\hline
\textbf{Average time (s)} &
  135.515 &  5.884 & 71.502 & \\
\hline
\end{tabular}
\caption{Scenario 1 (b)}
\label{s1b}
\end{subtable}

\caption{Mean absolute error (MAE) of the parameter estimates across simulations, with the mean and standard deviation of these estimates in parentheses, for each estimation method. The average computational time (in seconds) and the average number of simulated data points (\(\hat{\mathbb{E}}(N)\)) are also reported.}
\label{tabla-scen-1}
\end{table}

\begin{table}[H]
\begin{subtable}[t]{\textwidth}
\centering
\begin{tabular}{l|ccc|c}
\hline
\multirow{2}{*}{\textbf{True value}} & \multicolumn{3}{c|}{\textbf{MAE} \scriptsize \textbf{(Mean $\pm$ SD)}} & \multirow{2}{*}{\bm{$\hat{\mathbb{E}}(N)$}} \\
 & \textbf{Likelihood} & \textbf{EM} & \textbf{Inlabru} & \\
\hline
$\mu = 2$ &
  \makecell{0.139 \\[-0.4em] \scriptsize (2.056 $\pm$ 0.189)} &
  \makecell{0.204 \\[-0.4em] \scriptsize (2.164 $\pm$ 0.203)} &
  \makecell{0.138 \\[-0.4em] \scriptsize (1.958 $\pm$ 0.166)} &
  \multirow{6}{*}{1068} \\
\cline{1-4}
$k = 0.85$ &
  \makecell{0.037 \\[-0.4em] \scriptsize (0.840 $\pm$ 0.046)} &
  \makecell{0.057 \\[-0.4em] \scriptsize (0.794 $\pm$ 0.034)} &
  \makecell{0.026 \\[-0.4em] \scriptsize (0.842 $\pm$ 0.032)} &
  \\
\cline{1-4}
$\alpha = 1$ &
  \makecell{0.083 \\[-0.4em] \scriptsize (1.008 $\pm$ 0.104)} &
  \makecell{0.049 \\[-0.4em] \scriptsize (1.025 $\pm$ 0.057)} &
  \makecell{0.045 \\[-0.4em] \scriptsize (0.995 $\pm$ 0.057)} &
  \\
\cline{1-4}
$\beta = 0.02$ &
  \makecell{$6.617 \cdot 10^{-4}$ \\[-0.4em] \scriptsize ($0.020 \pm 8.312\cdot 10^{-4}$)} &
  \makecell{0.001  \\[-0.4em] \scriptsize ($0.019 \pm 7.083\cdot 10^{-4}$)} &
  \makecell{$6.135 \cdot 10^{-4}$ \\[-0.4em] \scriptsize ($0.020 \pm 7.791\cdot 10^{-4}$)} &
  \\
\hline
\textbf{Average time (s)} &
  100.334 & 7.360  &  88.462 & \\
\hline
\end{tabular}
\caption{Scenario 2 (a)}
\label{s2a}
\end{subtable}
\hfill
\begin{subtable}[t]{\textwidth}
\centering
\begin{tabular}{l|ccc|c}
\hline
\multirow{2}{*}{\textbf{True value}} & \multicolumn{3}{c|}{\textbf{MAE} \scriptsize \textbf{(Mean $\pm$ SD)}} & \multirow{2}{*}{\bm{$\hat{\mathbb{E}}(N)$}} \\
 & \textbf{Likelihood} & \textbf{EM} & \textbf{Inlabru} & \\
\hline
$\mu = 4$ &
  \makecell{0.382 \\[-0.4em] \scriptsize (3.717 $\pm$ 0.391)} &
  \makecell{0.177 \\[-0.4em] \scriptsize (4.054 $\pm$ 0.214)} &
  \makecell{0.217 \\[-0.4em] \scriptsize (3.833 $\pm$ 0.198)} &
  \multirow{6}{*}{962} \\
\cline{1-4}
$k = 0.6$ &
  \makecell{0.050 \\[-0.4em] \scriptsize (0.602 $\pm$ 0.064)} &
  \makecell{0.026 \\[-0.4em] \scriptsize (0.579 $\pm$ 0.024)} &
  \makecell{0.020 \\[-0.4em] \scriptsize (0.596 $\pm$ 0.025)} &
  \\
\cline{1-4}
$\alpha = 2.5$ &
  \makecell{0.290 \\[-0.4em] \scriptsize (2.576 $\pm$ 0.358)} &
  \makecell{0.109 \\[-0.4em] \scriptsize (2.531 $\pm$ 0.136)} &
  \makecell{0.106 \\[-0.4em] \scriptsize (2.496 $\pm$ 0.134)} &
  \\
\cline{1-4}
$\beta = 0.01$ &
  \makecell{$5.238 \cdot 10^{-4}$ \\[-0.4em] \scriptsize ($0.010 \pm 6.515\cdot 10^{-4}$)} &
  \makecell{$3.485 \cdot 10^{-4}$ \\[-0.4em] \scriptsize ($0.009 \pm 3.841\cdot 10^{-4}$)} &
  \makecell{$3.146 \cdot 10^{-4}$ \\[-0.4em] \scriptsize ($0.010 \pm 4.076\cdot 10^{-4}$)} &
  \\
\hline
\textbf{Average time (s)} &
 106.387 & 6.013 & 40.784 & \\
\hline
\end{tabular}
\caption{Scenario 2 (b)}
\label{s2b}
\end{subtable}

\begin{subtable}[t]{\textwidth}
\centering

\begin{tabular}{l|ccc|c}
\hline
\multirow{2}{*}{\textbf{True value}} & \multicolumn{3}{c|}{\textbf{MAE} \scriptsize \textbf{(Mean $\pm$ SD)}} & \multirow{2}{*}{\bm{$\hat{\mathbb{E}}(N)$}} \\
 & \textbf{Likelihood} & \textbf{EM} & \textbf{Inlabru} & \\
\hline
$\mu = 3$ &
  \makecell{0.234 \\[-0.4em] \scriptsize (3.084 $\pm$ 0.283)} &
  \makecell{0.355 \\[-0.4em] \scriptsize (3.332 $\pm$ 0.273)} &
  \makecell{0.229 \\[-0.4em] \scriptsize (2.825 $\pm$ 0.219)} &
  \multirow{6}{*}{1018} \\
\cline{1-4}
$k = 0.75$ &
  \makecell{0.054 \\[-0.4em] \scriptsize (0.752 $\pm$ 0.068)} &
  \makecell{0.079 \\[-0.4em] \scriptsize (0.671 $\pm$ 0.034)} &
  \makecell{0.026 \\[-0.4em] \scriptsize (0.757 $\pm$ 0.032)} &
  \\
\cline{1-4}
$\gamma = 3.5$ &
  \makecell{0.330 \\[-0.4em] \scriptsize (3.572 $\pm$ 0.434)} &
  \makecell{0.202 \\[-0.4em] \scriptsize (3.675 $\pm$ 0.173)} &
  \makecell{0.133 \\[-0.4em] \scriptsize (3.465 $\pm$ 0.166)} &
  \\
\cline{1-4}
$\sigma = 0.05$ &
  \makecell{0.002 \\[-0.4em] \scriptsize (0.050 $\pm$ 0.002)} &
  \makecell{0.003 \\[-0.4em] \scriptsize (0.047 $\pm$ 0.002)} &
  \makecell{0.001 \\[-0.4em] \scriptsize (0.050 $\pm$ 0.002)} &
  \\
\hline
\textbf{Average time (s)} &
 129.533 & 7.058 & 107.593 & \\
\hline
\end{tabular}

\caption{Scenario 3 (a)}
\label{s3a}
\end{subtable}
\begin{subtable}[t]{\textwidth}
\centering
\begin{tabular}{l|ccc|c}
\hline
\multirow{2}{*}{\textbf{True value}} & \multicolumn{3}{c|}{\textbf{MAE} \scriptsize \textbf{(Mean $\pm$ SD)}} & \multirow{2}{*}{\bm{$\hat{\mathbb{E}}(N)$}} \\
 & \textbf{Likelihood} & \textbf{EM} & \textbf{Inlabru} & \\
\hline
$\mu = 5$ &
  \makecell{0.213 \\[-0.4em] \scriptsize (5.041 $\pm$ 0.269)} &
  \makecell{0.211 \\[-0.4em] \scriptsize (5.097 $\pm$ 0.248)} &
  \makecell{0.241 \\[-0.4em] \scriptsize (4.813 $\pm$ 0.228)} &
  \multirow{6}{*}{963} \\
\cline{1-4}
$k = 0.5$ &
  \makecell{0.052 \\[-0.4em] \scriptsize (0.507 $\pm$ 0.081)} &
  \makecell{0.032 \\[-0.4em] \scriptsize (0.471 $\pm$ 0.024)} &
  \makecell{0.020 \\[-0.4em] \scriptsize (0.501 $\pm$ 0.025} &
  \\
\cline{1-4}
$\alpha = 6$ &
  \makecell{0.541 \\[-0.4em] \scriptsize (6.170 $\pm$ 0.816)} &
  \makecell{0.308 \\[-0.4em] \scriptsize (6.158 $\pm$ 0.353)} &
  \makecell{0.272 \\[-0.4em] \scriptsize (6.014 $\pm$ 0.340)} &
  \\
\cline{1-4}
$\sigma = 0.03$ &
  \makecell{$9.274 \cdot 10^{-4}$ \\[-0.4em] \scriptsize ($0.030 \pm 0.001$)} &
  \makecell{$9.988 \cdot 10^{-4}$ \\[-0.4em] \scriptsize ($0.029 \pm 0.001$)} &
  \makecell{$8.146 \cdot 10^{-4}$ \\[-0.4em] \scriptsize ($0.030 \pm 0.001$)} &
  \\
\hline
\textbf{Average time (s)} &
   129.115 & 6.516  & 53.081 & \\
\hline
\end{tabular}

\caption{Scenario 3 (b)}
\label{s3b}
\end{subtable}
\caption{Mean absolute error (MAE) of the parameter estimates across simulations, with the mean and standard deviation of these estimates in parentheses, for each estimation method. The average computational time (in seconds) and the average number of simulated data points (\(\hat{\mathbb{E}}(N)\)) are also reported.}
\label{tabla-scen-2-3}
\end{table}

\begin{table}[H]
\begin{subtable}[t]{\textwidth}
\centering
\begin{tabular}{l|ccc|c}
\hline
\multirow{2}{*}{\textbf{True value}} & \multicolumn{3}{c|}{\textbf{MAE} \scriptsize \textbf{(Mean $\pm$ SD)}} & \multirow{2}{*}{\bm{$\hat{\mathbb{E}}(N)$}} \\
 & \textbf{Likelihood} & \textbf{EM} & \textbf{Inlabru} & \\
\hline
$\mu = 3$ &
  \makecell{0.184 \\[-0.4em] \scriptsize (3.070 $\pm$ 0.227)} &
  \makecell{0.209 \\[-0.4em] \scriptsize (3.103 $\pm$ 0.243)} &
  \makecell{0.208 \\[-0.4em] \scriptsize (2.852 $\pm$ 0.207)} &
  \multirow{6}{*}{1069} \\
\cline{1-4}
$k = 0.75$ &
  \makecell{0.054 \\[-0.4em] \scriptsize (0.754 $\pm$ 0.067)} &
  \makecell{0.044 \\[-0.4em] \scriptsize (0.708 $\pm$ 0.030)} &
  \makecell{0.023 \\[-0.4em] \scriptsize (0.753 $\pm$ 0.029)} &
  \\
\cline{1-4}
$\alpha = 3.5$ &
  \makecell{0.268 \\[-0.4em] \scriptsize (3.590 $\pm$ 0.337)} &
  \makecell{0.142 \\[-0.4em] \scriptsize (3.612 $\pm$ 0.139)} &
  \makecell{0.112 \\[-0.4em] \scriptsize (3.481 $\pm$ 0.140)} &
  \\
\cline{1-4}
$\beta = 0.02$ &
  \makecell{$7.011 \cdot 10^{-4}$ \\[-0.4em] \scriptsize ($0.020 \pm 8902\cdot 10^{-4}$)} &
  \makecell{$6.623 \cdot 10^{-4}$ \\[-0.4em] \scriptsize ($0.020 \pm 8.275\cdot 10^{-4}$ )} &
  \makecell{$6.444 \cdot 10^{-4}$ \\[-0.4em] \scriptsize ($0.020 \pm 7.983\cdot 10^{-4}$)} &
  \\
\hline
\textbf{Average time (s)} &
 126.180 & 8.380  & 65.476 & \\
\hline
\end{tabular}
\caption{Scenario 4 (a)}
\label{s4a}
\end{subtable}
\hfill
\begin{subtable}[t]{\textwidth}
\centering
\begin{tabular}{l|ccc|c}
\hline
\multirow{2}{*}{\textbf{True value}} & \multicolumn{3}{c|}{\textbf{MAE} \scriptsize \textbf{(Mean $\pm$ SD)}} & \multirow{2}{*}{\bm{$\hat{\mathbb{E}}(N)$}} \\
 & \textbf{Likelihood} & \textbf{EM} & \textbf{Inlabru} & \\
\hline
$\mu = 5$ &
  \makecell{0.212 \\[-0.4em] \scriptsize (4.930 $\pm$ 0.256)} &
  \makecell{0.195 \\[-0.4em] \scriptsize (5.024 $\pm$ 0.244)} &
  \makecell{0.241 \\[-0.4em] \scriptsize (4.812 $\pm$ 0.227)} &
  \multirow{6}{*}{976} \\
\cline{1-4}
$k = 0.5$ &
  \makecell{0.052 \\[-0.4em] \scriptsize (0.505 $\pm$ 0.066)} &
  \makecell{0.022 \\[-0.4em] \scriptsize (0.486 $\pm$ 0.025)} &
  \makecell{0.020 \\[-0.4em] \scriptsize (0.499 $\pm$ 0.025)} &
  \\
\cline{1-4}
$\alpha = 6$ &
  \makecell{0.377 \\[-0.4em] \scriptsize (6.024 $\pm$ 0.563)} &
  \makecell{0.220 \\[-0.4em] \scriptsize (6.044 $\pm$ 0.278)} &
  \makecell{0.215 \\[-0.4em] \scriptsize (5.985 $\pm$ 0.270)} &
  \\
\cline{1-4}
$\beta = 0.01$ &
  \makecell{$5.947 \cdot 10^{-4}$ \\[-0.4em] \scriptsize ($0.010 \pm 7.529 \cdot 10^{-4}$)} &
  \makecell{$3.221 \cdot 10^{-4}$ \\[-0.4em] \scriptsize ($0.010 \pm 4.076 \cdot 10^{-4}$)} &
  \makecell{$3.272 \cdot 10^{-4}$ \\[-0.4em] \scriptsize ($0.010 \pm 4.104 \cdot 10^{-4}$)} &
  \\
\hline
\textbf{Average time (s)} &
  123.239 &  6.986  &  51.918 & \\
\hline
\end{tabular}
\caption{Scenario 4 (b)}
\label{s4b}
\end{subtable}
\caption{Mean absolute error (MAE) of the parameter estimates across simulations, with the mean and standard deviation of these estimates in parentheses, for each estimation method. The average computational time (in seconds) and the average number of simulated data points (\(\hat{\mathbb{E}}(N)\)) are also reported.}
\label{tabla-scen-4}
\end{table}

\begin{figure}[H]
    \centering
    \includegraphics[height=5.6cm]{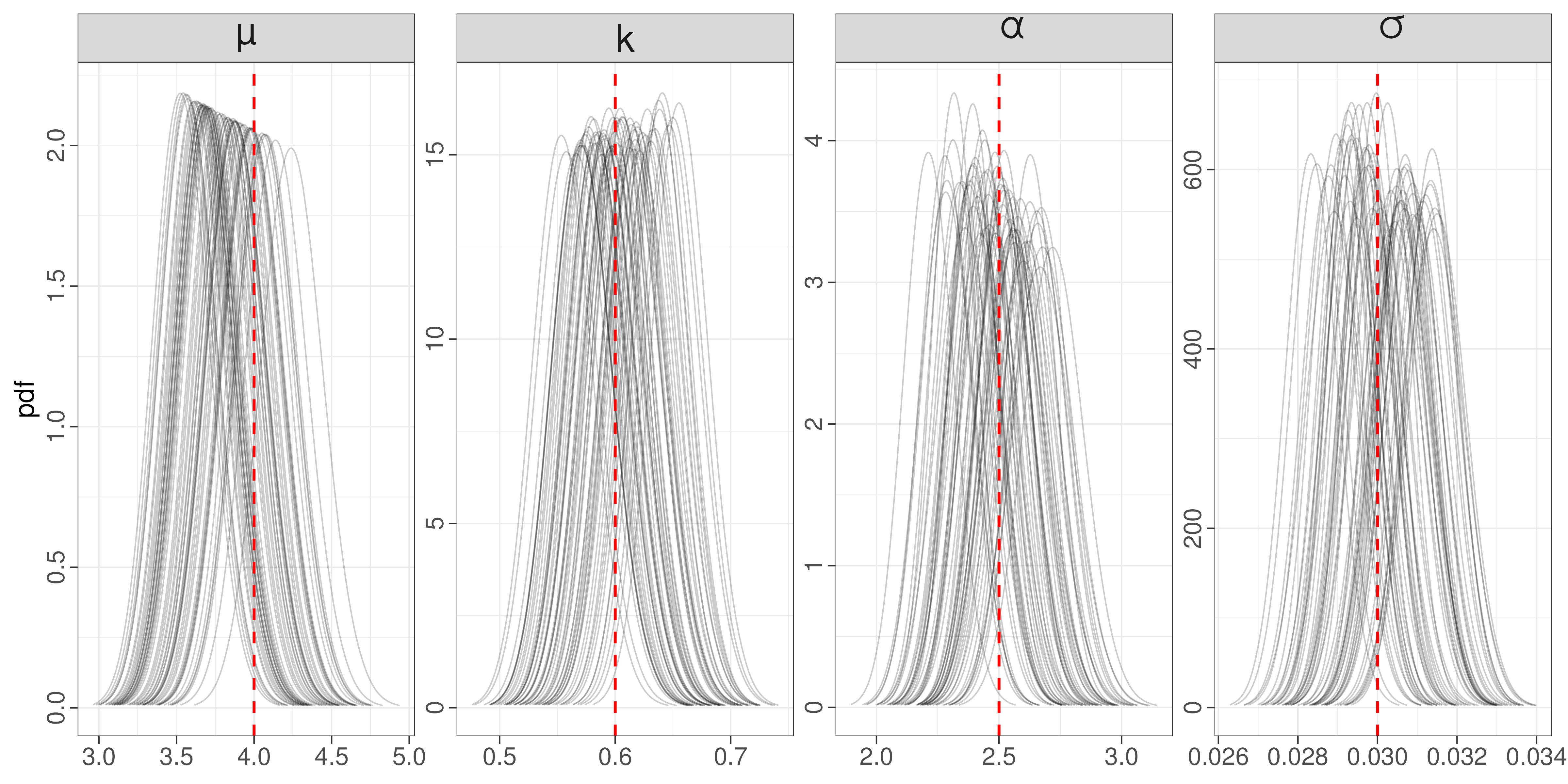}
    \vspace{-0.15cm}
    \caption{Posterior distributions of the parameters obtained by estimating with inlabru a sample of 50 simulations under scenario 1 (b). The red dashed line represents the true value of the parameter.}
    \label{posteriors_1b}
\end{figure}

The overall results indicate that all three methods are accurate in estimating the parameters across all combinations of triggering functions, as the mean of the parameters are quite close to their true values, and the standard deviations are small (Tables \ref{tabla-scen-1} to \ref{tabla-scen-4}). However, the direct maximum likelihood method and the inlabru method tend to be the most accurate, as the boxes in the boxplots usually encompass the true value, reflecting the high precision of the estimates, as seen in Figures \ref{fig-boxplot-1} and \ref{fig-boxplot-2}. \ab{We also note (see the boxplots) that the likelihood method appears to be more accurate in terms of point estimates than inlabru (the means are closer to the true values), although it is important to note that the INLA-based method provides the full posterior distribution of the parameters, and here we only show a summary of the posterior mean. This richer information can make INLA preferable despite its slightly less accurate point estimates. For instance, for scenario 1(b), we show in Figure \ref{posteriors_1b} the posterior distributions of the parameters corresponding to a representative sample of 50 simulations. In this case, most of the distributions are centred on the true values of the parameters, which indicates that the method is not only accurate and stable in terms of estimation, but also provides credible intervals that adequately reflect the uncertainty. Regarding the EM method, the boxplots of the parameters do not always encompass the true values, as seen, for example, in the case of parameter $k$ in scenarios 1 (a), 1 (b), 2 (a), 3 (a), and 4 (a), or for parameters $\mu$ and $\sigma/\beta$ in scenarios 1 (a), 2 (a), and 3 (a), although they are generally close to them.} 

Regarding further particularities, we observe that when the spatial parameter ($\sigma$ or $\beta$) is lower (indicating a stronger clustering effect among nearby events), as in scenarios 1 (b), 2 (b), 3 (b) and 4 (b), the MAE associated with this parameter is also lower for all methods (Tables \ref{s1b}, \ref{s2b}, \ref{s3b}, \ref{s4b}) compared to the corresponding scenarios 1 (a), 2 (a), 3 (a), and 4 (a), suggesting that spatial concentration improves estimation accuracy. In these cases,  parameter $k$ also tends to show a lower MAE. Regarding parameter $\mu$, the MAE is higher for inlabru and likelihood, unlike for EM, where the MAE is lower in these cases. However, when the temporal parameter ($\alpha$ or $\gamma$) is smaller (indicating longer temporal memory), as in scenarios 1 (a), 2 (a), 3 (a) and 4 (a), this parameter results in a lower MAE (Tables \ref{s1a}, \ref{s2a}, \ref{s3a}, \ref{s4a}), which may be due to the additional information provided by the extended influence of past events. 

Regarding the comparison of methods, in the (a) scenarios (Tables \ref{s1a}, \ref{s2a}, \ref{s3a}, \ref{s4a}), the EM method tends to have larger MAEs, whereas in the (b) scenarios (Tables \ref{s1b}, \ref{s2b}, \ref{s3b}, \ref{s4b}), the likelihood method tends to present the largest MAEs. In any case, inlabru method generally shows the lowest MAEs for all parameters and across all cases. Additionally, the EM method tends to overestimate $\mu$ and $\alpha/\gamma$, while underestimating $k$ and $\sigma/\beta$ in most of the cases considered. Inlabru also underestimates  parameter $\mu$ in some cases, such as scenarios 1 (b), 2 (b), 3 (b), and 4 (b). And the likelihood method tends to show larger standard deviations in the estimates, which is also reflected in the wider boxes in the boxplots.

In terms of computational time, the EM method is the fastest, as we did prior work by calculating the analytical expressions of the estimates. Therefore, this method is useful as long as these expressions are available. Otherwise, the estimation process can become slower due to the need for additional calculations to approximate the solutions at each step of the algorithm. Regarding inlabru and maximum likelihood methods, their average computational times become closer in some scenarios. Inlabru relies on priors instead of point estimates, while maximum likelihood requires calculating the approximation to the integral. This makes inlabru faster than maximum likelihood overall, and, as previously mentioned, it is often preferred because Bayesian inference can provide a more robust approach by incorporating prior knowledge and quantifying uncertainty, which can lead to more stable and accurate estimates, especially when working with sparse data or noisy observations (\ab{\citealt{rasmussen2013bayesian, 10.1785/0120140131, https://doi.org/10.1002/env.2566, holbrook2021scalable}}).

\begin{figure}[H]
\centering
\begin{minipage}{0.53\textwidth}
\centering
\vspace{0.3cm}
\begin{tabular}{c|ccc}
\hline
\multirow{2}{*}{\textbf{}} & \multicolumn{3}{c}{\textbf{MAE} \scriptsize \textbf{(Mean $\pm$ SD)}} \\
 & \textbf{Grid $\bm{10^3}$} & \textbf{Grid $\bm{15^3}$} & \textbf{Grid $\bm{35^3}$} \\
\hline
$\mu$ & 
  \makecell{0.212 \\[-0.4em] \scriptsize (2.096 $\pm$ 0.258)} & 
  \makecell{0.189 \\[-0.4em] \scriptsize (2.076 $\pm$ 0.223)} & 
  \makecell{0.181 \\[-0.4em] \scriptsize (2.068 $\pm$ 0.218)}\\
\cline{1-4}
$k$ & 
  \makecell{0.079 \\[-0.4em] \scriptsize (0.846 $\pm$ 0.098)} & 
  \makecell{0.053 \\[-0.4em] \scriptsize (0.844 $\pm$ 0.066)} & 
  \makecell{0.035 \\[-0.4em] \scriptsize (0.840 $\pm$ 0.043)} \\
\cline{1-4}
$\alpha$ & 
  \makecell{0.196 \\[-0.4em] \scriptsize (1.082 $\pm$ 0.262)} & 
  \makecell{0.143 \\[-0.4em] \scriptsize (1.044 $\pm$ 0.192)} & 
  \makecell{0.093 \\[-0.4em] \scriptsize (1.016 $\pm$ 0.119)} \\
\cline{1-4}
$\sigma$ & 
  \makecell{$1.827 \cdot 10^{-3}$ \\[-0.4em] \scriptsize (0.051 $\pm$ 0.002)} & 
  \makecell{$1.592 \cdot 10^{-3}$ \\[-0.4em] \scriptsize (0.050 $\pm$ 0.002)} & 
  \makecell{$1.417 \cdot 10^{-3}$ \\[-0.4em] \scriptsize (0.050 $\pm$ 0.002)} \\
\hline
\textbf{t(s)} & 16.447 & 28.978 & 194.603 \\ \hline
\end{tabular}
\vspace{0.4cm}
\captionof{table}{Comparison of results using the likelihood method for different grid resolutions. MAE, mean and standard deviation of parameter estimates across simulations of scenario 1 (a) are shown, with the average computational time.}
\label{compare_grid_sizes}
\end{minipage}%
\hfill
\begin{minipage}{0.43\textwidth}
\centering
\includegraphics[width=\linewidth]{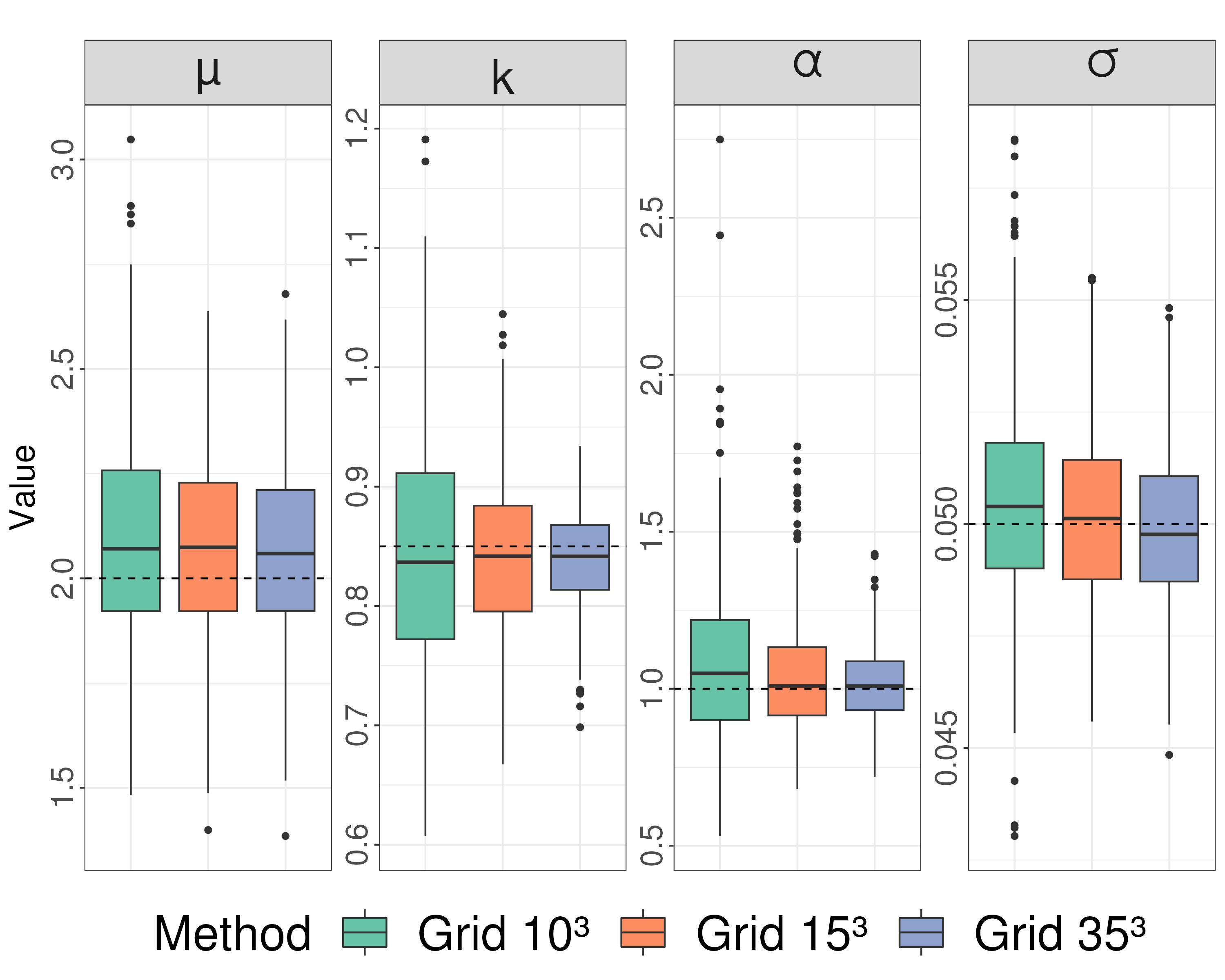}
\captionof{figure}{Boxplots of the estimated parameters across simulations of scenario 1 (a) using the likelihood method for different grid resolutions. The dashed line shows the true value of the parameters.}
\label{boxplot_grid_sizes}
\end{minipage}
\end{figure}

As mentioned, a limitation of the likelihood method is the choice of the grid for approximating the integral, which must be configured by the user, as it affects both computational time and the accuracy of the estimates. For example, in the case of scenario 1 (a), the results are obtained on a $25\times 25\times 25$ grid. If we change the resolution of the grid, we obtain the results shown in Table \ref{compare_grid_sizes} and 
Figure \ref{boxplot_grid_sizes}.

\vspace{-0.4cm}

We note in Table \ref{compare_grid_sizes} that the estimates remain accurate, but exhibit greater variability when the grid resolution is lower, as expected. This is because reducing the number of cells decreases the quality of the integral approximation. This behavior is also evident in the boxplots (Figure \ref{boxplot_grid_sizes}), as the grid resolution increases, the boxplots become narrower, with shorter whiskers.
\vspace{-0.15cm}
\vspace{-0.6cm}

\subsection{Additional case: intensity function with a non-constant background} \label{sec_extend}

We now present an extended, additional case, where we employ the parents-offspring method to simulate the point pattern data. We begin by modeling the number of parent events using a Poisson distribution with rate 500 (\ab{$\mu=5$}). The times at which these events occur are generated from a Beta(1, 2) distribution scaled to the interval $[0,100]$. The spatial locations of the parents are determined by a Gaussian Mixture Model (GMM) comprising three components. Each component has its own mean vector: $m_1 = [0.3, 0.7]$, $m_2 = [0.5, 0.5]$, and $m_3 = [0.7, 0.3]$. Additionally, the covariance matrices for each component are diagonal, with values $\text{diag}(\Sigma_1) = [0.01, 0.01]$, $\text{diag}(\Sigma_2) = [0.025, 0.01]$, and $\text{diag}(\Sigma_3) = [0.004, 0.004]$. The probabilities for the mixture components are $p_1 = 0.2$, $p_2 = 0.5$, and $p_3 = 0.3$. Once the parents are identified, the procedure outlined in Section \ref{sec:parents-offspring} is followed to generate the corresponding Hawkes process, using the following conditional intensity function

\begin{equation*}
    \lambda(s, t|\mathcal{H}_t)=\mu \cdot \mu_{\raisebox{-0.3ex}{\ab{\scriptsize S}}}(s) \cdot \mu_{\raisebox{-0.3ex}{\ab{\scriptsize T}}}(t)+ \sum_{t_i \in \mathcal{H}_t} k \cdot \alpha \cdot \exp\{-\alpha (t-t_i)\} \frac{1}{2\pi \sigma^2} \cdot \exp\{-\frac{(x-x_i)^2+(y-y_i)^2}{2\sigma^2}\}.
    \label{eq:last}
\end{equation*}

\begin{figure}[H]
\begin{minipage}[b]{0.5\textwidth}
    \centering
    \includegraphics[width=\textwidth, height=0.75\textwidth]{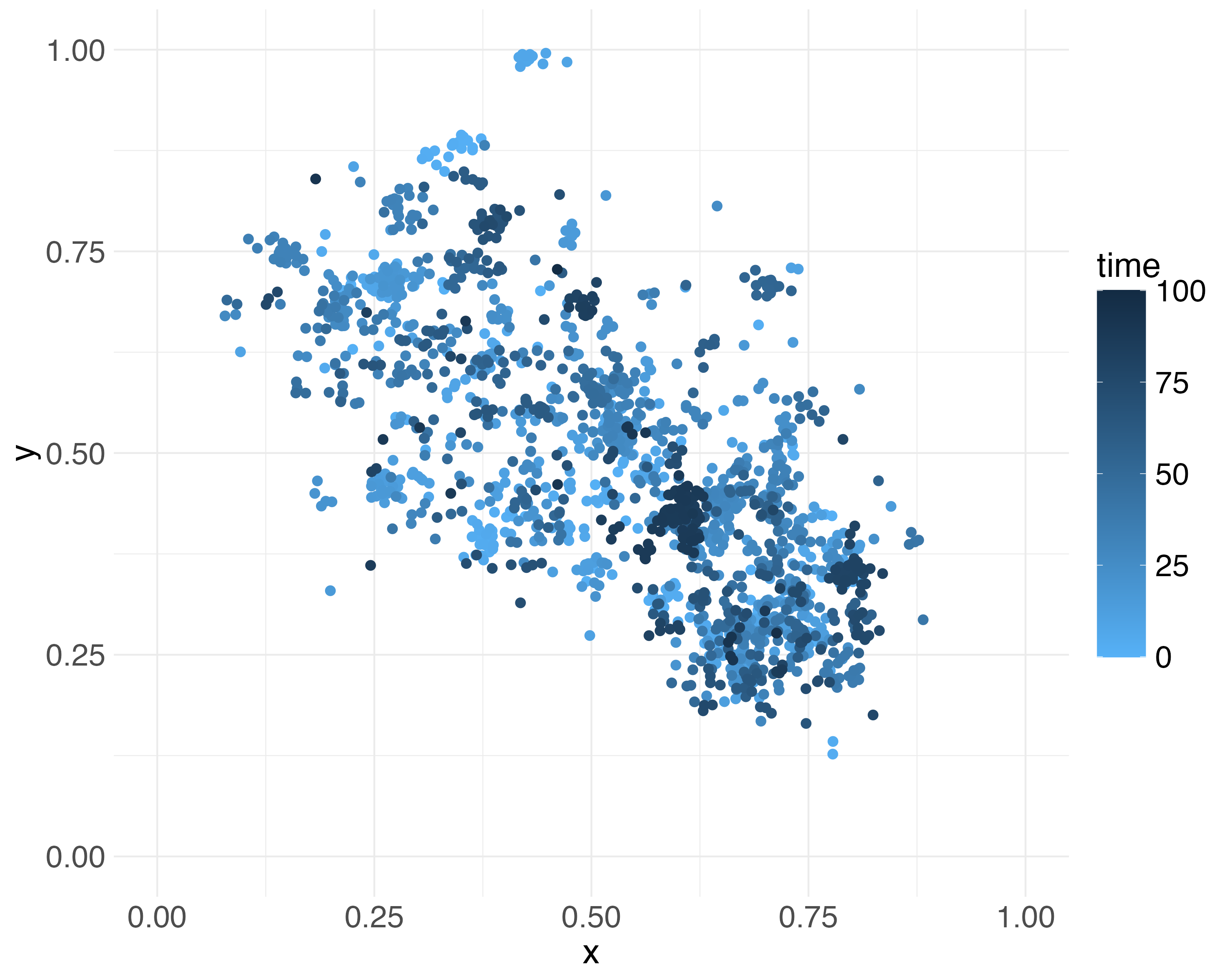}
  \end{minipage}%
  \begin{minipage}[b]{0.5\textwidth}
    \centering
    \includegraphics[width=\textwidth, height=0.75\textwidth]{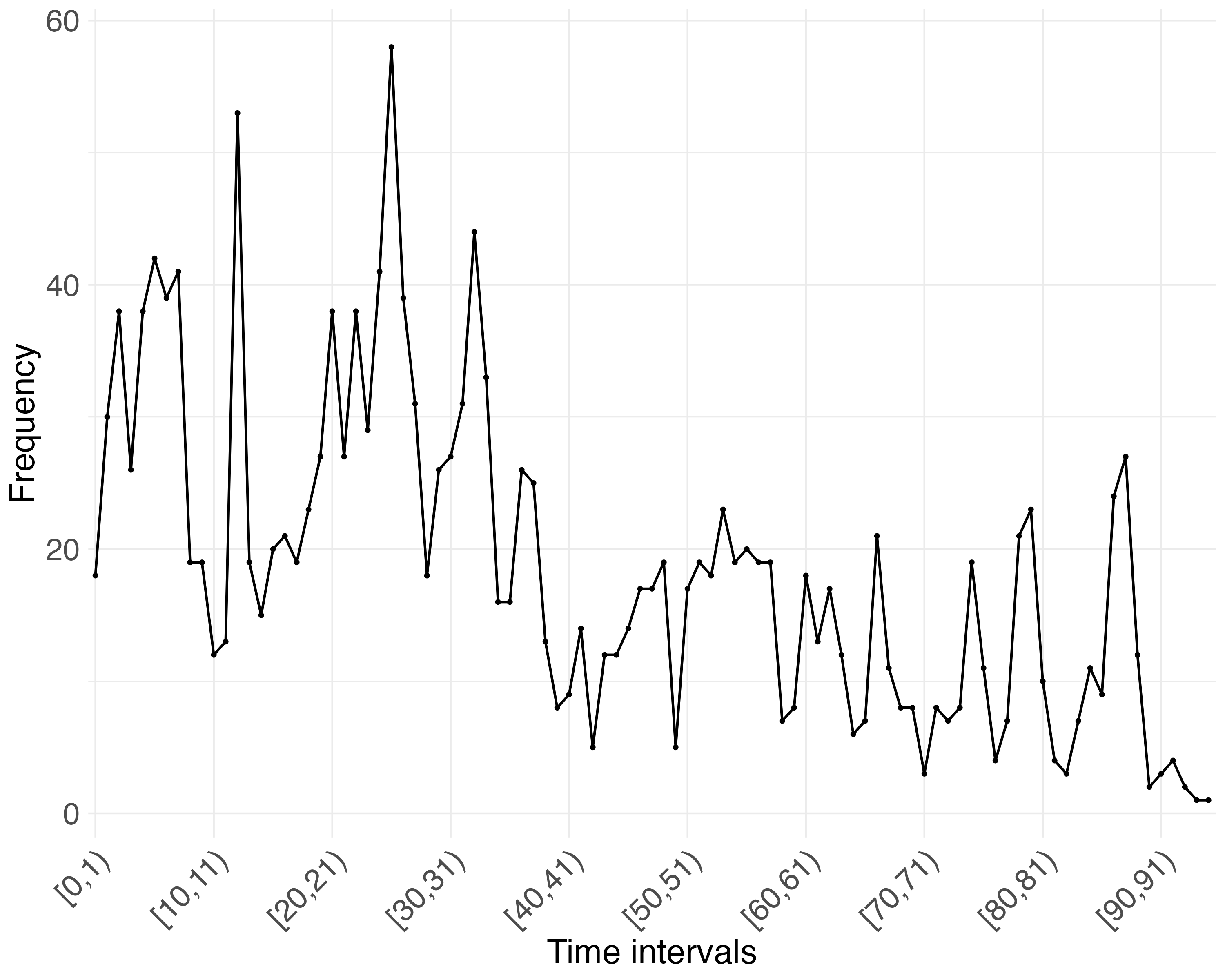}
  \end{minipage}%
 \caption{Extended case. On the left, the simulated point pattern using the parents-offspring method. On the right, the timeline of the events. The parameter configuration is $\mu=5, k=0.75, \alpha=3, \sigma=0.01$.}
  \label{data_extended}
\end{figure}

\begin{figure}[H]
    \centering
    \includegraphics[width=0.5\linewidth]{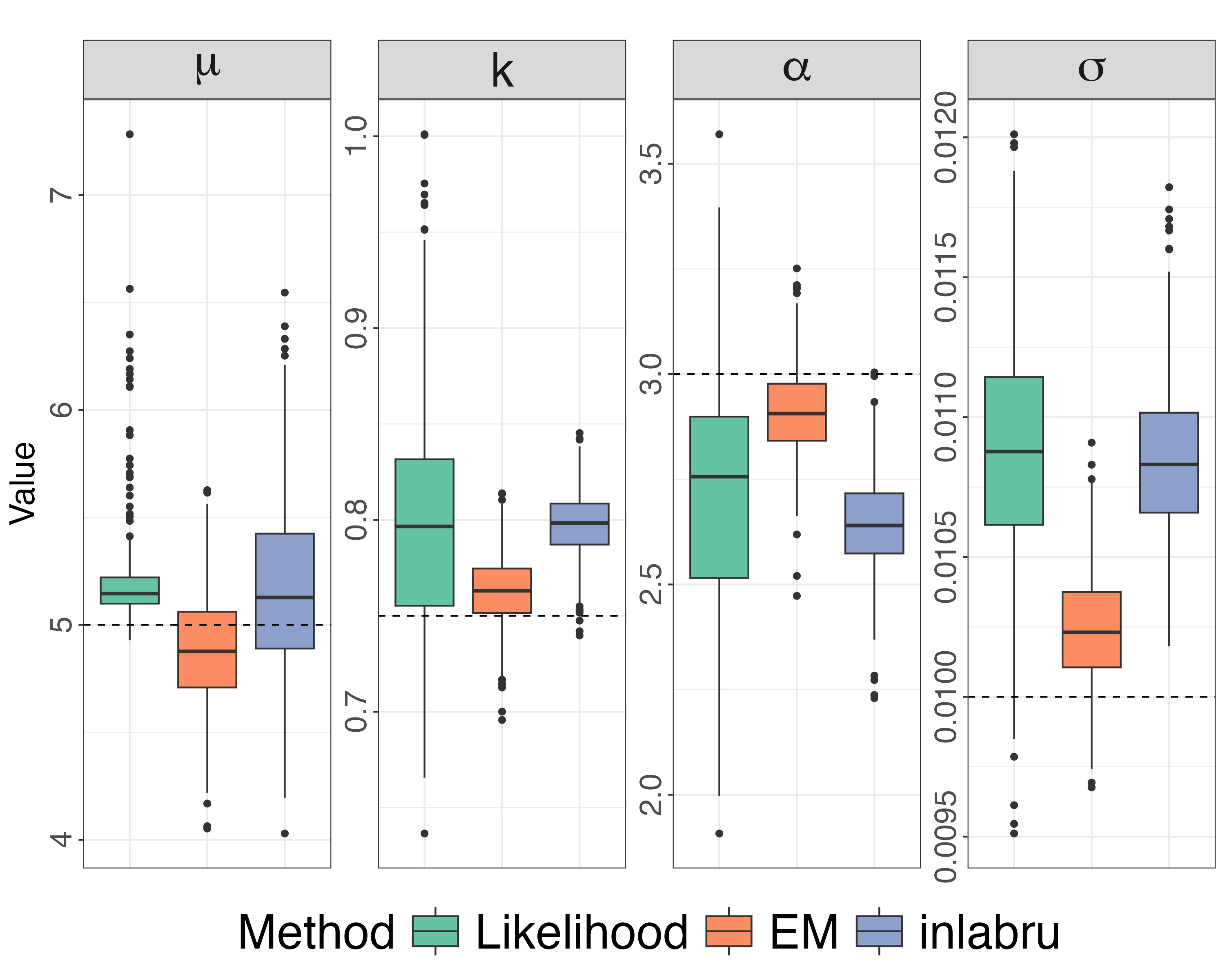}
    \caption{Boxplots by method of parameter estimation based on 300 simulations for the model with non-constant background, exponential triggering for time and Gaussian triggering for space, and for the set of parameters $\mu=5, k=0.75, \alpha=3, \sigma=0.01$. The dashed line represents the true value of the parameters.}
    \label{boxplot_extended}
\end{figure}

The parameter set used in this section is $\mu=5, k=0.75, \alpha=3, \sigma=0.01$. \ab{This configuration, which induces short-term temporal dependence and high spatial clustering at small distances, has been selected to produce concentrated clusters near the parent events in both time and space.} Moreover, we use a non-constant and separable background, which is divided into a constant parameter, a temporal function and an additional spatial function. The estimation of the spatial and temporal background components is carried out non-parametrically. \ab{Specifically, the estimation of the temporal background component is based on a one-dimensional density estimate using kernel smoothing applied to the event occurrence times, which allows capturing variations in the temporal intensity of the simulated data. To achieve this, a density function of the times is estimated on a regular grid, and each event is assigned a value from this function evaluated at its occurrence time. Subsequently, the obtained values are normalized to ensure that the density appropriately reflects the event rate within the considered time interval (\citealt{ZhuangMateu2019}). The estimation of the spatial background component is also performed using kernel smoothing over the observation window, which enables the capture of potential inhomogeneities in the spatial distribution of the events. For this purpose, the mean squared error (MSE) is used to select the optimal bandwidth for smoothing, which determines the level of detail in the estimation. Then, a two-dimensional density estimate is performed over a grid defined in the region of interest, obtaining a continuous field of spatial intensities. Each event in the pattern is assigned a value from this estimated intensity, interpolating its position within the grid.}
With these background functions, the goal is to demonstrate the flexibility to add additional variables or covariates into the Hawkes modeling framework. Figure \ref{data_extended} shows one of the simulated point patterns (left plot), along with a timeline of the generated events (right plot).
We can observe that small clusters have formed around certain events, the respective parents\ab{, as the parents-offspring method allows us to directly identify the clusters by fully reconstructing the genealogy.} Table \ref{tab:extended} and Figure \ref{boxplot_extended} show the corresponding results for the mean absolute error of the parameter estimates, with the mean and standard deviation of these estimates for the three inference methods, as well as the computation times under 300 simulations of the considered Hawkes model.

\begin{table}[H]
\centering
\begin{tabular}{l|ccc|c}
\hline
\multirow{2}{*}{\textbf{True value}} & \multicolumn{3}{c|}{\textbf{MAE} \scriptsize \textbf{(Mean $\pm$ SD)}} & \multirow{2}{*}{\bm{$\hat{\mathbb{E}}(N)$}} \\
 & \textbf{Likelihood} & \textbf{EM} & \textbf{Inlabru} & \\
\hline
$\mu = 5$ &
  \makecell{0.219 \\[-0.4em] \scriptsize (5.218 $\pm$ 0.253)} &
  \makecell{0.231 \\[-0.4em] \scriptsize (4.888 $\pm$ 0.260)} &
  \makecell{0.341 \\[-0.4em] \scriptsize (5.166 $\pm$ 0.409)} &
  \multirow{6}{*}{1993} \\
\cline{1-4}
$k = 0.75$ &
  \makecell{0.062 \\[-0.4em] \scriptsize (0.799 $\pm$ 0.063)} &
  \makecell{0.019 \\[-0.4em] \scriptsize (0.763 $\pm$ 0.019)} &
  \makecell{0.048 \\[-0.4em] \scriptsize (0.798 $\pm$ 0.017)} &
  \\
\cline{1-4}
$\alpha = 3$ &
  \makecell{0.311 \\[-0.4em] \scriptsize (2.714 $\pm$ 0.258)} &
  \makecell{0.114 \\[-0.4em] \scriptsize (2.910 $\pm$ 0.108)} &
  \makecell{0.357 \\[-0.4em] \scriptsize (2.643 $\pm$ 0.113)} &
  \\
\cline{1-4}
$\sigma = 0.01$ &
  \makecell{$8.750 \cdot 10^{-4}$ \\[-0.4em] \scriptsize (0.011 $\pm$ 4.442$\cdot 10^{-4}$)} &
  \makecell{$2.572 \cdot 10^{-4}$ \\[-0.4em] \scriptsize (0.010 $\pm$ 2.056$\cdot 10^{-4}$)} &
  \makecell{$8.439 \cdot 10^{-4}$ \\[-0.4em] \scriptsize (0.011 $\pm$ 2.712$\cdot 10^{-4}$)} &
  \\
\hline
\textbf{Average time (s)} &
  262.917 & 31.994 & 110.747 & \\
\hline
\end{tabular}
\caption{Mean absolute error (MAE) of the parameter estimates across simulations, with the mean and standard deviation of these estimates in parentheses, for each estimation method. The average computational time (in seconds) and the average number of simulated data points (\(\hat{\mathbb{E}}(N)\)) are also reported.}
\label{tab:extended}
\end{table}

In this case of a non-constant background, all three methods provide good estimates, as the estimated values are close to the true values, demonstrating their adaptability to modeling this type of scenario, which is also common in real-world applications. By looking at the boxplots, the methods tend to underestimate $\alpha$ and overestimate $k$ and $\sigma$. The EM method shows the best accuracy (as it tends to have the lowest MAEs for the parameters), in addition to being the fastest. Between the likelihood and inlabru methods, inlabru is faster, and although we have only shown a summary of the posterior mean, it provides a distribution of parameters, which may be preferred. Furthermore, the computational times increase for all three methods compared to the previous section, which is reasonable because the number of points generated is larger, although the fastest method continues to be EM, followed by inlabru.

\section{Conclusions and discussion}

This paper provides a comparison of a selection of simulation and estimation techniques for Hawkes point processes in space and time. With the help of the \texttt{R} codes provided, our goal is to serve as a guide for researchers entering the world of Hawkes processes. With this aim in mind, we have conducted an intensive simulation study comparing the performance of two simulation strategies and three inference methods under a variety of practical scenarios in which the triggering functions also vary. Regarding simulation methods, the acceptance-rejection method offers greater flexibility as it allows the implementation of more general Hawkes models with more complex intensity functions, \ab{and the parent–offspring method provides computational efficiency}. In terms of estimation, we conclude that for models involving analytically known and tractable functions, such as those explored in our study, all the methods work well.
In particular, regarding the maximum likelihood estimation method, it is very precise as it directly calculates the parameters by maximizing the complete likelihood function, making it highly suitable when one has a well-specified model and sufficient data, as in our case. However, it is the most computationally expensive method since it requires approximating the integral. \ab{The EM method is often less precise than full maximum likelihood estimation because it optimizes iteratively and its convergence can depend on parameters such as the stopping criterion or initialization. Like other optimization methods, it can converge to a local maximum, although the iterative nature of EM makes it particularly sensitive to this issue.} However, it is the fastest method, making it useful for models that do not require as much precision. Related to Bayesian inference, it provides a richer set of information by going beyond point estimation. This approach generates posterior distributions that include means, medians and other relevant characteristics, fully describing the probabilistic aspects and uncertainties associated with the parameters. For this reason, it is better suited for analyzing complex data or situations where initial conditions or model assumptions are uncertain or difficult to define. It is a good option to achieve a balance between computational time and accuracy.

It is also worth mentioning that there are other methods of statistical inference that have not being here explored. One is the partial likelihood method, as a modification of the likelihood for analyzing survival data. Instead of considering the joint probability of all events, as full likelihood does, the partial likelihood is calculated by conditioning on the time and location of each observed event (\citealt{molher1994,diggle2006,diggle2010partialb,diggle2013statistical,tamayo2014}). Despite its widespread use in survival analysis, the partial likelihood approach has several limitations. Firstly, it may lead to an efficiency loss compared to full maximum likelihood estimation, resulting in less precise parameter estimates and wider confidence intervals. Thus, its limitations in terms of efficiency, computational intensity and reliance on model assumptions make it less suitable in practice.
We also would like to mention the recent work by \citet{JonesTodd2022}, who suggest a Bayesian approach using the \texttt{R} package \texttt{stelfi}, leveraging TMB (Template Model Builder) (\citealt{Kristensen2016}) for fitting Hawkes process models. TMB is a C++ template library developed for efficiently implementing statistical models, with key features including automatic differentiation and smooth integration with \texttt{R}. However, despite its advantages, C++ programming can be challenging for individuals who are not acquainted with the language and integrating with \texttt{R} may pose difficulties for some users. Additionally, the availability of new features in TMB may depend on its development and updates, and the package up to now only addresses a scenario with a particular triggering, limiting its applicability in broader contexts. This is the reason why we have not included this method in our battery of parametric techniques.

Although this work has focused on parametric methods, a possible future line of research is to explore non-parametric methods for the estimation of spatio-temporal Hawkes processes, opening new possibilities for the analysis of data with complex or unknown structures.  \ab{Additionally, further extensions could also involve the consideration of other types of triggering, such as non-separable triggering functions, and the development of non-linear Hawkes process models, which may better capture intricate dependencies and interactions in the data.}





\bibliographystyle{apalike}

\bibliography{references}

\appendix

\section{Pseudocodes of the simulation methods}
\label{pseudo_append}

This appendix presents the pseudocodes for the simulation methods discussed in the body of the paper, the parent–offspring algorithm and the acceptance–rejection algorithm.

\begin{algorithm}[H]
\caption{Spatio-temporal simulation with parents and offspring approach}
\begin{algorithmic}[1]
\STATE \textbf{Initialization:}
   \STATE Initialize an empty list of events
   \STATE Set domains and parameters such as maximum number of iterations and convergence criterion
\STATE \textbf{Generation of parents:}
   \STATE Number of parents $n\sim Po(n_0)$, where $n_0$ is the average rate of the number of parents generated, as defined by the user
        \STATE Assign spatial and temporal locations to these parent events according to a probability distribution set by the user
        \STATE Add the parent events to the list
\STATE \textbf{Generation of offspring:}
        \STATE For each parent:
               \STATE Number of offspring $n_p\sim Po(k)$, where $k$ is the reproduction number
               \STATE Assign spatial and temporal locations to the offspring based on the triggering function
               \STATE Add the offspring to the list
\STATE \textbf{Branching Process:}
   \STATE Initialize a list of events containing all parent and offspring events
   \STATE Define a stopping criterion for the branching process
   \WHILE{the stopping criterion is not met and the maximum number of iterations is not reached}
        \FOR{each event in the list}
               \STATE Determine if the event can trigger the generation of offspring for the next generation
               \STATE If so, follow the offspring generation process
               \STATE Add the new events to the list
        \ENDFOR
   \ENDWHILE
\end{algorithmic}
\label{alg1}
\end{algorithm}

\begin{algorithm}[H]
\caption{Spatio-temporal simulation with acceptance-rejection method}
\begin{algorithmic}[1]
    \STATE \textbf{Generation of homogeneous Poisson point pattern:}
    \STATE \textbf{Input:} Spatial domain $S$, temporal domain $T$, intensity $\lambda_{\text{max}}$
    \STATE Simulate the number of events $n = N(S) \sim Po(\lambda_{\text{max}} \times |S \times T|)$
    \FOR{$i$ in $1$ to $n$}
        \STATE Sample a location $(s_i, t_i)$ uniformly from $S \times T$
    \ENDFOR
    \STATE \textbf{Thinning of a homogeneous Poisson process:}
    \STATE \textbf{Input:} Intensity function $\lambda(s, t)$
    \STATE Define an upper bound $\lambda_{\text{max}}$ for the intensity function
    \STATE Simulate a homogeneous Poisson process with intensity $\lambda_{\text{max}}$
    \STATE Thin the process as follows:
    \FOR{each point $(s, t)$ in the homogeneous Poisson process}
        \STATE Calculate $p = \frac{\lambda(s, t)}{\lambda_{\text{max}}}$
        \STATE Generate a sample $u\sim U(0, 1)$ 
        \IF{$u \leq p$}
            \STATE Retain the location $(s, t)$
        \ENDIF
    \ENDFOR
\end{algorithmic}
\label{alg2}
\end{algorithm}

\end{document}